\documentclass[12pt]{book}
\usepackage{latexsym}
\begin{document}

\title{INTRODUCTION\\ to {\Huge \bf \\EXTENDED\\
\vskip 0.5cm
ELECTRODYNAMICS}}
\author{Stoil Donev}
\date{}
\maketitle
\newpage
\thispagestyle{empty}
{\bf \Large Address of the author:}
\vskip 1cm
{\bf Dr. STOIL G. DONEV}\\
\indent INSTITUTE FOR NUCLEAR RESEARCH AND \\
\indent NUCLEAR ENERGY,\\
\indent BULGARIAN ACADEMY OF SCIENCES,\\
\indent Blvd. TZARIGRADSKO CHAUSSEE 72, 1784 SOFIA,\\
\indent BULGARIA\\
\indent TEL. OFFICE: (+359-2) 7431375; HOME: (+359-2) 732352,\\
\indent E-MAIL: SDONEV@INRNE.ACAD.BG

\newpage
\pagenumbering{roman}
\section{Foreword}
Classical Electrodynamics is probably the most fascinating and complete part
of the Classical Field Theory. Intuition, free thought, perspicuity and
research skill of many years finally brought about the synthesis of experiment
and theory, of physics and mathematics, which we have been calling for short
the {\it Maxwell equations} for the duration of a century and a half. From the
beginning of the second half of the 19th century till its end these equations
turned from abstract theory into daily practice, as they are today. Their
profound study during the first half of the 20th century brought forward a
new theoretical concept in physics known as {\it relativism}. Brave and
unprejudiced workers enriched and widened the synthesis achieved through the
Maxwell equations,
and created a new synthesis known briefly as {\it quantum electrodynamics}.
Every significant scientific breakthrough is based on two things: {\it respect for
the workers and their work and respect for the truth}. "May everyone be respected
as a personality, and nobody as an idol" one of the old workers used to say.
We may paraphrase that saying: "may every scientific truth be respected, but
no one be turned into dogma".

In this book I tried to follow the values this creed teaches, as far as my
humble abilities allow me to. Together with the analysis of the classical
electrodynamics and the quantum concept of the structure of the electromagnetic
field, the path followed brought me to the conclusion that a {\it soliton-like
solution of appropriate non-linear equations characterized by an intrinsic
periodical process is the most adequate mathematical model of the basic
structural unit of the field - the photon}. The fact that neither the Maxwell
equations nor the quantum electrodynamics offer the appropriate tools to find
such solutions, unmistakably emphasizes the necessity to look for new equations.
The leading physical ideas  in this search were the dual ("electro-magnetic")
nature of the field on the one hand and the local Energy-Momentum Conservation
Laws on the other. The realization that every such soliton-like solution
determines in an invariant way its own {\it scale factor}, as well as the
suitable interpretation of the famous formula of Planck for the relation
between the full energy $E$ of the photon and the frequency $\nu=1/T$ of the
beforementioned periodical process, which I prefer to write down as $h=E.T$,
advanced Planck's constant $h$ to the rank
of an estimator, separating the realistic soliton-like models of the photon
from among the rest. The resulting soliton-like solutions possess all integral
qualities of the photon, as described by quantum electrodynamics, but also a
structure, organically tied to an intrinsic periodical process, which in its
turn generates an intrinsic mechanical momentum - spin. I consider this soliton-
like oscillating non-linear wave much more clear and understandable than the
{\it ambiguous " particle-wave " duality}.

The dual 2-component nature of the field predetermined to a great extent the
generalization of the equations in the case of an interaction with another
continuous physical object, briefly called medium. The proposed physical
interpretation of the classical Frobenius equations for complete integrability
of a system of non-linear partial differential equations as a {\it criterium
for the absence of dissipation},
turned out to be relevant and was effectively used. The fruitfulness of the
new non-linear equations is clearly shown by the family of solutions, giving
(3+1)-dimensional interpretation of all (1+1)-dimensional 1-soliton and
multisoliton solutions. I have to say that the use of differential geometry
proved extremely useful.

This book is addressed to all who love theoretical physics and try to build up
their own point of view while in the same time show due respect for others'
opinions. The stress is laid on the conceptual and generic framework, while in
many cases the intermediate calculations were omitted. I did not propose
examples of complete description of actual systems, as this was not my
purpose. A feature of this book is the lack of citations. In my opinion the
reference list in the end will be enough for that purpose.

I would like to express my most sincere gratitude to all my friends and
colleagues, with whom I discussed to one extent or another the issues and
results herein presented. Each and every conversation was extremely valuable
to me both as an actual analysis of the issue and as a stimulus for its
further study. I would highly appreciate any remarks and opinions concerning
this book from anyone interested in the subject matter being analyzed.

\newpage
\section{To the Reader}
{\bf Dear Reader},

Opening the first pages of a new book, devoted to the part of science you like
and want to know as better as you can, the first and most natural question
you ask is: what new shall I learn if I read this book? Having in view the
huge quantity of monographs and other issues, devoted to every of the various
and numerous branches of physics, the natural respect to you requires a short
but sufficiently informative and fair response to this question. And this is
the contents of these preliminary notes. In other words, I'll try to explain
in short what is this {\itshape extended electrodynamics} and what aims to
achieve this extension of the well known from the University electrodynamics
of Faraday and Maxwell.

First of all, let me specify the kind of extension I mean. In theoretical
physics by means of mathematical relations and equations some of the really
existing objects and processes are modelled and described. An equation
separates those values of a given quantity, which are of some interest to that
person, who has written this equation. For example, the elementary equation
$x-1=0$ separates the value $1$ of the variable $x$. If, for some reasons, we
want to separate two values of the variable $x$, e.g. $(1,-1)$, we,
naturally, write down a new equation, which has two solutions: $1$ and
$(-1)$. The new equation will be $x^2-1=(x+1)(x-1)=0$. In this way we
extended the set of admissible values of the variable $x$, or we may say that
we have extended the equation $x-1=0$ to the equation $x^2-1=0$. The idea is
now quite clear:  {\itshape we write down new (algebraic or differential)
equation, whose solutions comprise as a subclass all solutions of the old
equation, but have also new ones}. The important point is the new solutions
to have the properties, desired by us.

The basic equations of Classical electrodynamics are the well known equations
of Faraday-Maxwell. So when we talk about {\itshape extended electrodynamics}
it is clear that just these equations we have to extend in the sense,
mentioned above. In order to motivate such an extension we should honestly
say two things: which properties of Maxwell equations we do not like
sufficiently, and what new properties will be required from the new solutions
of the extended equations. Let me briefly discuss this matter.

As it is well known, at zero electric current Maxwell's equations require the
components of the electric and magnetic vectors to satisfy the wave equation
(d'Alembert's equation). The solutions of this equation in the whole 3-space
have the following 2 properties: first, every solution can be represented by
a sum (finite or infinite) of a subclass of solutions, known as {\itshape
plane waves}; second, every solution, defined by some finite localized
and smooth enough initial condition, "blows up" radially and goes to infinity
with the velocity of light. The plane waves are characterized by the
condition, that there exists a system of canonical coordinates on
${\cal R}^3$ in which these solutions depend on one space variable only. This
means, that at every moment every such solution occupies the whole 3-space, or
an infinite its subregion. Therefore, they are {\itshape infinite} and their
integral energy is infinitely large. We note that this infinity does not come
from some singularity of the solutions. We set the question: do there exist
in Nature electromagnetic fields with properties adequate to these {\itshape
exact} solutions called 'plane waves'? The natural answer is negative, since
in order to create such an object it will be necessary to transform an
infinite quantity of some kind of energy into electromagnetic energy, which
will take infinite period of time.

On the other hand, according to the second property, every finite solution is
strongly time-unstable, so it could hardly be considered as a realistic
model of a real object. As for the static case, the components of the
electric and magnetic vectors satisfy the Laplace equation and, as it is well
known the solutions of this equation in the whole 3-space are singular, or
when they are finite and nonsingular, they are constant.We conclude, that the
vacuum Maxwell equations can not describe finite and time-stable, i.e.
{\itshape soliton-like} electromagnetic formations. And {\itshape our purpose
is to describe just such soliton-like electromagnetic formations}.

If we now recall the basic idealization of classical mechanics, the material
point, we'll see that it is a full antipode of the plane wave: the material
point {\itshape has no structure, occupies zero volume} and has finite
energy. Are there such objects in Nature? According to my opinion the real
objects are finite, i.e. at every moment of its existence every really
existing object occupies finite, comparatively small volume and it has
definite properties of constancy and stability. When they are subject to
external perturbations they survive or transform into other objects, obeying
definite conservation laws. Usually, a survival is connected with some change
in the behavior of the object as a whole, but it also keeps some of its
basic characteristics unchanged, otherwise, we could not say that the object
is the same, i.e its identification after the perturbation should be
possible.

At the beginning of our century it has become clear that the electromagnetic
field has a discrete nature, i.e. it consists of many non-interacting (or
very weakly interacting) objects, carrying energy, momentum and intrinsic
angular momentum. Moreover, {\itshape the integral energy of these objects
originates intrinsically from a periodic process with frequency of $\nu$} in
correspondence with the Planck's formula $E=h\nu$. The experiment shows
soon that these objects are finite, they move as a whole with the velocity of
light along straight lines, carry momentum $h\nu/c$ and intrinsic angular
momentum $h=ET=E/\nu$. The problem to describe such (free) objects, called
later photons, appeared. Because of the finite nature and time-stability
of photons it
is clear that the solutions of the wave equation can not give good enough
mathematical models of them. On the other hand, the material point (or
particle) can not also serve as a model, since there is no any sensible way
to assign the characteristic {\itshape frequency} to a free particle.
The frequency always comes from an outside elastic force, so it depends
strongly on this force and is not a proper characteristic of the particle.
The classical free particle moves always along straight lines with a constant
velocity. There have been some unsuccessful efforts to build an extended model
of the photon, but pressed by the quickly advancing experiment, the major
part of the physicists assume the prescription schemes of quantum
electrodynamics. A basic assumption in these schemes is that the photon is a
point-like object, so its frequency (or spin) is an integral characteristic
of the same kind as the proper mass and the electric charge. I do not share
this point of view, since {\itshape a free structureless object can not carry
the physical characteristic of intrinsic angular momentum and no periodic
process can be associated with its existence}. I think that {\itshape a
periodic process and an intrinsic angular momentum can be associated with a
free object only if it has structure}, and these characteristics may have
finite values only if the object is finite. The photon moves uniformly as a
whole, but it is not a point-like (or structureless) object, and its
existence is strongly connected with some intrinsic periodic process with a
constant frequency. This process occurs in the whole volume and in this way
generates the intrinsic angular momentum. Since at this stage we do not see
any other more-natural way to combine the known photon's features than the
notion of soliton-like objects, we turn to appropriate (3+1)-soliton-like
solutions of a definite system of nonlinear partial differential equations as
possible mathematical models.

As it was mentioned earlier, Maxwell's equations have no such solutions. On
the other hand, as an working tool in macrosituations they have proved their
wide applicability, so it does not seem reasonable to leave them off fully
and to search for new equations. Such a step would reject also the
corresponding well proved energetic relations for finite volumes, obtained by
these equations. Therefore an appropriate their extension seems natural and
reasonable provided that the new incorporated solutions have the desired
properties.

In the non-vacuum case, i.e. in the presence of energy-momentum exchange
between the field and some medium, there are also problems coming mainly from
the {\itshape hypothesis} that the Faraday induction law is valid for
{\itshape all} media, which leads to the notion that the field is able to
exchange energy-momentum {\itshape only with electrically charged particles}.
No doubt, such media exist, but the assumption that this is true for all
media, we do not accept. Further, the usual way to treat the field through
introducing the {\itshape polarization and magnetization} vectors and to
describe the media by means of corresponding permeability tensors and bound
currents and so on, may be good operational skills, but they do not lay on
approved physical principles and do not generate new ideas. In this case, as
well as in the vacuum case, we choose a different way, namely we extend the
equations in such a way, that to keep everything, that classical
electrodynamics is able to do, and to incorporate new solutions with good
enough properties. Let's say quite clear, that our purpose in this case is
the same as in the vacuum case, namely, {\itshape to describe soliton-like
field configurations with sufficiently clear physical interpretation}.

In the both cases, the leading idea for extension of the equations is an
analysis and appropriate formulation of the local energy-momentum
conservation laws in relativistic terms. In result we obtain {\itshape
nonlinear} equations. In the continuous medium case an additional and
entirely new moment is considered, namely, a physical interpretation of the
Frobenius integrability equations for Pfaff systems. Also, the classical
concept of a continuous medium is extended and understood as an arbitrary
continuous physical system, exchanging energy-momentum with the field. The
encouraging results we obtain in the both cases give us reasons for hope that
we are on the right way.

Finally, everyone, who decides to read this book, will follow the complete
version of the author's mental way and will get to the results obtained
following an easy and smooth way of reasoning. The real road, that I passed
through, was very different from what is given here. There were many turns,
unexpected traps and various positive and negative surprises, and all this
was taking place, when the fashion in theoretical physics was quite
different.  The deep belief in the conservation laws, in their universal
character, was the point of light, that was leading me through the jungle of
unknown possibilities and hard to evaluate hypotheses and helped me to
withstand the falling on one after another fashionable topics. Writing this
book I was doing my best to pay a maximum respect to every reader, to every
positive opponent. And, in order not to appear any doubts about my great
respect to the prominent men, who created Classical electrodynamics, I
would like to give here the creed that was the basic stimulus during this
period:  {\itshape the respect and esteem paid to the creators can not be
honest and genuine if they are not in correspondence with the respect and
esteem paid to the Truth}.

Now, {\bf Dear reader}, after you have got some idea of what could be
learned from this book, and if you have already made up your mind to become
well acquainted with the small harmonic world I tried to create, turn over
this page and be my fair corrective up to the last line of the last page.

\chapter {\Large \itshape From Classical Electrodynamics\\
to Extended Electrodynamics}
==========================================
\pagenumbering{arabic}

\section {\bf Basic Notions and Equations of \\
Classical Electrodynamics}

\subsection{Nonrelativistic approach}
In the mathematical model of the electromagnetic phenomena, called shortly
CLASSICAL ELECTRODYNAMICS (CED), the Newton's notions of space and time are
assumed, namely, a mathematical model of the real space is the real
3-dimensional mathematical space ${\cal R}^3$, considered as a 3-dimensional
Euclidean manifold with the standard Euclidean metric $g$, and a mathematical
model of the real time is the 1-dimensional mathematical space ${\cal R}$.
The standard orthogonal coordinates $(x,y,z)$ are usually used, so that the
metric tensor $g_{ik}$ has canonical components, i.e. the diagonal elements
are equal to 1, and all other components are zero.
Then the isomorphism between the tangent vectors and the 1-forms, defined by
$g$ and denoted by the same letter $g$, (or $g^{-1}$), looks as follows:
\[
g(V)=g\biggl(V^i \frac {\partial}{\partial x^i}\biggr)=
V^i g_{ik}dx^k=V_kdx^k,
\]
where $V_k=g_{ki}V^i$. This isomorphism is extended naturally (i.e.
component-wise) for arbitrary tensors (or tensor fields).
The metric $g$ defines also a volume element:
\[
\omega_0=\sqrt{|det g_{ik}|}dx\wedge dy\wedge dz=dx\wedge dy\wedge dz,
\]
a covariant derivative $\nabla$, which in standard coordinates reduces to the
usual derivative, and the Hodge $*$-operator
$*:\Lambda^p\bigl({\cal R}^3\bigr)\rightarrow\Lambda^{3-p}\bigl({\cal R}^3\bigr)$
according to the rule
\[
\alpha \wedge \beta = g(*\alpha,\beta)\omega_0,
\]
where $\alpha $ and $\beta $ are $p$ and $(3-p)$ forms respectively.
The operator $*$ is linear and in the canonical orthonormal basis \{dx,dy,dz\}
from the above defining relation we obtain
\[
*1=dx\wedge dy\wedge dz,\ *dx=dy\wedge dz,\  *dy=-dx\wedge dz,\  *dz=dx\wedge dy,
\]
\[
*dx\wedge dy=dz,\ *dx\wedge dz=-dy,\ *dy\wedge dz=dx,\ *dx\wedge dy\wedge dz=1.
\]
\noindent We note that the defining relation for the $*$ operator is
equivalent to the following: if $\alpha $ and  $\beta $ are $p$-forms then
\[
\alpha\wedge *\beta=g(\alpha,\beta)\omega_0.
\]
If {\bf d} is the {\itshape exteriour derivative}, then the operator
$\delta=(-1)^p*^{-1}{\bf d}*$ is called {\itshape coderivative} and we get
$(\delta \alpha)_{ik...}=-\nabla_j \alpha^j_{.ik...}$.  It is easily checked,
 that the operator $\delta$ is the dual to {\bf d} in the vector space of
forms with compact support with respect to the inner product
$(\alpha,\beta)=\int{\alpha\wedge *\beta}$, and for forms with different
degree is assumed $(\alpha,\beta)=0$.

\vskip 0.05cm
From physical point of view CED assumes the following:

1. The Electromagnetic (EM) field is a real object (or a set of real objects)
possessing energy and momentum.

2. The EM-field is able to interact (i.e. to exchange energy and momentum)
with material particles, possessing the characteristic {\it electric charge}
which takes positive, negative and zero values. Such charged particles are
called {\it field sources}.

3. The sources are called

a){\it free} - if under the action of the external field they can move throughout
the whole space, and

b){\it bounded} - if their motions are bounded inside comparatively small space
regions.

In order to describe the EM-field CED introduces the following quantities:

I. A couple of vector fields $(E,B)$, defined on  ${\cal R}^3$,
and a parametric dependence of $(E,B)$ on the time $t$ is admitted.

II. A scalar function $\rho (x,y,z,t) $, called {\itshape charge density},
such that the integral $\int{\rho\omega_0}$, computed over the region
$V\subset {\cal R}^3$, gives the entire electric charge in $V$.

III. A vector field ${\bf v}(x,y,z,t)$, describing the mechanical motion of
the charge-carriers, and the vector field ${\bf j}=\rho{\bf v}$
is called {\it electric current}.
\vskip 0.5cm
\noindent {\it Remark}. Further a vector field $V$ and the corresponding
1-form  $g(V)$ will be denoted by the same latter, since from the context it
will be clear if the object is tangent or co-tangent.

\vskip 0.5cm
An important concept in CED is the so called {\it flow of a vector field}
$U$ through a given 2-dimensional surface $S$. By definition, it is the
integral of the 2-form $*U$ over $S$: \ $\int_S *U$. It is important if the
surface is closed (most frequently homeomorphic to the 2-sphere $S^2$), or
it is not closed and has for a boundary  a given 1-dimensional manifold.

As a generalization of the experimental data in CED is assumed the following:

\vskip 0.3cm
$1^\circ $. {\it The flows of the electric $E$ and magnetic $B$ fields
through a closed 2-surface $S^2$, surrounding the 3-volume $V$ are defined
as follows}:

\[
\oint_{S^2} *E =4\pi \int_V \rho\omega_0,\quad \oint_{S^2} *B=0.
\]
\indent$2^\circ $. {\it If the 2-surface $S$ is not closed and its boundary
is the closed curve $l$, then the following relations hold}:

\[
\frac {d}{dt}\int_S*E=c\int_lB-
4\pi\int_S*{\bf j},\quad \frac {d}{dt}\int_S*B=-c\int_lE.
\]
\vskip 0.3cm
From these relations, making use of the notations $g^{-1}*{\bf d}g=rot$,
\linebreak $g^{-1}\delta g=div$ and the Stokes' formula
$$
\int_S{\bf d}\alpha = \int_l\alpha
$$
\noindent we get the differential equations
\begin{equation}
\frac 1c \frac{\partial E}{\partial t}=rotB - \frac {4\pi}{c}{\bf j},
\quad divB=0, 
\end{equation}
\begin{equation}
\frac 1c \frac{\partial B}{\partial t}=-rotE,\quad \ \ divE=4\pi\rho. 
\end{equation}
Because of the identity $ div_\circ rot=0$, from the first equation of (1.1)
and the second equation of (1.2) it follows the {\it continuity equation}
\begin{equation}
\frac {\partial \rho}{\partial t}=-div{\bf j},   
\end{equation}
\noindent the sense of which becomes clear from its integral form
\[
\frac {d}{dt}\int_V \rho\omega_0=\oint_{S_V} *{\bf j}.
\]
This relation shows, that any change of the electric charge inside the region
$V$ is caused by some processes of charge transport through the boundary of
$V$, i.e. {\it charges do not vanish and do not arise}.

Using the scalar product $g$ from the first equation of (1.1) and the first
equation of (1.2) we obtain
\[
\frac 1c g\biggl(E,\frac {\partial E}{\partial t}\biggr)+
\frac 1c g\biggl(B,\frac {\partial B}{\partial t}\biggr)=
g(E,rotB)-g(B,rotE)-\frac {4\pi}{c}g(E,{\bf j}).
\]
Since $g$ does not depend ot time and
\[
g(E,rotB)-g(B,rotE)=-div\bigl(E\times B\bigr)
\]
we get
\[
\frac {\partial}{\partial t} \frac {E^2+B^2}{8\pi}=
-{\bf j}.E-div{\bf S},
\]
where the vector
\[
{\bf S}=\frac {c}{4\pi} E\times B
\]
is the Poynting vector. This last relation describes the local balance
of the energy and momentum in the system
$EM$-{\it field and free current} in an unit space-time volume.

The equations (1,1), (1,2) are linear, so that if $(E_1,B_1)$ and $(E_2,B_2)$
are two solutions, then any linear combination
\[
E=aE_1 +bE_2,\  B=mB_1 +nB_2
\]
with constant coefficients is again a solution. The following question arises
naturally: do there exist constants $(a,b,m,n)$, such that the linear
combination
\[
E'=aE+bB,\ B'=mE+nB
\]
is again a solution? The answer to this question is positive
iff $m=-b, n=a$. The new solution will have energy density $w'$ and momentum
${\bf S'}$ as follows:
\[
w'=\frac {1}{8\pi}\biggl((E')^2 + (B')^2\biggr)=
\frac {1}{8\pi}(a^2 + b^2)\biggl(E^2 + B^2\biggr),
\]
\[{\bf S'}=(a^2+b^2)\frac {c}{4\pi} E\times B.
\]
Obviously, the new and the old solutions will have the same energy and
momentum if $a^2+b^2 =1$.

These observations show that besides the usual linearity, Maxwell's
equations admit also "cross"-linearity, i.e. linear combinations of $E$ and
$B$ of a definite kind define new solutions. Therefore, the difference
between the electric and magnetic fields becomes non-essential. The important
point is that with every solution $(E,B)$ of Maxwell's equations a
2-dimensional real vector space, spanned by the couple $(E,B)$, is
associated, and the linear transformations, which transform solutions into
solutions, are given by matrices of the kind
\[
\left\|\matrix{a   &b\cr
	       -b  &a\cr} \right\|.
\]
If these matrices are unimodular, i.e. if $a^2+b^2=1$, then the initial and
the transformed solutions have the same energy and momentum. It is well known
that matrices of this kind do not change the canonical complex structure
$J$ in ${\cal R}^2$: if the canonical basis of ${\cal R}^2$ is denoted by
$(e_1,e_2)$  then $J$ is defined by $J(e_1)=e_2$, $J(e_2)=-e_1$.

The above remarks suggest to consider $E$ and $B$ as two vector-components
of an ${\cal R}^2$-valued 1-form $\Omega$:
\[
\Omega=E\otimes e_1 + B\otimes e_2.
\]
So, the current $ {\bf j}$ becomes 1-form ${\cal J}={\bf j}\otimes e_1$ with values in
${\cal R}^2$, and the charge density becomes a function ${\cal Q}=\rho\otimes e_1$
with values in ${\cal R}^2$. Maxwell's equations take the form

\begin{equation}
\frac {1}{c} \frac {\partial \Omega}{\partial t}=
-\frac {4\pi}{c}{\cal J}-*{\bf d}J(\Omega),\quad
\delta \Omega=4\pi {\cal Q}, 
\end{equation}
\noindent where
$J(\Omega)=E\otimes J(e_1)+B\otimes J(e_2)=E\otimes e_2-B\otimes e_1$.
Note that according to the sense of the concept of current in
CED and because of the lack of magnetic charges, it is necessary to exist
a basis of ${\cal R}^2$, in which ${\cal J}$ and ${\cal Q}$ to have components
only along $e_1$. Nevertheless, this point of view shows that even at this
non-relativistic level the separation of the EM-field to "electric" and
"magnetic" is not adequate to the real situation.

\vskip 0.5cm
\subsection{Relativistic formulation}
We pass now to the
relativistic formulation of CED. We begin with the note that the relativism,
considered as a theoretical conception for understanding and modeling
the natural entities and processes, arises as a result of the analysis of
the invariance properties of Maxwell's equations with respect to linear
transformations of space-time coordinates $(x,y,z,\xi=ct)$. The assumption
for the linear character of the space-time transformations comes on the one
side from the linearity of the model space ${\cal R}^4$ and Maxwell's
equations, and on the other side, from the idea, that any straight-line
uniform motion should not affect the character and the course of {\it all}
physical processes. The latter formally means, that the parameters of the
admissible space-time transformations, interpreted as straight-line uniform
motions, should be determined by the relative constant velocity between two
frames. So the cartesian framings in ${\cal R}^4$ model the physical inertial
frames. The basic conclusion of the analysis carried out during the end of last
and the beginning of the current century consists in, that the invariance of
Maxwell's equations requires a {\it pseudo-eucledean space-time metric tensor
$\eta$
and, additionally, any component of E and $B$ depends linearly on all
components of $E$ and $B$ when a space-time transformation is performed.} (In
connection with this we note, that the "cross"-linearity mentioned above
admits "mixing" only for corresponding components of $E$ and $B$: $E_1$ with
$B_1$, $E_2$ with $B_2$ and $E_3$ with $B_3$.) This undoubtedly shows, that the
adequate mathematical object, describing the $EM$-field, must have 6 independent
components and its transformation properties should be determined entirely
by the admissible space-time transformations. In view of the 4-dimensionality
of space-time such objects are only the antisymmetric tensor fields of second
order. Because of the possible identification of the contravariant and
covariant tensor fields by means of  the pseudo-metric tensor  $\eta_{\mu \nu}$,
the obvious candidates for models of the $EM$--field are the
{\it differential  2--forms}.
So, starting with the 3-dimensional vector fields
$E$ and $B$ and with the pseudoeucledean structure  $\eta_{\mu\nu}$
of the model space  ${\cal R}^4$ we have to build a differential 2-form
$F\in \Lambda ^2 ({\cal R}^4 ,\eta)$. Further, the space-time
$({\cal R}^4 ,\eta)$, where $\eta _{\mu\nu}=-1$ for $\mu=\nu=1,2,3$;
$\eta_{44}=1$, and $\eta_{\mu\nu}=0$ for $\mu \neq \nu$
in standard coordinates $(x^1,x^2,x^3,x^4)=(x,y,z,\xi =ct)$,
will be denoted by $M$ and will be called Minkowski space. The pseudo-metric
$\eta_{\mu\nu}$ defines in the same way a volume element $\omega_0$
and the Hodge $*$-operator. We have
\[
\omega_0 =\sqrt {|det \eta_{\mu\nu}|} dx\wedge dy\wedge dz\wedge d\xi,
\]
\[
\alpha\wedge \beta =\eta (*\alpha,\beta)\omega_0\ \iff\ \alpha\wedge *\beta=
-\eta (\alpha,\beta)\omega_0,
\]
\[
**_p=-(-1)^{p(4-p)}id,\ *^{-1}_p=-(-1)^{p(4-p)}*_p,\ *\omega_0=1,\ *1=-\omega_0,
\]
\[
\begin{array}{ll}
*dx=dy\wedge dz\wedge d\xi           &*dx\wedge dy\wedge dz=d\xi \\
{*}dy=-dx\wedge dz\wedge d\xi        &*dx\wedge dy\wedge d\xi=dz \\
{*}dz=dx\wedge\ dy\wedge d\xi        &*dx\wedge dz\wedge d\xi=-dy \\
{*}d\xi=dx\wedge dy\wedge dz         &*dy\wedge dz\wedge d\xi=dx
\end{array}
\]
\[
\begin{array}{ll}
*dx\wedge dy=-dz\wedge d\xi          &*dy\wedge dz=-dx\wedge d\xi \\
{*}dx\wedge dz=dy\wedge d\xi         &*dy\wedge d\xi=-dx\wedge dz \\
{*}dx\wedge d\xi=dy\wedge dz         &*dz\wedge d\xi=dx\wedge dy.
\end{array}
\]
\indent An arbitrary 2-form $F$ on $M$ in standard coordinates looks as
follows:
\begin{eqnarray}
F={1\over 2} F_{\mu\nu}dx^\mu\wedge dx^\nu=F_{12}dx\wedge dy+F_{13}dx\wedge dz+F_{23}dy\wedge dz+ \nonumber \\
+F_{14}dx\wedge d\xi+F_{24}dy\wedge d\xi+F_{34}dz\wedge d\xi.\nonumber
\end{eqnarray}
\noindent Then for $*F$ we obtain
\begin{eqnarray}
*F=-\frac 12 \varepsilon_{\mu\nu\sigma\tau}F^{\sigma\tau}dx^\mu\wedge dx^\nu=F_{34}dx\wedge dy-F_{24}dx\wedge dz+F_{14}dy\wedge dz-\nonumber \\
-F_{23}dx\wedge d\xi+F_{13}dy\wedge d\xi-F_{12}dz\wedge d\xi. \nonumber
\end{eqnarray}
The definition of the components  $F_{\mu\nu}$ by means of the components of
$E$ and $B$ is made in the following way.
Let  $i$:${\cal R}^3\rightarrow {\cal R}^4$ be the standard immersion
$(x,y,z)\rightarrow (x,y,z,0)$. Then we define $i^*F$ and $i^*(*F)$ by
\begin{equation}
i^*F=*B,\quad i^*(*F)=*E,     
\end{equation}
\noindent where on the right-hand side of these equalities the Eucledean
$*$--operator is used. Relations (1,5) define $F$ uniquely, and we get
$$
F_{12}=B_3,\ F_{13}=-B_2,\ F_{23}=B_1,\ F_{14}=E_1,\ F_{24}=E_2,\ F_{34}=E_3.
$$
Recalling that in the static case with zero current Maxwell's equations reduce
to ${\bf d}E=0,\ {\bf d}*E=0,\ {\bf d}B=0, \ {\bf d}*B=0$,  and the well
known relation ${\bf d}i^*=i^*{\bf d}$, we obtain for this static case
$i^*{\bf d}F=0,\ i^*{\bf d}*F=0$. The map $i^*$ cancels all terms where $d\xi$
stays. Removing this map $i^*$, we get the equations ${\bf d}F=0,\ {\bf d}*F=0$,
so we keep all terms with $d\xi$ non-canceled, and
having in view the above component interpretation of  $F_{\mu\nu}$
we obtain exactly the left-hand sides of Maxwell's equations (1,2), (1,1)
respectively. We introduce now the 4-current $j^\mu$ by  $j^\mu=\rho u^\mu$,
where $u^\mu$ is the 4-velocity of the charged particles. Maxwell's equations
take the form
\begin{equation}
{\bf d}F=0,\quad {\bf d}*F=4\pi*j.                      
\end{equation}
These equations (1.6) may be written in various forms:
\[
\delta *F=0,\quad \delta F=4\pi j
\]
\[ \frac {\partial F_{\mu\nu}}{\partial x^\sigma}+\frac {\partial
F_{\nu\sigma}}{\partial x^\mu}+\frac {\partial F_{\sigma\mu}}{\partial
x^\nu}=0,\quad \nabla_{\sigma}F^{\sigma\nu}=-4\pi j^\nu.
\]
Of definite importance for the theory is the quantity
\begin{equation}
Q_\mu^\nu=\frac {1}{4\pi}\biggl[\frac 14
F_{\alpha\beta}F^{\alpha\beta}\delta_\mu^\nu-F_{\mu\sigma}F^{\nu\sigma}\biggr]=\\
\frac {1}{8\pi}\biggl[-F_{\mu\sigma}F^{\nu\sigma}-
(*F)_{\mu\sigma}(*F)^{\nu\sigma}\biggr] 
\end{equation}
\noindent since on the solutions of (1.6) the following relation holds:

\begin{equation}
\nabla_\nu Q_\mu^\nu=\frac {1}{4\pi}\biggl[F_{\mu\nu}(\delta F)^\nu+
(*F)_{\mu\nu}(\delta *F)^\nu\biggr]= F_{\mu\nu}j^\nu. 
\end{equation}
This relation describes the local energy-momentum balance in the system
$EM$--{\it field and free current}. The quantity $F_{\mu\nu}j^\nu$ is the Lorentz
force and it determines the energy-momentum, which the charge carriers get
from the field in an unit 4-volume. In regions with zero current $j^\mu =0$
we have $\nabla_\nu Q_\mu^\nu=0$, so we may create integral conserved
quantities. Because of its importance for the theory we shall consider this
point more in detail.

In classical field theory we build integral conserved quantities, i.e.
time-independent quantities, by means of a symmetric second rank tensor
$Q_{\mu\nu}$ with zero divergence $\nabla_\nu Q_\mu^\nu=0$
by making use of isometries, i.e. symmetries of the metric tensor,
in the following way. By definition a symmetry of a tensor field $g$ is a map
 $f:M\rightarrow M$, which keeps this tensor field unchanged: $f^* g=g$.
When an one-parameter group of symmetries $f_t$, defined by the generator $X$,
(or the vector field X) is given, then $X$ is called {\it local symmetry} of
$g$ and the following relation holds
$$
L_X g=lim_{t\rightarrow 0}\frac {f_t^* g-g}{t}=0,
$$
which means that the local symmetries of $g$ are those vector fields $X$ along
the integral lines of which $g$ stays unchanged. The expression on the right
is called {\it Lie-derivative} of $g$ along $X$. The local symmetries of the
metric tensor are also called {\it Killing} vector fields. The equation
$L_X g=0$, where $g$ is given, looks as follows
\[
\nabla_\mu X_\nu+\nabla_\nu X_\mu=0,
\]
where $\nabla$ is the corresponding symmetric Riemannean connection.
If now $Q_{\mu\nu}$ is a conservative tensor field, i.e.
$\nabla_\nu Q_\mu^\nu=0$, and $X$ is a local isometry, we obtain
\[
\nabla_\nu (Q_\mu^\nu X^\mu)=(\nabla_\nu Q_\mu^\nu)X^\mu+
Q^{\mu\nu}\nabla_\nu X_\mu =Q^{\mu\nu}\nabla_\nu X_\mu.
\]
Because of the symmetry of $Q$, in the sum $Q^{\mu\nu}\nabla_\mu X_\nu$
only the symmetric part of $\nabla_\mu X_\nu$ may contribute, but this symmetric
part is zero since $X$ is a local isometry.
In this way with every local isometry $X$ of the metric the 1-form
 $Q_{\mu\nu}X^\mu dx^\nu$ is associated, and this 1-form has zero divergence.
The last means that the 3-form $*(Q_{\mu\nu}X^\mu dx^\nu)$ is closed, so
according to the Stokes theorem, the integral of this 3-form over ${\cal R}^3$
will not depend on time. Of course, these considerations make sense only for
finite valued such 3-integrals, i.e. for {\it finite} field functions.

In order to complete the energy-momentum balance picture we have to point out
how the charged particles of the 4-current use the gained from the field
energy-momentum to change their behaviour, i.e. we
have to write down the equations of motion of these charge carriers.

Assuming that only interaction between the particles and the field takes
place, the energy-momentum tensor of the particles is defined by
\[
K_\mu^\nu=\mu_o c^2 u^\nu u_\mu,
\]
where $\mu_o$ denotes the invariant mass density of the particles. So, the
full local energy-momentum conservation law requires
$\nabla_\nu (Q_\mu^\nu +K_\mu^\nu)=0$.
Since particles do not vanish and do not arise, which formally means that the
mass continuity equation $\nabla_\nu (\mu_o u^\nu)=0$ holds, we obtain
\begin{equation}
\mu_o c^2 u^\nu\nabla_{\nu} u_\mu=-F_{\mu\nu}j^\nu.    
\end{equation}
\noindent This equation for the vector field $u$ describes the mechanical
evolution of the charge-carriers. We note that this is a compact form of a
{\it nonlinear} system of partial differential equations for the components
of $u$, while Maxwell's equations are linear. This shows that the current
$j_\mu=\rho u_\mu$ {\it cannot be defined independently}, i.e. it strongly
depends on the field $F$. Therefore, the variational procedure for verifying
Maxwell's equations with "given and not subject to variation" current does not
seem to be quite correct. Otherwise, we lose the full energy-momentum
conservation law, which is hardly preferable.

\vskip 0.5cm
\noindent {\it Remark}.\ It is clear that the particles with mass distribution
$\mu_o$ and charge distribution $\rho$, are considered as {\it sources} of the
field $F$ according to the usual interpretation of equations (1.6) amd (1.7),
and, therefore, they {\it cannot be considered as test particles}, i.e. as not
disturbing the field. If the field is not disturbed, i.e. if it does not exchange
energy-momentum with the particles, then $\nabla_\nu Q_\mu^\nu=0$
and the corresponding solution of the field equations, which is meant to
define "Lorentz force", has to satisfy the current-free Maxwell equations
${\bf d}F=0,\ \delta F=0$, as in the static Coulomb case. In this case the
dynamical equations for the particles appear as an additional assumption, and
talking about {\it full} energy-momentum conservation law is hardly sensible
from the point of view of the field, nevertheless, it may get some sense from
the point of view of the particles.
\vskip 0.5cm
We consider now the {\it conformal invariance} of pure field Maxwell equations.
In accordance with (1.6) if $j=0$ we get
\begin{equation}
{\bf d}F=0,\quad {\bf d}*F=0.
\end{equation}
The only factor not permitting a full invariance of these equations is
the $*$-operator, more exactly, its restriction on 2-forms. On a 2n-dimensional
riemannian manifold the restriction of $*$ on $n$-forms is always conformally
invariant because 2 conformal metrics $g$ and $\tilde{g}=f^2 g, f(x)\neq 0,
x\in M$, generate the same $*_n$ operator. In our case $n=2$ and $g=\eta$, so
\[
\tilde{*}F=\frac 12 F_{\mu\nu}\tilde{*}(dx^\mu\wedge dx^\nu)=
-\frac 12 F_{\mu\nu}\tilde{\eta}^{\mu\sigma}\tilde{\eta}^{\nu\tau}\varepsilon_{\sigma\tau\alpha\beta}\sqrt{|det\tilde{\eta}_{\rho\kappa}|} dx^{\alpha}\wedge dx^{\beta}=
\]
\[
-\frac 12 F_{\mu\nu}f^{-4}\eta^{\mu\sigma}\eta^{\nu\tau}\varepsilon_{\sigma \tau\alpha\beta}f^4 \sqrt{|det\eta_{\rho\kappa}|}dx^{\alpha}\wedge dx^{\beta}=*F.
\]
If we recall the expression (1.7) for the energy-momentum tensor, we'll see
that the $*$-operator is applied there also on 2-forms only, which shows, that
it is conformally invariant too. Moreover, the expressions (1.7), (1.8)
clearly show, that the following relations hold:
\begin{equation}
Q_{\mu}^{\nu}(F)=Q_{\mu}^{\nu}(*F),\quad \nabla_{\nu}Q_{\mu}^{\nu}(F)=
\nabla_{\nu}Q_{\mu}^{\nu}(*F).
\end{equation}
These two equalities determine a full equivalence from energetic point of
view between $F$ and $*F$. This important fact will be substantially used
later when we'll be writing down the new equations of Extended
Electrodynamics (EED).

From pure algebraic point of view the $EM$-field, i.e. the 2-form $F$, has
two invariants

\begin{equation}
I_1=\frac 12 F_{\mu\nu}F^{\mu\nu}=B^2-E^2,\quad I_2=\frac 12 F_{\mu\nu}(*F)^{\mu\nu}=2B.E .
\end{equation}
These are the coefficients ( up to a sign) of the two 4-forms
\[
F\wedge *F=-\eta (F,F)\omega_{\circ}=-\frac 12 F_{\mu\nu}F^{\mu\nu}\omega_{\circ},
\]
\[
F\wedge F=-F\wedge **F=\eta(F,*F)\omega_{\circ}=\frac 12 F_{\mu\nu}(*F)^{\mu\nu}\omega_{\circ}.
\]
The following relations hold
\[
(4\pi)^2Q_{\mu\nu}Q^{\mu\nu}=I_1^2+I_2^2,\quad (4\pi)^2Q_{\mu\sigma}Q^{\nu\sigma}=\frac 14 \bigl[I_1^2+I_2^2\bigr]\delta_{\mu}^{\nu}.
\]
The equations for the eigen values of $F$ and $*F$
\[
det\|F_{\mu\nu}-\lambda \eta_{\mu\nu}\|=0,\quad det\|(*F)_{\mu\nu}-\lambda^* \eta_{\mu\nu}\|=0
\]
look as follows
\[
\lambda^4 +I_1 \lambda^2-\frac14 I_2^2=0,\quad (\lambda^*)^4-I_1(\lambda^*)^2-\frac14 I_2^2=0.
\]
For the corresponding eigen values we obtain
\[
\lambda_{1,2}=\pm \sqrt{-\frac12 I_1 +\frac12 \sqrt{I_1^2+I_2^2}} ,\quad
\lambda_{3,4}=\pm \sqrt{-\frac12 I_1 -\frac12 \sqrt{I_1^2+I_2^2}} ,
\]
\[
\lambda^*_{1,2}=\pm \sqrt{\frac12 I_1 +\frac12 \sqrt{I_1^2+I_2^2}} ,\quad
\lambda^*_{3,4}=\pm \sqrt{\frac12 I_1 -\frac12 \sqrt{I_1^2+I_2^2}} .
\]
Multiplying the equation $F^{\mu}_{.\nu} \xi^\nu=\lambda\xi^\mu$ on the left
by $-F_\mu^{.\sigma}$ and adding on the two sides of the equation obtained
$\frac14 F_{\alpha\beta}F^{\alpha\beta}\delta_\mu^\nu$, we get
\[
Q^\mu_\nu\xi^\nu =\gamma \xi^\mu,\quad \gamma=\biggl[\frac14 F_{\alpha\beta}F^{\alpha\beta}+\lambda^2\biggr]=\biggl[\frac12I_1+\lambda^2\biggr].
\]
This shows, that the eigen vectors of $F$ are eigen vectors of $Q$ too, and
the eigen values $\lambda$ of $F$ and $\gamma$ of $Q$ are related by the
above condition. The corresponding relation between $\gamma$ and $\lambda^*$
reads
\[
\gamma=\biggl[\frac 14 (*F)_{\alpha\beta}(*F)^{\alpha\beta}+
(\lambda^*)^2\biggr]=\biggl[-\frac12 I_1+(\lambda^*)^2\biggr].
\]
As for the eigen vectors of $F$, $*F$ and $Q$ we mention just the
{\it isotropic}
case, i.e. when $I_1=I_2=0$. Clearly, if this is the case, then all eigen
values are equal to zero and it can be shown, that there exists
{\it just one
and common for F, *F, and Q isotropic eigen direction, defined by the
isotropic vector} $\zeta, \zeta^2=0$, and all other eigen vectors are
space-like. Moreover, there exist two 1-forms $A$ and $A^*$, such that the
following presentation holds
\begin{equation}
F=A\wedge \zeta,\quad *F=A^*\wedge \zeta.
\end{equation}
Obviously, the 1-forms $A$ and $A^*$ are
defined up to additive factors colinear to $\zeta$. We show now that these
two 1-forms are spacelike, mutually orthogonal, they have equal magnitudes
and are orthogonal to $\zeta$:  $A^2=(A^*)^2<0,\ A.A^*=0,\
A.\zeta=A^*.\zeta=0.$
In fact,
\[
0=*(A^* \wedge A^* \wedge \zeta)=*(A^* \wedge *F)=
-(A^*)^\sigma F_{\sigma\mu}dx^\mu=
-(A^*)^\sigma A_\sigma \xi_\mu dx^\mu,
\]
\[
0<4\pi Q_4^4=-F_{4\nu}F^{4\nu}=-
(*F)_{4\nu}(*F)^{4\nu}=-A^2\zeta_4\zeta^4=-(A^*)^2\zeta_4\zeta^4,
\]
\[
0=I_1=\frac12 F_{\mu\nu}F^{\mu\nu}=
\frac12 (A_\mu \zeta_\nu-A_\nu \zeta_\mu)(A^\mu \zeta^\nu-A^\nu \zeta^\mu)=
\]
\[
-(A.\zeta)^2=-\frac12 (*F)_{\mu\nu}(*F)^{\mu\nu}=(A^{*}.\zeta)^2.
\]
We see that in this case of zero inzvariants  $I_1=I_2=0$ there exists a new
invariant, namely the square of $A$ and $A^*$. Besides, from the first row of
equalities is seen that $A$ is an eigen vector of $*F$ and $A^*$ is an eigen
vector of $F$. Here are two useful relations:
\begin{equation}
det\|F_{\mu\nu}\|=det\|(*F)_{\mu\nu}\|=\frac14 (I_2)^2,\quad det\|(F\pm*F)_{\mu\nu}\|=(I_1)^2 .
\end{equation}
Finally we note that all subdeterminants of third order are proportional to
$E.B$, so, the highest order non-zero subdeterminants in this case may be those
of second order.

\vskip 0.5cm
\subsection{Continuous media}
All considerations made up to here
are characterized by the assumption, that in regions with $\rho=0$ the EM-
field propagates in space without losing energy and momentum, i.e. there is
no interaction with the medium, which we call in such a case {\it
electromagnetic vacuum}. CED extends its applicability to media, which
interact with the field, exchanging energy-momentum at every point and moment.
The majority of the known really existing media are built of a great number of
closely connected, i.e. strongly interacting, neutral and electrically
charged particles. An exact description of the mechanical motion of each one
of these particles in presence of the external EM-field is practically
impossible, so we have to assume simplified models of the various media.
Doing this, it is important not to forget all conditions and scales under which the
simplifying assumptions of a given model work.

In order to adapt the already developed mathematical machinery of the theory
to describe what happens in presence of such media, which are briefly called
{\it macroscopic bodies}, the approximation {\it physically small volume} is
introduced in the following way. If $l$ denotes the average distance among
the particles, creating a given medium, if $\Delta V$ denotes the physically
small volume and if $L$ denotes some typical linear scale of the macroscopic
object, we want the following relations to hold

\begin{equation}
l^3\ll\Delta V\ll L^3 .
\end{equation}
Further we assume these conditions satisfied, and all media, satisfying them
will be shortly called {\it macromedia}.

From practical point of view important are those macromedia, which can be
{\it electrified} and {\it magnetized} when placed in external EM-fields.
Such media are called {\it dielectrics}. The additional electrifying is due to
the presence of {\it bound charges} in these media. Subject to the action
of the external field these charges perform limited in small regions
displacements. Such
displacements cause appearance of additional charges, currents and dipole
moments. After an averaging over the volume $\Delta V$, they are denoted
respectively by $\rho_b$-{\it bound charge density}, ${\bf j}_b$-{\it bound
current density}, and $P$-{\it polarization vector}. The additional magnetization is due to
the circle-like displacements of the charges, generating in this way new
magnetic moments. The averaging of these new magnetic moments over the volume
$\Delta V$ defines the {\it magnetization vector} ${\cal M}$.

In analogy with the case {\it free charges in vacuum} the following relations
among these new quantities are assumed:
\begin{equation}
\rho_b=-divP,\quad {\bf j_b}=crot{\cal M}+\frac {\partial P}{\partial t}.
\end{equation}

After replacing in Maxwell equations (1.1) and (1.2) {\bf j} and $\rho$ by
$({\bf j}+{\bf j}_b)$ and $(\rho +\rho_b)$ respectively, and having in view
(1.16) we obtain the Maxwell's equations for continuous media

\begin{equation}
\frac 1c \frac{\partial D}{\partial t}=rotH - \frac {4\pi}{c}{\bf j},\quad divB=0,
\end{equation}
\begin{equation}
\frac 1c \frac{\partial B}{\partial t}=-rotE,\quad \ \ divD=4\pi\rho,
\end{equation}
where
\begin{equation}
H=B-4\pi{\cal M},\quad D=E+4\pi P.
\end{equation}

When passing from one medium to another, the dielectric properties of which
strongly differ from those of the first one, it is naturally to expect a
violation of the continuous properties of $H$ and $D$. Therefore it is necessary
to define the behaviour of these quantities on the corresponding boundary
surfaces. To this end, two new quantities are introduced: {\it surface density
of the electric charge} $\sigma$ and {\it surface density of the current} $i$.
Then the analysis of the above equations brings us to the following relations:

\[
(D_n)_2-(D_n)_1=4\pi\sigma, \quad (E_n)_2-(E_n)_1=0,
\]
\[
(H_n)_2-(H_n)_1=\frac {4\pi}{c} i,\quad (B_n)_2-(B_n)_1=0,
\]
where the index "n" denotes the normal to the boundary surface component of
the corresponding vector at some point.

Assuming that the quantity of electromagnetic energy, transformed to mechanical
work or heat during 1sec. in the volume $V$ is equal to $\int_V({\bf j}.E)dV$,
and making use of Maxwell's equations for medium, we get

\begin{equation}
({\bf j}.E)=-\frac {1}{4\pi}\biggl[\biggl(E.\frac {\partial D}{\partial t}\biggr)+
\biggl(H.\frac {\partial B}{\partial t}\biggr)\biggr]-div\biggl[\frac{c}{4\pi}
E\times H\biggr].
\end{equation}
Replacing now $D=E+4\pi P$ and $H=B-4\pi M$ in this relation we obtain
\[
({\bf j}.E)=-\frac {\partial}{\partial t}\frac{E^2+B^2}{8\pi}-div\biggl[\frac{c}{4\pi}E\times B\biggr]-
\]
\[
- \biggl[E.\frac{\partial P}{\partial t}-{\cal M}.\frac{\partial B}{\partial t}\biggr]+c.div\biggl[E\times {\cal M} \biggr].
\]
These relations describe the local energy-momentum balance.

The 8 equations (1.17)-(1.18) have to determine 15 functions
$E_i,B_i,H_i$, $D_i,j_i$. Clearly, more relations among these functions
are needed, in order to determine them. The usual additional relations
assumed are of the kind
\[
P^i=P^i\biggl(E^j,\frac {\partial E^j}{\partial x^k},...;B^j,\frac {\partial B^j}{\partial x^k},...\biggr),\quad
{\cal M}^i={\cal M}^i\biggl(E^j,\frac {\partial E^j}{\partial x^k},...;B^j,\frac {\partial B^j}{\partial x^k},...\biggr).
\]

The most frequently met assumption is $P=P(E),\ {\cal M}={\cal M}(B)$
together with the requirement $P(0)=0,\ {\cal M}(0)=0$. A series development
gives
\[
P^i=\kappa^i_j E^j+\frac12 \kappa^i_{jk}E^j E^k+. . .;\quad {\cal M}^i=\alpha^i_j B^j+\frac12 \alpha^i_{jk}B^j B^k+. . .
\]
The tensors $\kappa^i_j,\ \kappa^i_{jk}, . . . $ are called {\it polarization
tensors} (of corresponding rank), and $\alpha^i_j,\alpha^i_{jk},. . . $
are called {\it magnetization tensors } (of corresponding rank). For $D^i$
and $H^i$ we obtain respectively
\[
D^i=E^i+4\pi(\kappa^i_j E^j+\frac12\ \kappa^i_{jk}E^j E^k+. . .)=(\delta^i_j +4\pi \kappa^i_j)E^j+. . .
\]
\[
H^i=B^i-4\pi(\alpha^i_j B^j+\frac12 \alpha^i_{jk}B^j B^k+. . .)=(\delta^i_j- 4\pi \alpha^i_j)B^j-. . .
\]
If the medium is homogeneous and isotropic and the EM-field is weak, the
nonlinearities in these developments are neglected, so
$$
D^i=(1+4\pi\kappa)\delta^i_jE^j =\varepsilon^i_j E^j=\varepsilon\delta^i_jE^j
$$
and
$$
H^i=(1-4\pi \alpha)\delta^i_jB^j =\alpha^i_j B^J=\alpha\delta^i_jB^j.
$$

The constants $\varepsilon$ and $\mu=\alpha^{-1}$
are called
{\it dielectric} and
{\it magnetic } permeabilities respectively. In case of nonisotropic media
the two tensors $\varepsilon^i_j$ and $\mu^i_j=(\alpha^i_j)^{-1}$ are used.

In the relativistic formulation of the electrodynamics of continuous media
besides the 2-form $F$, a new 2-form $S$ is introduced, namely
\[
S={\cal M}_3dx\wedge dy-{\cal M}_2 dx\wedge dz+{\cal M}_1dy\wedge dz-
P_1 dx\wedge d\xi-P_2 dy\wedge d\xi-P_3 dz\wedge d\xi
\]
as well as a new current
\[
j^\mu_b=(\frac 1c {\bf j}_b,\rho_b).
\]
With these notations the equations (1.16) acquire the following compact form
\begin{equation}
\frac {\partial S^{\sigma\nu}}{\partial x^\sigma}=-J^\nu_b.
\end{equation}
If we introduce now the 2-form $G=F-4\pi S$,
then equations (1,17)-(1,18) look as follows
\begin{equation}
\delta *F=0,\quad \delta G=4\pi j.
\end{equation}
The two relations  $D^i=\varepsilon^i_jE^j$ and $B^i=\mu^i_jH^j$  may be
unified in one relation of the kind
\begin{equation}
G_{\mu\nu}=R_{\mu\nu}^{..\alpha\beta}F_{\alpha\beta},\ \mu<\nu,\ \alpha<\beta .
\end{equation}
Obviously, $R_{\mu\nu}^{..\alpha\beta}=-R_{\nu\mu}^{..\alpha\beta},\
R_{\mu\nu}^{..\alpha\beta}=-R_{\mu\nu}^{..\beta\alpha}.$
Comparing now the relativistic relation (1.23) and the non-relativistic two
relations, we obtain
\[
R_{i4}^{..kl}=0, \ R_{kl}^{..j4}=0,\ R_{i4}^{..j4}=\varepsilon_i^j,
\]
\[
R_{kl}^{..mn}=\tilde{\varepsilon}_{klr}\chi^r_s \tilde{\varepsilon}^{smn},\ \chi^r_s=(\mu^r_s)^{-1},\ k<l,\ m<n.
\]
The equations  $\varepsilon^i_j=\varepsilon^j_i,\ \mu^i_j=\mu^j_i$ lead to
$R_{\mu\nu}^{..\alpha\beta}=R_{\alpha\beta}^{..\mu\nu}$. It is immediately
verified that
\[
R_{\mu\nu}^{..\alpha\beta}+R_{\mu\alpha}^{..\beta\nu}+R_{\mu\beta}^{..\nu\alpha}=0.
\]
The  $(6\times 6)$ matrix $R_{\mu\nu}^{..\alpha\beta}$ looks as follows:
\[
R_{\mu\nu}^{..\alpha\beta}=\left\|\matrix{
\chi_3^3   &-\chi_2^3  &\chi_1^3      &0               &0               &0\cr
-\chi_3^2  &\chi^2_2   &-\chi_1^2     &0               &0               &0\cr
\chi_3^1   &\chi_2^1   &\chi^1_1      &0               &0               &0\cr
0          &0          &0             &\varepsilon_1^1 &\varepsilon_1^2 &\varepsilon_1^3 \cr
0          &0          &0             &\varepsilon_2^1 &\varepsilon_2^2 &\varepsilon_2^3 \cr
0          &0          &0             &\varepsilon_3^1 &\varepsilon_3^2 &\varepsilon_3^3
\cr}\right\|.
\]
For the invariant $R=R_{\mu\nu}^{..\mu\nu}$ we obtain
$$
R=2(\varepsilon_1^1+\varepsilon_2^2+\varepsilon_3^3+\chi_1^1+\chi_2^2+\chi_3^3).
$$
These algebraic properties of the tensor  $R_{\mu\nu}^{..\alpha\beta}$
are the same as those of the Riemann curvature tensor. Since for vacuum we
have  $\varepsilon_i^j=\chi_i^j=\delta_i^j$
for $R_{\mu\nu}^{..\alpha\beta}$ we get
$$
R_{\mu\nu}^{..\alpha\beta}=\delta_\mu^\alpha \delta_\nu^\beta - \delta_\mu^\beta\delta_\nu^\alpha,
$$
or
$$
R_{\mu\nu,\alpha\beta}=\eta_{\mu\alpha} \eta_{\nu\beta} - \eta_{\mu\beta}\eta_{\nu\alpha},
$$
which is exactly the induced by $\eta$ metric in the bundle of 2-forms over
the Minkowski space-time. In other words, the presence of an active
continuous medium, i.e. non-trivial functions $\varepsilon_i^j(x^\nu)$ and
$\chi_i^j (x^\nu)$, could be described by an appropriate {\it curved} metric
in the space $\Lambda^2(M)$.

Now we are going to consider the problem of energy-momentum distribution
of the field in presence of an active medium. Recall that in case of vacuum,
these quantities are described by the energy-momentum tensor
$$
Q_\mu^\nu=\frac {1}{4\pi}\biggl[\frac 14 F_{\alpha\beta}F^{\alpha\beta}\delta_\mu^\nu-F_{\mu\sigma}F^{\nu\sigma}\biggr]=\\
\frac {1}{8\pi}\biggl[-F_{\mu\sigma}F^{\nu\sigma}-(*F)_{\mu\sigma}(*F)^{\nu\sigma}\biggr].
$$
The natural generalization of this tensor in presence of a new 2-form $S$, or $G$,
looks as follows
\begin{equation}
W_\mu^\nu=\frac {1}{8\pi}\biggl[\frac 12 F_{\alpha\beta}G^{\alpha\beta}\delta_\mu^\nu-F_{\mu\sigma}G^{\nu\sigma}-G_{\mu\sigma}F^{\nu\sigma}\biggr].
\end{equation}
Using the identity, which holds for any two 2-forms in the Minkowski space
\begin{equation}
\frac 12 F_{\alpha\beta}G^{\alpha\beta}\delta_\mu^\nu=F_{\mu\sigma}G^{\nu\sigma}-(*G)_{\mu\sigma}(*F)^{\nu\sigma},
\end{equation}
for $W_\mu^\nu$ is obtained
$$
W_\mu^\nu=\frac {1}{8\pi}\biggl[-F_{\mu\sigma}G^{\nu\sigma}-(*F)_{\mu\sigma}(*G)^{\nu\sigma}\biggr]=
\frac {1}{8\pi}\biggl[-G_{\mu\sigma}F^{\nu\sigma}-(*G)_{\mu\sigma}(*F)^{\nu\sigma}\biggr].
$$
Obviously, $W_{\mu\nu}=W_{\nu\mu}$, and if $S_{\mu\nu}$ $\rightarrow 0$,
or equivalently, $G=F$, we get $W_\mu^\nu$$\rightarrow Q_\mu^\nu$. Here are
the explicit expressions of $W_\mu^\nu$ by means of the components of the
 3-vectors $E,B,D,H$:
\[
W_i^j=\frac{1}{8\pi}\biggl[E_i D_j +E_j D_i+B_i H_j+B_jH_i+\delta_i^j(B.H-E.D)\biggr],
\]
\[
W_i^4=\frac{1}{8\pi}\biggl[(E\times H)_i + (B\times D)_i\biggr],\quad W_4^4=\frac{1}{8\pi}(E.D+B.H)
\]
It is easily verified the following relation
\begin{equation}
\nabla_\nu W_\mu^\nu=\frac {1}{8\pi}\biggl[F_{\mu\nu}(\delta G)^\nu+G_{\mu\nu}(\delta F)^\nu+
(*F)_{\mu\nu}(\delta *G)^\nu+(*G)_{\mu\nu}(\delta *F)^\nu\biggr].
\end{equation}
If we require at $j=0$ in (1.22) the following local conservation law to
hold
\begin{equation}
\nabla_\nu W_\mu^\nu =0
\end{equation}
and make use of the above introduced definitions, we get the equation
\begin{equation}
4\pi S_{\mu\nu}(\delta S)^\nu=F_{\mu\nu}(\delta S)^\nu-(*F)_{\mu\nu}(\delta *S)^\nu.
\end{equation}
This relation determines in what way and how much of the field energy-momentum
is utilized by the medium, characterized by $S_{\mu\nu}$. Since the equations
are 8, and the unknown functions are 12, the 4 equations (1.28), although
{\it nonlinear}, could be used as additional field equations for $F_{\mu\nu}$
and $S_{\mu\nu}$. In this way the number of the equations becomes equal to
the number of the unknown functions.

As we mentioned earlier, the symmetry of the energy-momentum tensor
$W_\mu^\nu$ is necessary to define correctly integral conserved quantities,
using the 10 Killing vector fields of Minkowski space. The tensor $W_\mu^\nu$,
given by (1.24), has the following additional properties:
$$
W(F,G)=W(G,F),\quad W(F,G)=W(*F,*G),
$$
$$
\nabla_\nu W_\mu^\nu (F,G)=\nabla_\nu W_\mu^\nu (G,F),\quad \nabla_\nu W_\mu^\nu (F,G)=\nabla_\nu W_\mu^\nu (*F,*G).
$$
Moreover, since the $*$-operator is applied only on 2-forms, the tensor
$W_\mu^\nu$ is {\it conformally invariant}. Finally, if the free current $j$
is not equal to zero, then from (1.26) and (1.28) and Maxwell's equations it
follows that the energy-momentum, transferred over to the medium in an unit
4-volume is determined by the well known expression $F_{\mu\nu}j^\nu$.
We should not forget also, that if $G_{\mu\nu}=F_{\mu\nu}$ we obtain
$W_\mu^\nu=Q_\mu^\nu$.

All these properties of $W_\mu^\nu$ suggest that it is a good candidate
for energy-momentum tensor of the EM-field in presence of a continuous
medium. As for the new condition (1.27), its adequacy to the real processes
of energy-momentum redistribution in the medium under consideration has to be
verified every time we are going to use it.

\section {\bf Solutions to Maxwell's Equations}
\vskip 0.05cm
\subsection{The solutions as models of physical objects and processes}
The equations are mathematical relations among several quantities, some
of which are known and some are unknown. To solve an equation means to find
those values of the unknown quantities, which make this equation into
identity. Clearly, the values found, i.e. the solutions, will depend on the
values of the known  quantities, as well as on the very equation. If the
number of the independent equations is not equal to the number of the
unknowns we say that the problem is {\it underdetermined} or
{\it overdetermined} depending on whether the number of equations is
 {\it less} or {\it larger} than the number of the unknowns respectively.
One of the important problems is that of {\it uniqueness}, i.e. under what
conditions a given equation has only one solution. This problem is based on
the fact that an equation may have many, even infinitely many, solutions.
For example, the equation $f'(x)=0$ is satisfied by every constant function;
the equation $\frac {\partial f}{\partial x}=0$, where the unknown function
$f$ depends on two independent variables $(x,y)$, has infinitely many solutions,
which are parametrized by one differentiable function of one variable:
$f(x,y)=g(y)$.

In mathematics we frequently write down relations, which have sense in various
classes of functions, even in various classes of mathematical objects. For
example, the equation ${\bf d}\alpha =0$ admits as solutions differential forms
of various degree as well as differential forms with compact or noncompact
support on various manifolds. This is true for a very
large class of differential operators, therefore, it is necessary the set of
objects, where we are going to look for a solution, to be pointed out
sufficiently in detail. It may happen, that in the set of objects, where we
look for a solution, the equation considered has no solutions. For example,
the widely used Laplace equation $\Delta f=0$ has no non-constant solutions
in the class of {\it finite} functions $f:{\cal R}^3 \rightarrow {\cal R}$;
the widely used {\it wave} equation $\Box f=0$ has infinitely many
(1+1)-soliton-like solutions, and {\it has no} (3+1)-soliton-like solutions.

So, every  differential equation separates a class of functions (or objects),
namely those, which have the {\it local} property to satisfy this differential
equation. Not every solution has properties desired by us, so any separation
of some subclass of solutions with definite additional properties is made by
means of imposing additional conditions.

In theoretical physics we make {\it mathematical models of physical objects
and processes}. We assume the idea, that {\it the really existing objects are
finite and time-stable}. This means, that at any moment of their life they
occupy {\it finite and sufficiently small} regions of the 3-dimensional space
and if there are no perturbations from outside, they would live sufficiently long
time, i.e. they are time-stable. From mathematical point of view this requires
that, with respect to the space variables $(x,y,z)$ the corresponding
mathematical objects {\it are everywhere smooth and have compact support}.
The time evolution of the object is usually determined by an equation,
defining a definite relation among the various derivatives of the components
of the corresponding mathematical object with respect to the time and space
coordinates, i.e. {\it locally}.

The real physical objects have {\it structural
properties}, as well as properties as a whole, i.e. {\it integral} properties.
If we are interested only in the integral properties and behaviour of the
physical object, we talk about {\it material points}, and the behaviour of
such objects is described by {\it ordinary differential equations}. The local
description of an object, having a dynamical structure, requires {\it partial
differential equations}, as well as rules, pointing out the connection between
the local and integral characteristics and properties of the object.

If we are able to identify a physical object during a finite period of time,
this means, that with respect to our means of identification this object
shows {\it definite properties of constancy}. If these properties of constancy
are measurable, then the introduced by corresponding measurement procedures
quantities will have constant in time values, so we can talk about {\it
conservation laws}, which are specific for the object considered. There exist
physical quantities, which can be introduced for various physical objects.
The importance of a physical quantity is in a direct dependence on its
{\it universality}, i.e. how much broad is the class of objects, admitting
this quantity as a characteristic.

At a definite level and under definite conditions it is possible some objects
to be transformed into other objects. If such a transformation takes place the
natural question arises:{\it what is the behaviour of those physical quantities,
which characterize the initial objects, as well as the final ones, is there
any quantitative connection between the initial and final values of these
quantities?}. A positive response to this question would be of great
importance for theoretical physics since it would allow some definite
mathematical relations to be written down immediately and some {\it true}
conclusions, based on these relations and concerning the real objects and
processes, to be made.

The study of the real objects and processes from this point of view has
resulted into formulation of the so called {\it conservation principles for
energy, momentum and the full angular momentum}. (A principle is any rule
which is proved to hold for all known situations and which is extended as
a hypothesis for the unknown situations). The physicists have succeeded in
defining these quantities in such a way, that the time constancy of their
values to be an adequate expression of the constancy properties of the real
objects and to point out definite necessary conditions, which all real
processes and transformations of objects have to respect and obey. We would
like to note specially, that the {\it universality} of these quantities and
principles is of primary importance.

Another very important property of these quantities is their {\it additiveness},
which enables us to build integral conservation laws, making use of the
corresponding local conservation laws, i.e. time independent characteristics
of the system as a whole.

Let us now connect these conclusions with the above mentioned circumstance,
that the partial differential equations, which we use to describe the existence
and evolution of the extended natural objects, admit usually many solutions.
Most frequently, the model differential equation is a mathematical expression
of one or several characteristic or typical local features of the object
(or process), but not of all of them. Therefore, some of the solutions may
posses properties, which do not correspond to the real integral properties
of the object under consideration. So, additional conditions, such as new
equations, inequalities, etc., of local or integral character have to be
formulated. In such a case, in order to separate the non-adequate solutions,
we shall always try to combine the above introduced notion for {\it finite
and time stable object} and the availability of conserved quantities for
any object, observing the following rule:

\vskip 0.5cm
{\bf In order some solution of a given differential equation to be considered
as a realistic model of a real object it is necessary this solution to be
finite, time-stable and the corresponding integral energy, momentum and full
angular momentum to be well defined finite quantities}.

\vskip 0.5cm
\subsection{Spherically symmetric solutions}
As we mentioned in
section 1.1.2 the symmetry of a tangent object (i.e. a section of some tensor
degree of the tangent and cotangent bundles) is a smooth map (usually a
diffeomorphism) of the base space onto itself, such that the object does not
change. We recall that in the non-relativistic frame the base space is
${\cal R}^3$, while in the relativistic frame the base space is ${\cal R}^4$.
Since the spherically symmetric solutions of Maxwell's equations are very
important and lead to the notion of {\it elementary charge}, we are going to
consider this point more in detail in nonrelativistic, as well as, in
relativistic terms.

The intuitive notion of spherical symmetry consists in, that at the same
distance from a given point the properties of the object under consideration
are the same. The more accurate notion in the frame of the mathematical model
requires:

1. Pointing out the point of the base manifold, which is the center of
symmetry.

2. A definition of {\it distance}.

A procedure for separation of those maps, which keep the symmetry center
stable as well as the distance between any two points unchanged.

In the frame of non-relativistic considerations the base space ${\cal R}^3$,
is furnished by the Eucledean distance ${\Delta l}$ between any two points
with standard coordinates $(x_1,y_1,z_1)$ and $(x_2,y_2,z_2)$
$$
\Delta l=\sqrt {(x_2-x_1)^2+(y_2-y_1)^2+(z_2-z_1)^2},
$$
or, equivalently, by the metric tensor $g$, which in these coordinates has the
well known canonical components $g_{ii}=1, g_{ij}=0\  if  \ i\neq j $.
Then the local symmetries$X$ of $g$ are determined by the relation $L_Xg=0$.
This last relation defines a system of differential equations for the
components of the vector field $X$. Besides the standard translations
 $\bigl[ {\partial}/{\partial x},
{\partial}/{\partial y},{\partial}/{\partial z}\bigr]$, these equations
have in spherical coordinates $(r,\theta,\varphi)$ the following solutions:

$$
X_1=\sin\varphi \frac{\partial}{\partial \theta}+\cot\theta \cos\varphi
\frac{\partial}{\partial \varphi},
\ X_2=-\cos\varphi \frac{\partial}{\partial
\theta}+\cot\theta \sin\varphi \frac{\partial}{\partial \varphi},
\ X_3=-\frac{\partial}{\partial \varphi}.
$$

These, of course, are the generators of the group $SO(3)$. So, we want our
$EM$-field, i.e. the couple $(E,B)$ of vector fields, to be symmetric with
respect to the local isometries $X_1,X_2,X_3$, i.e. the following equations
to hold
$$
L_{X_i}E=\left[X_i,E\right]=0,\ L_{X_i}B=\left[X_i,B\right]=0
$$
The solutions of these equations are
\[
E=q(r;t)\frac{\partial}{\partial r},\ B=m(r;t)\frac{\partial}{\partial r}.
\]
Since $g_{11}=1$, then the corresponding 1-forms are $E=q(r;t)dr,\
B=m(r;t)dr$. Obviously, ${\bf d}E=0,\ {\bf d}B=0$, i.e. $rotE=0,\ rotB=0$.
From Maxwell's equations with zero current it follows now that $E$ and $B$ do
not depend on time, i.e. {\it all spherically solutions of the current-free
Maxwell's equations are static}.

Let's now pay some attention to the following intermediate result:
Every spherically symmetric, i.e. $SO(3)$- invariant , 1-form $\alpha$ on
${\cal R}^3$ is of the kind $\alpha=f(r)dr$, so it is closed:
${\bf d}\alpha=0$. Then, if $\omega$ is $SO(3)$-invariant 2-form on
${\cal R}^3$, and since the action of the isometries commutes with the
$*$-operator, defined by the same metric, we conclude that $*\omega$ is also
$SO(3)$-invariant, hence it is closed. Historically, the things have started
with a specially chosen 1-form, namely the Coulomb force $F$ for a unit test
charge, so it necessarily it is closed, and the special properties of $F$
come from the observation that $*F$ is also closed. From modern point of view
the $SO(3)$-invariant 2-form approach seems more natural and clearer.

What has rest to be done is to solve the two equations ${\bf d}*E=0,
\ {\bf d}*B=0$. For the $*$-operator in spherical coordinates we get
\[
{\bf d}*E={\bf d}*q(r)dr={\bf d}\bigl[q(r)r^2\sin\theta d\theta \wedge d\varphi\bigr]=
\frac {\partial (qr^2)}{\partial r}\sin\theta dr\wedge d\theta \wedge d\varphi=0.
\]
We obtain $q(r)=const/r^2=q_0/r^2$, so
$$
E=\frac{q_0}{r^2}\frac{\partial}{\partial r}, \
B=\frac{m_0}{r^2}\frac{\partial}{\partial r}.
$$

These results may be given a topological interpretation. In fact, the
differential forms
$$ E=\frac {q_0}{r^2}dr,\quad
*E=q_0\sin\theta d\theta\wedge d\varphi
$$
are defined on the space ${\cal R}^3\setminus\{0\}$, i.e out of the point
$r=0$. Maxwell's equations require both of them to be closed, but while the
1-form $E=q_0/r^2 dr$ is exact, the two form
$*E=q_0 sin\theta d\theta\wedge d\varphi$
is not exact, since $\int_{S^2}{*E}=q_0$. Therefore,
the 2-form $*E$  represents the nontrivial  cohomology class of the space
${\cal R}^3\setminus \{0\}$, moreover, it is the only spherically symmetric
representative  of this class.

If now we favour the topological character of the field configuration
obtained, we should define the electric charge as the (appropriately
paramet\-rized) cohomology class of the space ${\cal R}^3\setminus\{0\}$, and
the field, as
a representative of this class (we chose the spherically
symmetric representative $\omega=q_0sin\theta d\theta \wedge d\varphi$
of this class, but this is not obligatory).  The
available eucledean metric $g$ makes it possible to introduce the
$*$-operator and the notion of spherical symmetry.  Then the 1-form
$*\omega=q_0r^{-2}dr$, which now depends on the metric chosen, is exact and
coincides with the notion of a field of an elementary source in CED. Note
that, if we choose another metric $\tilde g$ on ${\cal R}^3\setminus\{0\}$,
then the relation ${\bf d}*\omega=0$ may not hold. So, the notion of field in
CED is strongly connected with the eucledean metric, while the notion of
charge has topological origin.

This picture favours the topological aspect of the electric charge and gives
some benefits for the relativistic formulation of CED. In fact, the closed
2-form $\omega$ is pulled back naturally as a closed 2-form on
$({\cal R}^3\setminus \{0\})\times {\cal R}$, and the pseudoeucledean metric
$\eta$ guarantees that $*_{\eta}\omega$ is closed, even exact. Of course, the
consideration of all solutions of the two equations ${\bf d}\omega=0,\linebreak
\ {\bf d}*\omega=0$ as admissible models of real $EM$-fields is an additional
hypothesis, but it is a natural one in the frame of this approach.

In the frame of the relativistic formalism the spherically symmetrical problem
requires to solve the equations ${\bf d}F=0,\ {\bf d}*F=0$ together with the
symmetry relations $L_{X_i}F=0,\ L_{X_i}*F=0$, which reduce to the first
relation only since the $*-$operator is spherically symmetric. The most
general form of a spherically symmetric $F$ looks as follows:
\[
F=f(r,\xi)dr\wedge d\xi + h(r,\xi)\sin\theta d\theta \wedge d\varphi.
\]
Maxwell's equations require
\[
F=\frac {C_1}{r^2}dr\wedge d\xi + C_2 \sin\theta d\theta \wedge d\varphi,
\]
where $C_1$ and $C_2$ are constants. The additional requirements
\[
F_{r\rightarrow \infty}\rightarrow 0,\quad \oint_{S^2}*F=q_0
\]
determine $C_1=q_0/4\pi, \ C_2=0$.
\vskip 0.5cm

\subsection{Local and integral interaction energy of spherically symmetric
fields. The Coulomb's force}

Let us consider the following situation: two electric charges, occupying two
3-dimensional regions $\{O_1\}$ and $\{O_2\}$, considered as "open balls".
The boundaries of these balls are the two non-overlapping spheres $S_1$ and
$S_2$, and the distance between the centers of these two spheres is $R$. The
non-trivial topology of the space, where the two representatives $\omega_1$
and $\omega_2$ of the corresponding cohomology classes are defined is
${\cal R}^3\setminus \{O_1,O_2\}$. Our purpose is to define {\itshape
interaction} between the two fields $\omega_1$ and $\omega_2$ in such a way,
that the calculated integral interaction energy to be equal to the well known
classical expression $W=q_1 q_2/r$. Clearly, the local interaction energy
expression has to be a 3-form and symmetric with respect to the two
interacting fields. We define this 3-form as follows:
\begin{equation}
w=\frac{1}{4\pi}\omega_1 \wedge *\omega_2 .
\end{equation}
The two fields $\omega_1$ and $\omega_2$ are the only spherically symmetric
representatives of the two cohomology classes. Since the eucledean metric is
also spherically symmetric, the corresponding 1-forms $*\omega_1$ and
$*\omega_2$ are also closed 1-forms and even exact. Therefore, there are two
functions $f_1$ and $f_2$, such that ${\bf d}f_1=*\omega_1$ and
${\bf d}f_2=*\omega_2$. The 3-form $w$ turns out to be exact, in fact
\[
4\pi w=\omega_1\wedge *\omega_2 = \frac 12 \omega_1\wedge *\omega_2 +\frac 12
\omega_2 \wedge *\omega_1=
\]
\[
=\frac 12 \bigl[\omega_1 \wedge {\bf d}f_2 + \omega_2 \wedge {\bf d}f_1\bigr]=
\frac12 \bigl[{\bf d}(f_2\omega_1)+{\bf d}(f_1\omega_2)\bigr]=
\frac12 {\bf d}\bigl[f_2\omega_1+f_1\omega_2\bigr].
\]
We integrate now this expression over the region $D$ out of the two open
balls. Obviously, the boundary $S_D$ of $D$ is
$S_D=S_{\infty}^2 \cup S_1 \cup S_2$. Then, making use of the Stokes theorem,
we obtain
\[
W=\int_D w = \frac{1}{8\pi} \int_D {\bf d}\bigl[f_2\omega_1+f_1\omega_2\bigr]=
\]
\[
=-\frac{1}{8\pi} \biggl[\int_{S^2_{\infty}}(f_1 \omega_2 + f_2\omega_1)\ +\
\int_{S_1}(f_1\omega_2+f_2\omega_1)\ +\ \int_{S_2}(f_1\omega_2+f_2\omega_1)\biggr].
\]
We introduce two spherical coordinate systems
$(r,\theta,\varphi)$ and $(\bar{r},\bar{\theta},\bar{\varphi})$
originating at the centers of $S_1$ and $S_2$. We can write
\[
\omega_1=q_1 \sin\theta d\theta \wedge d\varphi,\quad f_1=-\frac {q_1}{r},
\]
\[
\omega_2=q_2 \sin\bar{\theta}d\bar\theta \wedge d\bar\varphi,\quad f_2=-\frac{q_2}{\bar{r}}
\]
Since $f_1$ and $f_2$ get zero values on the infinite sphere, we have to
calculate the integrals over the two spheres $S_1$ and $S_2$. Having in view
the relations
\[
\int_{S_2}\omega_1=0,\ \int_{S_1}\omega_2=0,\ f_1\vert_{S_1}=const,\ f_2\vert_{S_2}=const
\]
we obtain
\[
W=-\frac{1}{8\pi} \biggl[\int_{S_1} f_2\omega_1 + \int_{S_2}f_1\omega_2 \biggr]
=\frac{q_1q_2}{8\pi}\biggl[\int_{S_1}\frac {\sin\theta d\theta\wedge d\varphi}{\bar{r}}+
\int_{S_2}\frac{\sin\bar{\theta} d\bar{\theta}\wedge d\bar{\varphi}}{r}\biggr]=
\]
\[
=\frac{q_1 q_2}{8\pi}\biggl[\frac{1}{R_1^2}\int_{S_1}\frac{dS_1}{\bar{r}}+
\frac{1}{R_2^2}\int_{S_2}\frac{dS_2}{r}\biggr],
\]
where $dS_1=R_1^2 \sin\theta d\theta\wedge d\varphi$ and
$dS_2=R_2^2\sin\bar{\theta} d\bar{\theta}\wedge d\bar{\varphi}$
are the two surface elements. Since the function $1/r$ is {\itshape harmonic}
out of the point $r=0$, then using the mean-value theorem for a harmonic
function, we get
\[
\frac{1}{4\pi R_1^2}\int_{S_1}\frac{dS_1}{\bar{r}}=\frac1R,\quad
\frac{1}{4\pi R_2^2}\int_{S_2}\frac{dS_2}{r}=\frac1R.
\]
Finally,
\[
W=\frac{q_1q_2}{R},
\]
in correspondence with the purpose we set. This result shows that the
differential ${\bf d}W$, where $W$ is considered as a function of the two
centers $r_1=0, r_2=0$, has nothing to do with $*\omega_1$, or $*\omega_2$,
which are not defined at these points at all.

In order to introduce the Coulomb's force in a correct way we first note that
its real sense is to define what part of the full energy and momentum of the
two fields is transformed into mechanical energy and momentum of the two
particles as a consequence of the local interaction of the two fields.
This is an integral effect, since the very concept of electric charge has
an integral character. Formally, we attack this problem in the following way.
First we consider the trivial bundle
$({\cal R}^3\times q_1 S^2,\pi,{\cal R}^3,S^2)$.
In the natural coordinates $(x,y,z;\theta,\varphi)$ we consider the 3-form
\[
\Omega_{12}=q_1 sin\theta d\theta \wedge d\varphi\wedge \pi^*(*\omega_2)=
q_1 sin\theta d\theta\wedge
d\varphi\wedge\frac{q_2(xdx+ydy+zdz)}{\sqrt{(x^2+y^2+z^2)^3}},
\]
defined on the whole bundle space. Now the Coulomb's force can be defined as
the {\itshape fiber integral} of $\Omega_{12}$ as follows:
\[
\int_{S^2}\Omega_{12}=q_1q_2 \frac{(xdx+ydy+zdz)}{\sqrt{(x^2+y^2+z^2)^3}}
\]
The so obtained Coulomb's force could be interpreted as a characteristic of
the field $\omega_2$ when $q_1=1$, but it shouldn't be identified in any way
with $*\omega_2$, although they look the same.

Let's consider briefly the problem of general spherically symmetric solutions
of Maxwell's equations. Since in this case the time derivatives vanish we
have the system
\[
rotE=0,\ rotB=0,\ divE=0,\ divB=0,
\]
or, all the same
\[
{\bf d}E=0,\ {\bf d}B=0,\ \delta E=0,\ \delta B=0.
\]
Now, the Laplace operator $\Delta$ is defined by
\[
\Delta={\bf d}\delta + \delta {\bf d},
\]
so, we get
$$
\Delta E=0,\ \Delta B=0.
$$
(We should not forget that these are just {\it necessary} conditions, i.e.
the last two equations may have solutions, which do not satisfy the static
Maxwell equations.) We see, that the components of $E$ and $B$ are
{\it harmonic} functions. According to the theory of these functions, they
can not have local exteremums inside the regions of harmonicity, and if the
region is the whole space ${\cal R}^3$ and the function is in addition
bounded, it is a {\it constant}. These two properties of the solutions of
Laplace equation do not recommend them as possible models of free real
objects, which are finite and have to be described by finite functions of the
three space variables $(x,y,z)$, i.e. such, that necessarily to achieve
their maximum values, since they are zero out of some finite subregions of
${\cal R}^3$. Only in some topologically non-trivial regions they could be of
some interest in this respect.

Finally we note, that from relativistic point of view the concept "static
field" is not invariant, since the Lorentz transformations transform any
"static field" into "non-static" one.

\subsection{Wave equations and wave solutions}
From physical point of view when we talk about {\it waves} we mean
{\it propagation of some disturbance, or perturbation, in a given medium}. It is also
assumed that the perturbation does not alter the characteristic properties of
the medium, and the time-evolution of the perturbation depends on the medium
properties as well as on the specificities of the very perturbation. The waves
are divided to 2 classes: {\it elementary} ({\it linear}) and {\it
intrinsically coordinated} ({\it nonlinear}). The elementary waves are
observed in homogeneous media and are generated by perturbing the
equilibrium state of the medium through small quantities of
external energy and momentum. The important properties of linear waves come
from the condition, that during the propagation of the initial disturbance
throughout the medium the structure of the medium does not change
irreversibly, and the various such propagating perturbations do not interact
with each other substantially. From mathematical point of view this means
that the evolutionary equations, which are partial differential
equations, describing such phenomena, are linear, so any linear combination
with constant coefficients of solutions gives again a solution. In other
words, the set of solutions of such equations is a real (finite or
infinite dimensional) vector space.

The intrinsically coordinated, or nonlinear, waves disturb more deeply the
medium structure, but the corresponding changes of the medium structure stay
reversible. When subject to several such perturbations, the medium responses
to the various disturbances is different in general, so the medium
reorganization requires more complicated intrinsic coordination. All this
demonstrates itself in various ways, depending on the medium properties and
the initial perturbation. What we observe from outside is, that some important
properties of the initial perturbations are changed in result of the
interaction. In some cases we observe a time-stable coordination among the
responding reactions of the medium and if the corresponding formation is
finite, we may consider it as a new object. If this object keeps its energy
and momentum we frequently call the corresponding medium {\it vacuum}.
 Clearly, such objects can exist only in appropriate media. In
such cases, studying the objects, we get some information about the medium
itself.  From mathematical point of view these waves are described by
nonlinear equations, so that a linear combination of solutions is not, as a
rule, a new solution. The huge variety of various such cases could hardly be
looked at from a single point of view, except some most general features.

It is important to note, that in the both cases, linear and nonlinear, the
perturbations are bearable for the medium in the sense, that they do not
destroy it. We are not going to consider here unbearable perturbations.

One common for every kind of waves characteristic is the {\it
polarization}. The polarization determines the relation between the direction
of propagation (at some point of the medium) and the direction of deviation
from the equilibrium state of the medium point considered. If these two
directions are parallel we say that the polarization is longitudinal, and if
these directions are orthogonal we say the polarization is transverse. In
general the polarization depends on the space-time point, i.e. it is a local
characteristic. When the wave passes through some region of the medium, the
points inside this medium commit some displacements along some
(usually closed) trajectories. If these trajectories are straight lines we
say that the polarization is {\it linear}, if they are circles we say the
polarization is circular, etc. It is important to note that the polarization
is an intrinsic property of the wave, therefore it ia an important for the
theory characteristic. In particular, the mathematical character of the
object (scalar, tensor, spinor, differential form, etc.), describing the
wave, depends substantially on it. If the wave is linear, and the
corresponding equation admits solutions with various polarizations, then
summing up solutions with appropriate polarizations we can obtain a solution
with a beforehand defined polarization.

Other common characteristics of the waves are the {\it propagation velocity},
determining the energy transfer from point to point of the medium, and
{\it the phase surface}, built of all points, being in the same state with
respect to the equilibrium state at a given moment. We are going to make use
of these characteristics further.

Let us consider now the Maxwell's equations in regions far from sources:
$\rho=0$. We have
\[
\frac 1c \frac {\partial E}{\partial t}=rotB,\ divB=0,\
\frac 1c \frac {\partial B}{\partial t}=-rotE, \ divE=0.
\]
From the first and the second equations we obtain
\[
rot(rotB)-\frac 1c \frac{\partial}{\partial t}rotE=
grad(divB)-\Delta B +\frac {1}{c^2} \frac {\partial^2 B}{\partial t^2}=\\
-\Delta B +\frac {1}{c^2} \frac {\partial^2 B}{\partial t^2}=0.
\]
From the third and the fourth equations we obtain
\[
-\Delta E +\frac {1}{c^2} \frac {\partial^2 E}{\partial t^2}=0.
\]
These are the well known {\it wave equations}, and we are going to consider
some of their properties. First we note, that these equations are just {\it
necessary conditions for every solution of the vacuum Maxwell's equations}.
Therefore, they may have solutions, which do not satisfy Maxwell's equations.
Second, every component of $E$ and $B$ satisfies the same equation and does
not depend on the rest ones.
Third, these are second order differential equations of hyperbolic type.

We are interested in the following: {\it Do the vacuum Maxwell's equations
admit finite and time-stable solutions, so that such solutions to serve as
models of real objects?}. The positive answer to this question would be a
serious virtue from the point of view of their adequacy as model equations
for an important class of real objects, while the negative answer would make
us searching for new equations, having solutions with the desired properties.
At the beginning of the last century (about 1818), i.e. more than 40 years
before the appearance of Maxwell's equations this problem has been
essentially solved by Poisson, and because of its importance we shall
consider it in some more detail.

Let's denote by $u$ any component of the vector fields $E$ and $B$. Then $u$
satisfies the wave equation. We are interested in the behaviour of $u$ at
$t>0$, if at $t=0$ the function $u$ satisfies the initial conditions
\[
u|_{t=0}=f(x,y,z),\quad \frac{\partial u}{\partial t}\biggl|_{t=0}=F(x,y,z).
\]
Further we assume that the functions $f(x,y,z)$ and $F(x,y,z)$ are finite,
i.e. they are different from zero in some finite region
$D\subset{\cal R}^3$, which corresponds to the above introduced concept of a
real object. Besides, we assume also that $f$ is continuously differentiable
up to third order, and $F$ is continuously differentiable up to the second
order. Under these conditions Poisson proved that a unique solution
$u(x,y,z;t)$ of the wave equation is defined, and it is expressed by the
initial conditions $f$ and $F$ by the following formula:
\begin{equation}
u(x,y,z,t)=\frac{1}{4\pi c}\left\{\frac{\partial}{\partial t}\Biggl[\int_
{S_{ct}}\frac{f(P)}{r}d\sigma_r \Biggr]+\int_{S_{ct}}\frac{F(P)}{r}
d\sigma_r \right\},
\end{equation}
where $P$ is a point on the sphere $S$ centered at the point $(x,y,z)$ and a
radius $r=ct$, $d\sigma_r$ is the surface element on $S_{r=ct}$.

The above formula $(1.30)$ shows the following. In order to get the solution
at the point $(x,y,z)$, being at an arbitrary position with respect to the
region $D$, where the initial condition, defined by the two functions $f$ and
$F$, is concentrated, it is necessary and sufficient to integrate these
initial conditions over a sphere $S$, centered at $(x,y,z)$ and having a
radius of  $r=ct$. Clearly, the solution will be different from zero only if
the sphere $S_{r=ct}$ crosses the region $D$ at the moment $t>0$.
Consequently, if $r_1=ct_1$ is the shortest distance from $(x,y,z)$ to $D$,
and $r_2=ct_2$ is the longest distance from $(x,y,z)$ to $D$, then the
solution will be different from zero only inside the interval $(t_1,t_2)$.

From another point of view this means the following. The initially
concentrated perturbation in the region $D$ begins to expand radially , it
comes to the point $(x,y,z)$ at the moment $t>0$, makes it "vibrate" (
i.e. our devices show the availability of a field) during the time interval
$\Delta t=t_2-t_1$, after this the point goes back to its initial condition
and our devices find no more the field. Through every point out of $D$ there
will pass a wave, and its forefront reaches the point $(x,y,z)$ at the moment
$t_1$ while its backfront leaves the same point at the moment $t_2$. Roughly
speaking, the initial condition "blows up" radially and goes to infinity with
the velocity of light.

This mathematical result shows that {\it every} finite nonstatic solution of
Maxwell's equations in vacuum is time-unstable, so these equations {\it have
no} smooth enough time-dependent solutions, which could be used as models of
real objects. As for the static solutions, as it was mentioned earlier, they
also can not describe real objects.

These explicit results state clearly, that if we want to describe
3-dimensio\-nal time-dependent soliton-like electromagnetic formations (or
configurations) it is necessary to leave off Maxwell's equations and to look
for new equations for $E$ and $B$, or for $F_{\mu\nu}$.

In relativistic notations the vacuum Maxwell's equations
\[
{\bf d}F=0,\ \delta F=0
\]
naturally, require $F$ to satisfy the equation
\[
({\bf d}\delta +\delta {\bf d})F=\Delta F=0,
\]
which in standard coordinates $(x^1,x^2,x^3,x^4)=(x,y,z,\xi=ct)$ gives the
usual wave equations for the components $F_{\mu\nu}$:
\[
g^{\alpha\beta}\frac{\partial^2 F_{\mu\nu}}{{\partial x^\alpha}{\partial x^\beta}}=
\frac{1}{c^2}\frac{\partial^2 F_{\mu\nu}}{\partial t^2}-
\frac{\partial^2 F_{\mu\nu}}{\partial x^2}-
\frac{\partial^2 F_{\mu\nu}}{\partial y^2}-
\frac{\partial^2 F_{\mu\nu}}{\partial z^2}=0.
\]
Of course, not every solution of $\Delta F=0$ is a solution to Maxwell's
equations.

\vskip 0.5cm
\subsection{Plane electromagnetic waves}
The exact solutions of Maxwell's equations in the whole 3-space, known as
{\it plane electromagnetic waves}, are interesting not as some models of
really  existing objects, but as an convenient way to introduce some
important characteristics of a class of EM-fields. The standard (i.e. the
most widely spread) way to define such a solution is the following: {\it
there is a rectangular coordinate system $(x,y,z)$, in which this solution
depends on one space variable only}. The time-dependence of the solution is
determined by the equations. Right now we note, that such a solution, if it
exists, will be {\it infinite}! In fact, if $z$ is the only coordinate, on
which the solution depends, even if the dependence on $z$ is finite, i.e.{\it
localized and without singularities}, with respect to the rest two
coordinates $(x,y)$ this solution is {\it constant}; even at $x\rightarrow
\infty,\ y\rightarrow \infty$ the values of the components of $E$ and $B$, or
$F_{\mu\nu}$, do not change. This simply means that the initial condition
occupies the whole 3-space, or its infinite subregion, with finite values for
the components of $E$ and $B$. Since the integral energy $W$ of every
solution does not depend on time, we calculate it, making use of the initial
condition, and obtain
\begin{equation}
W=\frac{1}{8\pi}\int{(E^2+B^2)}dxdydz=\infty.
\end{equation}
Let us now see how the plane wave looks like in the corresponding coordinate
system, where $E$ and $B$ depend on $z$ and $t$ only. Since the derivatives
with respect to $x$ and $y$ will be zero, from the wave equations we get
\[
E=\Bigl[E_1(ct+\varepsilon z),E_2(ct+\varepsilon z),E_3(ct+\varepsilon z)\Bigr],
\]
\[
B=\Bigl[B_1(ct+\varepsilon z),B_2(ct+\varepsilon z),B_3(ct+\varepsilon z)\Bigr],\
\varepsilon =\pm 1.
\]
Now the equations $divE=0$ and $divB=0$ require $E_3=const$ and $B_3=const$.
Let us put these constants equal to zero since we do not interest in constant
solutions. The Maxwell's equations reduce to
\[
\frac 1c \frac {\partial B_1}{\partial t}=\frac {\partial E_2}{\partial z},\
\frac 1c \frac {\partial E_2}{\partial t}=\frac {\partial B_1}{\partial z},
\]
\[
\frac 1c \frac {\partial E_1}{\partial t}=-\frac {\partial B_2}{\partial z},\
-\frac 1c \frac {\partial B_2}{\partial t}=\frac {\partial E_1}{\partial z}.
\]
These equations have the following solution:
\[
E=\Bigl[E_1(ct+\varepsilon z),E_2(ct+\varepsilon z),0\Bigr]=\Bigl[u(ct+\varepsilon z),p(ct+\varepsilon z),0\Bigr],
\]
\[
B=\Bigl[B_1(ct+\varepsilon z),B_2(ct+\varepsilon z),0\Bigr]= \Bigl[\varepsilon p(ct+\varepsilon z),-\varepsilon u(ct+\varepsilon z),0\Bigr].
\]
For the Poynting's vector we obtain
 $4\pi S=\Bigl[0,0,-\varepsilon c(u^2+p^2)\Bigr]$.
This solution, obviously, has the properties
\[
E.B=0,\ E^2-B^2=0,
\]
i.e. {\it the field has zero invariants}. Now we show in relativistic terms
how the requirement for zero invariants $I_1=I_2=0$ determines the solution
{\it plane wave}. According to subsection {\bf 1.1.2} at zero invariants the
energy-momentum tensor $Q_{\mu\nu}$, defined by (1.7), has only zero eigen
values and unique isotropic eigen direction, defined by the couple of
opposite isotropic vectors $ \pm V=\varepsilon V$. Making use of the
representation (1.13) for $F$ and $*F$ we obtain for $Q_{\mu\nu}$
\[
Q_\mu^\nu=-A_\sigma A^\sigma V_\mu V^\nu=-(A^*)_\sigma(A^*)^\sigma V_\mu V^\nu.
\]
On the solutions of Maxwell's equations we shall have
\[
0=\nabla_\nu Q_\mu^\nu=-A^2 V^\sigma\nabla_\sigma V_\mu-
V_\mu\nabla_\sigma(A^2 V^\sigma).
\]
This relation shows, that the integral lines of the vector field $V$ are {\it
isitropic geodesics}, i.e. straight lines. Let's now choose the coordinates
$(x,y,z,\xi)$ in such way that the integral lines of $V$ to lie entirely in
the plane $(z,\xi)$. Since $V_4\neq 0$ always, we can suppose $V_4=1$. Then
in these coordinates we shall have $V=(0,0,\varepsilon,1)$ and
\[
F_{12}=F_{34}=0,\  F_{13}=\varepsilon F_{14},\ F_{23}=\varepsilon F_{24},
\]
\[
A=(F_{14},F_{24},0,0),\
A^*=(-F_{23},F_{13},0,0)=(-\varepsilon A_2,\varepsilon A_1,0,0).
\]
Clearly, in these notations $A$ and $A^*$ are the relativistic equivalents of
$E$ and $B$ respectively.

Denoting  $F_{14}=u$, $F_{24}=p$, for ${\bf d}F=0$ and $\delta F=0$
we get
\[
{\bf d}F=\varepsilon(p_x-u_y)dx\wedge dy\wedge dz +
(p_x-u_y)dx\wedge dy\wedge d\xi+
\]
\[
\varepsilon (u_\xi-\varepsilon u_z)dx\wedge dz\wedge d\xi+
\varepsilon (p_\xi -\varepsilon p_z)dy\wedge dz\wedge d\xi =0,
\]
\[
\delta F=(u_\xi -\varepsilon u_z)dx+(p_\xi -\varepsilon p_z)dy+
\varepsilon(u_x + p_y)dz+(u_x + p_y)d\xi=0.
\]
From these equations it follows $u_x+p_y=0$ and $u_y-p_x=0$ and from these
last relations we get the equations $u_{xx}+u_{yy}=0,\ p_{xx}+p_{yy}=0$,
i.e. $u$ and $p$ are harmonic functions with respect to the variables
$(x,y)$. Since we are searching for {\it finite} solutions in the {\it whole
space}, from the well known properties of the harmonic functions it follows
that $u$ and $p$ do not depend on $(x,y,)$. Thus,
\[
u=u(\xi+\varepsilon z),\ p=p(\xi+\varepsilon z).
\]
It is seen that the dependence of the field components on the unique
space-variable $z$ stays arbitrary, i.e. {\it can not be determined by the
Maxwell's equations}. Obviously, if $(k,\varphi_0)$ and $(l,\psi_0)$ are two
couples of real numbers, then
$u(k\xi+\varepsilon kz+\varphi_0)$ and $p(l\xi+\varepsilon lz+\psi_0)$
define again a solution. On the other hand the numbers $k,l$ define two
constant isotropic vectors
${\bf k}=(0,0,\varepsilon k_3,k_4),\  {\bf l}=(0,0,\varepsilon l_3,l_4),\
k_3=k_4=k,\  l_3=l_4=l$,
so we can write down
$u(k\xi+\varepsilon kz+\varphi_0)=u(k_\mu x^\mu +\varphi_0);\
p(l\xi+\varepsilon lz+\psi_0)=p(l_\mu x^\mu +\psi_0)$.
The vectors $({\bf k},k),({\bf l},l)$, or just their space-like parts
${\bf k}$ and ${\bf l}$ are called {\it wave vectors} of the two independent
solutions
\[
F(u)=\varepsilon udx\wedge dz +udx\wedge d\xi,
\ F(p)=\varepsilon pdy\wedge dz +pdy\wedge d\xi,
\]
and the quantities $\varphi_0,\psi_0$ are called {\it phase constants}.
The quantities $(k_\mu x^\mu +\varphi_0),\ (l_\mu x^\mu+\psi_0)$ are called
{\it phases}. Every of the two solutions is called {\it linearly polarized
plane wave}.

It seems important to note that {\it the general plane wave is a sum, or a
linear combination, of two entirely independent (except the common direction
of propagation) linearly polarized plane waves}. Because of the linearity of
Maxwell's equations a sum of linearly polarized plane waves with different
directions of propagation is also a solution, but it is no more a plane wave.
To propagate as a whole along a definite direction is one of the specific
properties of the plane waves but this is not enough to use it as a model
solution for real objects because of their infinity: infinite volume,
infinite energy, infinite momentum and angular momentum.

A special interest for the theory is the choice of the two functions $u$ and
$p$ as elementary periodic functions, namely
\[
u=U_0cos(k_\mu x^\mu+\varphi_0),\ p=P_0cos(l_\mu x^\mu+\psi_0),
\]
since most of the real EM-fields show definite properties of {\it
periodicity}.  In such case the quantities $k_4/c$ and $l_4/c$ are usually
called {\it frequency} $\nu$, (or {\it circular frequency} $\omega=2\pi\nu$),
and the quantity $\lambda=|{\bf k}|^{-1}$ is called {\it wave length}. So we
can write down (e.g. for $u$)
\[
\frac{1}{|{\bf k}|}=\lambda=\frac c\nu =cT,
\]
where $T=1/\nu$ is called {\it period}. Such waves are usually
called {\it harmonic}. We'd like to note specially, that this {\it
time-periodicity is a consequence of the specially chosen initial condition,
namely that the function $u$, considered as a function of one variable, is a
periodic function of the space-variable $z$}. This periodicity is {\it
admissible} by the Maxwell's equations, but it is not a necessary consequence
of these equations. So, the introduced "wave" characteristics come from a
special class of initial conditions and nothing more. Finally we note that
the general plane wave is determined by 4 real parameters
$k,l,\varphi_0,\psi_0$ and 2 arbitrary functions of one (and the same)
independent variable.

Two linearly polarized harmonic plane waves of the kind
\[
E_1=\Bigl[U_0 cos(\nu t+\varepsilon \frac z\lambda +\varphi_0),\ 0,\ 0 \Bigr],\
B_1=\Bigl[0,\ -\varepsilon U_0cos(\nu t+\varepsilon \frac z\lambda +\varphi_0),\ 0 \Bigr];
\]
\[
E_2=\Bigl[0,\ P_0 sin(\nu t+\varepsilon \frac z\lambda +\varphi_0),\ 0 \Bigr],\
B_2=\Bigl[\varepsilon P_0 sin(\nu t+\varepsilon \frac z\lambda +\varphi_0),\ 0,\ 0 \Bigr]
\]
are called {\it consistent}. Summing them up we obtain again a harmonic plane
wave
\[
E=\Bigl[U_0 cos(\nu t+\varepsilon \frac z\lambda +\varphi_0),\
P_0sin(\nu t+\varepsilon \frac z\lambda +\varphi_0),\ 0 \Bigr],
\]
\[
B=\Bigl[\varepsilon P_0 sin(\nu t+\varepsilon \frac z\lambda +\varphi_0),\
-\varepsilon U_0cos(\nu t+\varepsilon \frac z\lambda +\varphi_0),\ 0 \Bigr].
\]
We see that the consistent linearly cycling $E$ and $B$ of the two harmonic
and linearly polarized waves create an illusion for circulating $E$ and
$B$ of their superposition, i.e. the couple of orthogonal vectors $(E,B)$
takes part in two motions: {\it advancing} along $z$ and {\it circulating} in
the plane $(x,y)$ left-wise or right-wise with the frequency of $\nu$. In
this way we get an impression about an elliptically (or circularly at
$U_0=P_0$) polarized plane wave.

Note that the harmonic plane wave occupies the whole 3-space, its
energy-density is the constant quantity $w\sim (U^2+P^2) $, its full energy is
infinite since the 3-dimensional integral of a constant over the whole space
${\cal R}^3$ is infinity.

Finally we note the transverse character of any plane wave, which is seen
from the transverse direction of $E$ and $B$ with respect to the direction of
propagation.

\subsection{Potentials}
According to the general notion of potential this is a scalar or some tensor
field, from which by differentiating (once or more) the {\it physical} field,
 i.e. the field that is physically measured, is obtained. The most frequent
case is a "single" differentiating, but there are examples of "double"
differentiating. Such is the case of gravitational field in the frame of
General Relativity, where the curvature field, which is identified with the
physical field, is obtained from the metric tensor through double
differentiating. As we shall see now, some of the solutions of Maxwell's
equations can also be obtained by double differentiating of some of the
solutions of the wave equation. This is the method of Hertz potentials.
In fact, if $A_1$ and $A_2$ are two solutions (vector fields) of the vector
wave equation $\Box A=0$, then the expressions
\[
E=rot(rotA_1)-\frac{\partial}{\partial \xi}rotA_2,\
B=rot(rotA_2)+\frac{\partial}{\partial \xi}rotA_1
\]
define a solution to Maxwell's equations. In relativistic notations we have
\[
(\Box G)_{\mu\nu}=-\Bigl[({\bf d}\delta + \delta {\bf d})G\Bigr]_{\mu\nu}=0.
\]
Then
\[
F={\bf d}\delta G=-\delta {\bf d}G
\]
is a solution to Maxwell's equations $\delta F=0,\ {\bf d}F=0$. In fact,
since ${\bf d} \circ {\bf d}=0$,\ $\delta \circ \delta =0$ and
$\Box \circ {\bf d}={\bf d}\circ \Box,\ \Box \circ \delta=\delta \circ \Box$,
then
\[
\delta F=\delta ({\bf d}\delta G)=(\Box - {\bf d}\delta )\delta G=
\delta \Box G=0,
\ {\bf d}F={\bf d}({\bf d}\delta G)=0.
\]
These solutions are used for description of the Hertz vibrator's radiation.
The spherical wave
\[
(A_1,A_2,A_3)=\Biggl[\frac {a_1(ct-r)}{r},\ \frac {a_2(ct-r)}{r},\
\frac {a_3(ct-r)}{r}\Biggr],
\]
which is a solution of the wave equation, but {\it is not} a solution to the
Maxwell's equations, is used in the above shown way to build a solution of
Maxwell's equations. Clearly, the standard choice of the components $a_i$ as
elementary harmonic functions brings a singularity at the point $x=y=z=0$.

The general solution of Maxwell's equations with non-zero and {\it not
depending} on the field electric current, is obtained by means of introducing
the 4-potential, i.e. an 1-form $A=A_\mu dx^\mu$ and defining $F={\bf d}A$,
which is {\it always} possible, since in CED we have in all cases
${\bf d}F=0$. The 1-form $A$ is determined up to a term of the kind
${\bf d}f$, since $A$ and $A+{\bf d}f$ give the same $F$. Then
\[
\delta F=\delta {\bf d}A=(\Box -{\bf d}\delta)A=\Box (A+{\bf d}f)-
{\bf d}\delta (A+{\bf d}f).
\]
We see that choosing $f$ as a solution to the equation
$-\delta {\bf d}f=\Box f=\delta A$
we can redefine $A$ as $A'=A+{\bf d}f$. Clearly $A$ and $A'$ define the same
$F$, and $\delta A'=0$. So, the 1-form $A'$ satisfies the equation
$\delta{\bf d}A'=\Box A'=4\pi j$, the solutions of which are well known when
the current $j$  {\it does not depend} on $A'$. Of course, when the change in
the mechanical energy of the charge-carriers is taken into account, and these
carriers do not "appear" and "disappear", then according to subsection
(1.1.2), equation (1.9), the 4-current $j_\mu=\rho u_\mu$ depends on the
field $F$, and the equations become nonlinear. Therefore, it is an illusion
to think, that the 4-potential approach solves the problem completely: in all
really significant cases the 4-current depends substantially on the field $F$
in accordance with equation (1.9), which, in turn, follows from the
energy-momentum conservation law and could hardly be put into some doubt.
This fact shows some inadequacy of the thesis for the universal character of
the 4-potential approach, its (always possible) introduction does not lead to
a complete solution of the entire problem. In all cases the energy-momentum
conservation requires nonlinear inter-relations between the current and the
field.

\section {\bf Amplitude and Phase}
\vskip 0.05cm
\subsection{Amplitude and phase of a plane wave}

The importance of the concepts of {\it amplitude} and {\it phase} in the
electromagnetic theory is out of any doubt, but sufficiently general and
universal definitions of these concepts in CED are still missing. Our purpose
in this section is to consider some new ways to introduce these concepts into
theory through a pure algebraic and coordinate free approach in both,
nonrelativistic and relativistic formalisms. We consider first the case of a
plane wave solution.

In the corresponding coordinate system the plane wave solution is
\[
F=\varepsilon U_0 cos(k_\mu x^\mu+\varphi_0)dx\wedge dz +
U_0 cos(k_\mu x^\mu+\varphi_0) dx\wedge d\xi+
\]
\[
\varepsilon P_0 sin(l_\mu x^\mu+\psi_0) dy\wedge dz +
P_0 sin(l_\mu x^\mu+\psi_0) dy\wedge d\xi,
\]
or, in terms of $E$ and $B$
\[
E=\Biggl[U_0cos(k_\mu x^\mu+\varphi_0),
\ P_0sin(l_\mu x^\mu+\psi_0),\ 0\Biggr],\
\]
\[
B=\Biggl[\varepsilon P_0 sin(l_\mu x^\mu+\psi_0),
-\varepsilon\ U_0 cos(k_\mu x^\mu+\varphi_0),\ 0\Biggr].
\]
As we mentioned, the quantity
$k_\mu x^\mu +\varphi_0=\nu t+\varepsilon z/\lambda+\varphi_0$
is called {\it phase} of the plane wave. As for the amplitude, according to
the usual sense of this concept, it is the {\it magnitude}, or the {\it
maximum value} of a given quantity. In our case we have a couple $(E,B)$ of
vector fields, so it seems natural to define the amplitude by the relation
$$
\sqrt{E^2+B^2}=\sqrt{U_0^2+P_0^2},
$$
i.e. a square root of the energy-density. For the vector product $E\times B$
we obtain
$$
E\times B=\Bigl[(0,0,-\varepsilon(U_0^2+P_0^2)\Bigr].
$$
This is a constant vector.

Now, the triple $(E,B,E\times B)$ defines a basis
of the tangent (or cotangent) space at every point, where the field is
different from zero (we assume $E\neq 0, B\neq 0$). Moreover this is an
orthogonal basis. We denote this basis by ${\cal R}_1$, so we can write
${\cal R}_1=(E,-\varepsilon B,-\varepsilon E\times B)$. From the properties
of the plane wave solutions we obtain $|E|=|B|$. But, the physical dimension
of the third vector $E\times B$ is different from that of the first two. So,
we introduce the factor $\alpha$
\[
\alpha=\frac{1}{\sqrt{\frac{E^2+B^2}{2}}}.
\]
Making use of $\alpha$, we introduce the basis
$$
{\cal R}=\Bigl[\alpha E,-\varepsilon\alpha B,-\varepsilon\alpha^2E\times B\Bigr].
$$
Hence, at every point we've got two bases:\ ${\cal R}$ and the
coordinate basis ${\cal R}_0=\Bigl[{\partial_x},{\partial_y},{\partial_z}\Bigr]$,
as well as the
corresponding co-bases ${\cal R}^*$ and ${\cal R}_0^*=(dx,dy,dz)$.  We are
interesting in the invariants of the corresponding transformation matrix
${\cal M}$ from ${\cal R}_0^*$ to ${\cal R}^*$.  It is defined by the
relation ${\cal R}_0^*.{\cal M}={\cal R}^*$.  So, we obtain
\[
{\cal M}=\left\|\matrix{
\alpha E_1   &-\varepsilon\alpha B_1 &-\varepsilon\alpha^2 (E\times B)_1 \cr
\alpha E_2   &-\varepsilon\alpha B_2 &-\varepsilon\alpha^2 (E\times B)_2 \cr
\alpha E_3   &-\varepsilon\alpha B_3 &-\varepsilon\alpha^2 (E\times B)_3
\cr}\right\|.
\]
We shall try to express
the amplitude and the phase of the plane wave as functions of the invariants
of this matrix. So, in all cases , where this is possible, the invariant
character of the so defined phase and amplitude will be out of doubt.  As it
 is well known, in general, every square $n\times n$-matrix ${\cal L}$ has
$n$ invariants $I_1,I_2,...,I_n$, where $I_k$ is the sum of all principle
minors of order $k$. The invariant $I_1({\cal L})={\cal L}_{11}+...+{\cal
L}_{nn}$ is the sum of all elements on the principle diagonal, and the
invariant $I_n=det({\cal L})$ is the determinant of the matrix. In our case
$n=3$, so for the invariant $I_2$ we get \[ I_2=\det\left\|\matrix{ m_{11}
&m_{12}\cr m_{21} &m_{22} \cr}\right\|+ \det\left\|\matrix{ m_{11} &m_{13}\cr
m_{31} &m_{33}
\cr}\right\|+
\det\left\|\matrix{
m_{22} &m_{23}\cr
m_{32} &m_{33}
\cr}\right\|.
\]
We compute $I_3({\cal R})$.
\[
I_3({\cal R})=\det\left\|\matrix{
\alpha E_1 &-\varepsilon \alpha B_1 &-\varepsilon \alpha^2(E\times B)_1\cr
\alpha E_2 &-\varepsilon \alpha B_2 &-\varepsilon \alpha^2(E\times B)_2\cr
\alpha E_3 &-\varepsilon \alpha B_3 &-\varepsilon \alpha^2(E\times B)_3
\cr}\right\|=\alpha^4(E\times B)^2.
\]
Now the amplitude ${\cal A}$ of the plane electromagnetic wave may be defined
as follows:
\[
{\cal A}=\sqrt[2]{\alpha^{-2} I_3({\cal R})}=
\sqrt[2]{\alpha^2 (E\times B)^2}=\alpha|E\times B|.
\]

\vskip 0.05cm
\subsection{Amplitude and phase of a general field}
The invariant character of the above given definition of the plane wave
amplitude suggests its natural extending to an arbitrary field. So, if the
couple $(E,B)$ represents the field, we introduce the matrix
${\cal M}({\cal R})$
of the basis
${\cal R}=\Bigl[\alpha E, -\alpha B, -\alpha^2 (E\times B)\Bigr]$
and define the amplitude ${\cal M}$ of the field by
\begin{equation}
{\cal A}(E,B)=\sqrt[2]{\alpha^{-2} I_3({\cal R})}=\alpha|E\times B|.
\end{equation}

We go further now to define the phase in the general case. We'll make use of
the matrix of the basis
$$
{\cal R}=\Bigl[\alpha E,-\alpha B, -\alpha^2E\times B\Bigr],
$$
defined by the general field $(E,B)$.
The invariants
\[
I_1({\cal R})=\alpha E_1-\alpha B_2-\alpha^2 (E\times B)_3,
\]
\[
I_2({\cal R})=-\alpha^2 (E\times B)_3+
	       \alpha^3 \Bigl[E(E.B)-B(E.E)\Bigr]_2+
	       \alpha^3 \Bigl[E(B.B)-B(E.B)\Bigr]_1,
\]
\[
I_3({\cal R})=\alpha^4 (E\times B)^2,
\]
obviously, are physically dimensionless. When the inequality
\[
\frac 12\biggl|I_1({\cal R})-1\biggr|\leq 1,
\]
holds, then the function $arccos$ is defined on the expression on the left.
In these cases, by definition, the phase $\varphi$ of the field $(E,B)$ shall
be defined by
\begin{equation}
\varphi=arccos\Biggl[\frac 12\Bigl[I_1({\cal
R})-1\Bigr]\Biggr]
\end{equation}
For the plane wave solution
\[
E=\Bigl[u(ct+\varepsilon z),p(ct+\varepsilon z),0\Bigr],
\ B= \Bigl[p(ct+\varepsilon z),-u(ct+\varepsilon z),0\Bigr]
\]
we get
\[
I_1({\cal R})=I_2({\cal R})=
\frac{2u}{\sqrt{u^2+p^2}}+1=\frac{2E_1}{|E|} +1,
\]
and for a circularly polarized plane monochromatic wave we get
$\varphi=k_\mu x^\mu+const$.

Let's now see when the basis ${\cal R}$ is normed, i.e. when
$$
 |\alpha E|=1,\ |\alpha B|=1,\ \alpha^2 |E\times B|=1.
$$
From the first two equations it follows $|E|=|B|$, and from the third
equation it follows $E.B=0$. Moreover, the relations $|E|=|B|,\ E.B=0$
follow from the third equation only: $\alpha^2|E\times B|=1$. So, the normed
character of ${\cal R}$ leads to its orthonormal character, consequently,
$\det{\cal M}({\cal R})=1$. Vice versa, the requirement
$\det{\cal M}({\cal R})=1$ leads to the orthonormal character of ${\cal R}$.
We obtain, that the requirement $\det{\cal M}({\cal R})=1$ is equivalent to
the null character of the field: $I_1=B^2-E^2=0,\ E.B=0$.

The relations obtained suggest to define and consider the following 4-linear
map:
\[
R(x,y,v,w)=\det\left\|\matrix{
x_1 &y_1 &(v\times w)_1\cr
x_2 &y_2 &(v\times w)_2\cr
x_3 &y_3 &(v\times w)_3
\cr}\right\|.
\]
The following relations hold:
\[
 R(x,y,v,w)=(x\times y).(v\times w)=\Bigl[y\times(v\times w)\Bigr].x=
\Bigl[(v\times w)\times x\Bigr].y,
\]
\[
 R(x,y,v,w)=-R(y,x,v,w),
\]
\[
 R(x,y,v,w)=-R(x,y,w,v),
\]
\[
 R(x,y,v,w)+R(x,v,w,y)+R(x,w,y,v)=0,
\]
\[
 R(x,y,v,w)=R(v,w,x,y),
\]
\[
 R(x,y,x,y)=(x\times y)^2.
\]
We note that this 4-linear map has all algebraic properties of the Riemannian
curvature tensor, therefore in the frame of this section, we shall call it
{\it algebraic curvature}. For the corresponding 2-dimensional curvature
$K(x,y)$, determined by the two vectors $(x,y)$ we obtain
\[
K(x.y)=\frac {R(x,y,x,y)}{x^2 y^2 -(x.y)^2}=\frac{(x\times y)^2}{x^2 y^2(1-cos^2 (x,y))}=
\frac{x^2 y^2 sin^2(x,y)}{x^2 y^2 sin^2(x,y)}=1.
\]
Let $(e_1,e_2,e_3)$ be a basis. We compute the corresponding Ricci tensor
$R_{ik}$ and the scalar curvature ${\bf R}$.
\[
R_{ijkl}=R(e_i,e_j,e_k,e_l)=(e_i\times e_j).(e_k\times e_l),
\]
\[
R_{ik}=\sum_l {R_i^l}_{kl}=(e_i \times e_1).(e_k \times e_1)+
(e_i \times e_2).(e_k \times e_2)+(e_i \times e_3).(e_k \times e_3),
\]
\[
{\bf R}=\sum_i R^i_i =2\Bigl[(e_1 \times e_2)^2 +(e_1 \times e_3)^2 +
(e_2 \times e_3)^2\Bigr].
\]
For our basis ${\cal R}_1$ we obtain the following non-zero components:
\[
R_{12,12}=4\frac{E^2.B^2}{(E^2+B^2)^2} sin^2(E,B),
\]
\[
R_{13,13}=R_{12,12}.\frac{2E^2}{E^2+B^2},\
R_{23,23}=R_{12,12}.\frac{2B^2}{E^2+B^2},
\]
and for the scalar curvature we get
\[
{\bf R}(E,B)=24\frac{E^2 B^2}{(E^2+B^2)^2}sin^2(E,B).
\]

After this short retreat let's go back to the  quantities {\it phase} and
{\it amplitude}. The above mathematical consideration suggests to try to
relate these two concepts with the notion of curvature in purely formal
sense, namely as a 2-form with values in the bundle $L_{T({\cal R}^3)}$ of
linear maps in the tangent bundle. Most generally, a 2-form $R$ with values
in the bundle $L_{T({\cal R}^3)}$ looks as follows
\[
R=\frac 12 R_{ijl}^{k} dx^i \wedge dx^j \otimes \frac{\partial}
{\partial x^k}\otimes dx^l.
\]
We have to determine the coefficients $R_{ijl}^{k}$,
i.e. we have to define a $(3\times 3)$-matrix ${\cal R}$ of 2-forms. We
define this matrix in the following way:
\[
{\cal R}=\left\|\matrix{
\alpha E_1 dy\wedge dz  &\alpha B_1 dy\wedge dz  &\alpha^2 (E\times B)_1 dy\wedge dz\cr
\alpha E_2 dz\wedge dx  &\alpha B_2 dz\wedge dx  &\alpha^2 (E\times B)_2 dz\wedge dx\cr
\alpha E_3 dx\wedge dy  &\alpha B_3 dx\wedge dy  &\alpha^2 (E\times B)_3 dx\wedge dy
\cr}\right\|.
\]
The columns of this matrix are the 2-forms $*E,\ *B,\ *(E\times B)$,
multiplied by the factor $\alpha$ at some degree in order to obtain
physically dimensionless quantities.

We aim to define the amplitude and the phase of the field $(E,B)$, making use
of this matrix. The amplitude ${\cal R}$ of the field we define by
\begin{equation}
{\cal A}=\frac{1}{3\alpha}R_{ijkl}R^{ijkl}= \frac{2}{3\alpha}\Biggl[1+2\frac
{(E\times B)^2}{(E^2+B^2)^2}\Biggr].
\end{equation}

In order to define the phase we first consider the 2-form  $tr\circ R$.
We get
\[
tr\circ {\cal R}=\alpha E_1dy\wedge dz +\alpha B_2 dz\wedge dx +
\alpha ^2 (E\times B)_3 dx\wedge dy.
\]
The square of this 2-form is
\[
(tr\circ {\cal R})^2=\alpha ^2\Biggl[(E_1)^2+(B_2)^2+\alpha ^2\bigl[(E\times
B)_3\bigr]^2\Biggr].
\]
Now the phase $\varphi$ of the field we define by
\begin{equation}
\varphi=arccos\Biggl[\sqrt{\biggl|\frac {(tr\circ R)^2-1}{2}\biggr|}\ \Biggr].
\end{equation}
whenever the right-hand expression is well defined.

Let's compute this quantity for a plane wave, moving along the
$z$-coor\-dinate from $-\infty$ to $+\infty$:
\[ E=\biggl[u(z-ct),\ p(z-ct),\ 0\biggr],
\ B=\biggl[-p(z-ct),\ u(z-ct),\ 0\biggr].
\]
We get
\[
(E\times B)=\biggl[0,\ 0,\ (u^2 +p^2)\biggr],\ \alpha =\frac {1}{\sqrt{u^2+p^2}},\
\]
\[
tr\circ R=\frac{u}{\sqrt{u^2+p^2}} dy\wedge dz +
\frac{u}{\sqrt{u^2+p^2}} dz\wedge dx +dx\wedge dy,\
(tr\circ R)^2=1+\frac{2u^2}{u^2+p^2},
\]
\[
{\cal A}=\sqrt{u^2+p^2},\ \varphi=arccos\Biggl(\frac{|u|}{\sqrt{u^2+p^2}}\Biggr)
\]
In the case of a plane harmonic wave $\varphi=k_\mu x^\mu +\varphi_0$.
Note that since in this coordinate system the components of the plane wave do
not depend on the coordinates $x$ and $y$, the corresponding 2-form
$tr\circ R$ is {\it closed}: ${\bf d}(tr\circ R)=0$.

If we work in relativistic terms, it is necessary to introduce some changes
in the matrix of 2-forms. First, we add one more column and one more row.
Second, instead of $*E,\ *B,\ *(E\times B)$ it is more convenient to use
their dual with respect to the pseudoeucledean $*$-operator
$*_4(*E),\ *_4(*B),\ *_4(*E\times B)$. So, the matrix ${\cal R}$ takes the
form

\[
\left\|\matrix{
\alpha E_1 dx\wedge d\xi  &\alpha B_1 dx\wedge d\xi  &\alpha^2 (E\times B)_1 dx\wedge d\xi   &\alpha^2 (E\times B)_1 dx\wedge d\xi\cr
\alpha E_2 dy\wedge d\xi  &\alpha B_2 dy\wedge d\xi  &\alpha^2 (E\times B)_2 dy\wedge d\xi   &\alpha^2 (E\times B)_2 dy\wedge d\xi\cr
\alpha E_3 dz\wedge d\xi  &\alpha B_3 dz\wedge d\xi  &\alpha^2 (E\times B)_3 dz\wedge d\xi   &\alpha^2 (E\times B)_3 dz\wedge d\xi\cr
0                         &0                         &0                                      &0
\cr}\right\|.
\]
Respectively, we obtain

\[
tr\circ R=\alpha E_1 dx\wedge d\xi +\alpha B_2 dy\wedge d\xi +
\alpha^2(E\times B)_3 dz\wedge d\xi,
\]
\[
\frac12 R_{\mu\nu\alpha\beta}R^{\mu\nu\alpha\beta}=-2,
\]
\[
(tr\circ R)^2=-\alpha ^2\Biggl[(E_1)^2+(B_2)^2+\alpha ^2\bigl[(E\times B)_3\bigr]^2\Biggr].
\]
In these terms the definitions for the {\it amplitude} ${\cal A}$ and the
{\it phase} $\varphi $ will look as follows

\begin{equation}
{\cal A}=-\frac{1}{4\alpha}R_{\mu\nu\alpha\beta}R^{\mu\nu\alpha\beta},\
\varphi=arccos\Biggl[\sqrt{\biggl|\frac{|(tr\circ R)^2|-1}{2}\biggr|}\ \Biggr].
\end{equation}
For the solution plane wave the 2-form $tr\circ R$ is

\[
tr\circ R=\frac{u}{\sqrt{u^2+p^2}} dx\wedge d\xi +
\frac{u}{\sqrt{u^2+p^2}} dy\wedge d\xi +dz\wedge d\xi
\]
and in the general case it is not closed. Since $cos\varphi=u/\sqrt{u^2+p^2}$,
then the equation ${\bf d}(tr\circ R)=0$ requires (in this coordinate system)
the following conditions to be fulfilled:
$\varphi_x =\varphi_y ,\ \varphi_z =0$.

\vskip 1cm
With these remarks on the amplitude and phase of the $EM$-field in CED we
come to a close of our short review of the basic principles, concepts and
relations  in Maxwell's theory. As it is clear from what we presented so
far our purpose is not to describe positive prescriptions for getting results
in this theory. We have tried to pay attention to those moments of the
theory, which show directly or indirectly, its frame of applicability. Doing
this, we get a clearer notion of how and what to alter in view of what kind
of objects we want to describe. In the next section we summarize those points
of of Maxwell's theory in order to have clearer and more definite motivation
for the appropriate to our aims changes in the theory.

\section {\bf Why and What to Change in  \\
Classical Electrodynamics}
\vskip 0.05cm
If the question {\it why do we want to change something in CED} is raised,
we respond shortly in the following way: {\it because we want to enrich CED
with new areas of applicability, extending in a natural way the class of
admissible solutions, aiming to describe (3+1)-dimensional soliton-like
objects in the pure field case, as well as in the presence of active external
fields (media)}.

\vskip 0.5cm
At the end of the last and the beginning of this century it has become clear
that some experimentally established facts can not be understood and
explained in the frame of Faraday-Maxwell's electrodynamics. For example,

1.\ Why the initiation of photochemical reactions depends on the color and
not on the intensity of light?

2.\ Why the velocities of the photoelectrons does not depend on the intensity
of light?

3.\ Why the shortwave radiation is emitted from bodies at high temperature?

4.\ Why the shortwave radiation is chemically more active than the longwave
one?

More generally: {\it why the influence of light on matter depends
qualitatively on its color and not on its intensity}?

These and other experimentally established facts motivate a serious analysis
of Maxwell's electrodynamics. In result, Planck and, later, Einstein set fort
the discrete point of view on the structure of the electromagnetic field. The
later experiments of Compton proved the truth of this viewpoint and the
notion of {\it photon} as an elementary electromagnetic formation (natural
object) was created. Soon the photons were provided with integral
characteristics like {\it frequency, energy, momentum, spin}. The Planck
constant $h$ turned into an omnipresent parameter, serving to separate the
real photons from other theoretically admissible electromagnetic formations.
The Planck's formula $E=h\nu$ is the essential criterion for reality of the
objects considered. This formula clearly says that only those elementary
electromagnetic formations can really exist, the existence of which is
strongly bound up with the availability of an intrinsic periodic process
with a period of $T=1/\nu$, and the corresponding to this process integral
action $E.T$ is exactly equal to the Planck constant: $h=E.T$. May be this
limitation seems to be very strong, but we have no reasons not to trust it so
far. Anyway, it is clear that the Planck formula separates a class of natural
objects being characterized by the elementary action of $h$. Moreover, this
intrinsic periodic process and the fact, that every photon moves as a whole
uniformly by the same velocity $c$, no matter what its frequency is, support
the notion that they are finite soliton-like objects. Otherwise it is
hardly understandable where the quantity {\it frequency} can come from as a
characteristic of a {\it free and uniformly moving point-like, i.e.
structureless, object}.

These and other circumstances set a clear chalenge before the those days
theoretical physics: description of 3-dimensional, finite  soliton-like
objects, having all integral properties of the free photons. The established
through time quantum-probabilistic approach does not solve this problem since
it is built on other principles and purcues other objectives. Modern quantum
electrodynamics, although its serious achievements in describing the atomic
phenomena, also works with the assumption "structureless photon" and is
interested mainly in its integral propeties.

In view of our intention to build a soliton-like model of the free photon a
serious analysis of the initial and basic principles of classical
electrodynamics had to be made. The purpose of this ananlysis is to find what
to preserve and what to change, i.e. to find those points of the
Faraday-Maxwell theory, the appropriate change of which will be of use and
will not bring us to any undesirable complications.
Being fully aware of the  significance  of Faraday-Maxwell theory in
physics and in all our knowledge of the natural world, we present our
conclusions fairly and in a transperant way as far as we are able to do this.
As we said earlier, in doing this we follow the rule that {\itshape the
respect and esteem paid to the creators can not be honest and genuine
if they are not in correspondence with the respect and esteem paid to the
Truth}.

\vskip 0.5cm

\noindent{\bf 1}.\ The elementary spherically symmetric and topologically
nontrivial solution (see 1.2.2) for the electric field $E$ ($B=0$), which
could be identified with the only spherically symmetric representative of the
corresponding cohomology class, defines in fact the electric charge as a
topological invariant. The important point here, we'd like to mention, is the
static character of the solution-field obtained, so where a test particle,
placed somewhere in the field, should take {\it momentum} from in order to
change its own momentum according to the momentum conservation law, i.e. the
equations of motion, is not quite clear. Since $B=0$, the Poynting vector
$S=E\times B$, traditionally considered as describing the local
momentum-transport of the field, is equal to zero. Clearly, the static
character of the field is an illusive feature and does not give an adequate
picture of a real situation.  But it is an {\it exact}, so a trustworthy,
solution! Because of the radial direction of the field strength, i.e. of the
particles' momentum change, it seems hardly possible to avoid the notion,
that some {\it real objects move radially and carry momentum to and from the
source}. Since there is no momentum accumulation at the source object the
same momentum has to be carried "to" and "from" for a unit time. As for the
intrinsic mechanism of momentum exchange between the field and the particle
CED tells nothing, it gives just the final integral effect. For the
general static case, according to (1.2.3), because of the well known
properties of the solutions to Laplace equation, they are not suitable for
models of real objects. So, our general conclusion is that {\it no static
solution of Maxwell's equations presents a sufficiently adequate picture of
real objects and processes} .

\vskip 0.5cm
\noindent {\bf 2.}\ The computation of the full interaction energy for two
spherically symmetric fields in (1.2.3) clearly shows, that while the
interaction of the two fields is a local fact and takes place at every point
where the two fields are different from zero, the interaction energy is an
integral characteristic and in no case should be identified with some
potential. From this point of view the Coulomb force, describing this
interaction as a change of the mechanical momentum of the charge-carriers,
can not be a local characteristic, since it takes into account the
contribution to this process of the interaction at all points. The "close"
form of $*\omega$ and $dW$, where $W$ is considered as a function of the
points, where the two charges stay, is not a sufficient ground the {\it
integral} characteristic "interaction force" to be identified with $*\omega$.
As for the potential $U$, introduced by the relation $dU=*\omega$, it is a
local characteristic too, so $dU\neq dW$ always.

\vskip 0.5cm
\noindent {\bf 3.}\ The considerations in (1.2.4) are of basic significance
for us: every localized finite initial condition determines unique solution
of the wave equation, the future time-behaviour of which could shortly be
characterized as a "radial blow-up". So, the same is true for the pure field
Maxwell equations. An important feature of the 3-dimensional case is the
availability of "forefront" and "backfront", which simply means, that every
point of the space will "feel" the field a finite period of time, after which
it will "forget" what happened. This result leaves no chance and hopes for
making use of Maxwell equations to obtain soliton-like solutions, they have
no such solutions in the whole space. Although undesirable, this conclusion
is unavoidable. This is the mathematical reality and we have to accept it
with the corresponding respect.

\vskip 0.5cm
\noindent{\bf 4.}\ As we noted in (1.2.5) the popular and well liked solution
of the pure field Maxwell equations {\it plane wave} is {\it unphysical,
unrealistic}, since it is {\it infinite: \ it possesses infinite integral
energy and occupies an infinite 3-volume}. Besides, all periodic-wave
solutions are defined by appropriate initial conditions, so the wave
characteristics like {\it period, frequency} and {\it wave vector} come from
these initial conditions, i.e. they are admissible by the equations but are
not necessary for all solutions. So, other kind of solutions, having no such
characteristics, are also admissible. But, electromagnetic radiation without
such characteristics, has never been observed and is hardly possible. If this
is true, what to do with such solutions in view of the desired adequacy
between the theory and the reality. Note that the representation of the plane
wave $u(z-ct)$ through simple monochromatic plane waves (the so called
wave packets) does not solve the problem since these packets are not
time-stable objects.

\vskip 0.5cm
\noindent{\bf 5.}\ The 4-potential approach is not of interest
from our point of view, since for the pure field case it reduces the problem
to a subclass of solutions of the wave equation for the components $A_\mu$ of
this 4-potential, namely those, satisfying the additional condition
$\nabla_\mu A^\mu =0$, and so soliton-like solutions are excluded. In the
presence of a current the problems with the nonlinear dependence of the
current on the field, mentioned in (1.2.6), appear and are hardly avoidable
in general. From our viewpoint the 4-potential approach is important rather
to legalize the gauge fields in theoretical fields, although at an elementary
(linear) abelean level $U(1)$. Unfortunately, even for nonabelean gauge
fields, leading to the nonlinear Yang-Mills equations, (3+1)-soliton-like
solutions are not found, moreover, there are some studies, showing that in
the pure field case there are no such solutions. Let's not forget, that the
so called {\it instantons}, i.e. solutions to the equations  $*F=F$, are
possible only at {\it positively defined} space-time metric and {\it zero
energy-momentum tensor}, so they could be hardly considered as models of real
objects. As for the {\it monopoles}, they want additional field, interacting
with the gauge field in a special way.

\vskip 0.5cm
\noindent{\bf 6.}\ In the case of continuous media CED adds to the free
current $j^\mu$ an additional current $j^\mu_{b}$, called {\it bound}. The
two new vectors $P$ and ${\cal M}$ are introduced (see 1.1.3) in the same way
as the vectors $E$ and $B$ are connected to the free current. So, the number
of the unknown functions becomes greater than the number of the equations,
which makes the things difficult to overcome as the historic development
shows. The introduction of coefficients-material constants by means of series
developments seems to be a very useful practical skill, but it can not serve
as a promising theoretical idea. The inherited inertia in thinking that
energy-momentum exchange with the field can occur {\it only if charge
carrying particles are present} still bears upon our minds. The
{\it hypothesis} that the Faraday induction law is universally (i.e. for {\it
all media, including vacuum}) valid has turned almost into a dogma, which, by
the way, forbids energy-momentum exchange through $*F$. Even if some brave
investigators admit an energy-momentum exchange between the field and the
medium to occur through $*F$, they begin to talk about magnetic charges (in
analogy with the electric case) having similar to the electric
charges properties.

From our point of view the real and important moment is the very {\it
energy-momentum exchange}, and how this exchange is realized is an additional
problem, depending on the special case considered. So, {\it the most
important step in our approach is to find an appropriate model equation for
this exchange, because this is the universal characteristic property of
every interaction in nature}. All further specializations depend on the case
under consideration.

\vskip 0.5cm
Hence, in view of what was said so far and in view of the purposes we set, we
come to the following conclusion. The algebraic construction {\it a couple of
vector fields} $(E,B)$ on  ${\cal R}^3$, or a {\it differential 2-form on
the Minkowski space-time}, is in general adequate to the field as far
as it reflects well enough its algebraic and general polarization
properties. Maxwell's equations do not reflect adequately enough the local
properties of the field, therefore not all solutions can represent
satisfactory models of real objects. So, our choice is to preserve, though in
an altered form, the basic algebraic picture of the field in relativistic
terms, but we'll replace Maxwell's equations with new nonlinear equations, the
physical sense of which is to define how the local energy-momentum exchange
between the field and some other continuous physical object (medium or field)
is carried out. The reason to turn to the local energy-momentum conservation
laws reflects our point of view that they are {\it more hopeful} and {\it
more universal}.

\vskip 0.5cm
Let us outline in few words our notion of the objects we want to describe. As
we mentioned in (1.2.1) they must be {\it extended}, but {\it finite} and {\it
time-stable}. Besides, their existence must be strongly connected with an
{\it internal and intrinsic dynamics, in particular - periodic process}. The
characteristics of the internal dynamics must be in strong relation with the
integral characteristics of the object. It is necessary to find invariant
quantities, separating the really existing objects from all theoretically
admissible. If some interaction processes take place, a transformation of
these objects to each other or to new ones, obeying definite conservation
laws, should be possible. Various internal structures at different levels,
as well as creation of stable structures out of these objects, should be also
possible. A stability with respect to external disturbances must be
available, so that such a disturbance to result finaly in the behaviour
of the object as a whole: {\it uniform motion} when there are no external
disturbances and {\it accelerated motion} in presence of a bearable
external disturbance. From mathematical point of view this means that at any
moment the functions, describing our object(s), shall be different from zero
inside a finite {\it connected} 3-dimensional region with trivial or
non-trivial topology, while the time-evolution should be determined by the
dynamic field equations.

Passing to the formulation of the basic principles and their mathematical
adequacies of what will be called further {\it Extended Electrodynamics}
(EED), we are fully aware of, that at this initial stage of our investigation
the following two things should be obeyed. First, from purely pragmatic point
of view, it seems better to preserve as admissible all solutions to Maxwell's
equations as exact solutions to the new equations and to use them whenever it
is possible. Second, the local energy-momentum conservation laws will
probably define "weak" equations, so additional conditions (initial data or
some new equations), reflecting some new specific features of the objects
under consideration, have to be imposed on the solutions. In such cases we
shall make use of the considerations and conclusions in (1.2.1).  We prefer
to work in relativistic terms, since we consider this language more adequate
to the physical situations we want to describe.

Finally, we assume the {\it general covariance principle}. i.e. the
understanding that  physical sense may have those concepts and statements,
which do not depend on the local coordinates used. Accordingly, we'll aim at
a coordinate-free formulation of the basic statements and equations in all
cases when this is possible, paying no attention to the simple Minkowski
space background used.

\section {\bf Extended Electrodynamics}
\vskip 0.05cm
\subsection{Physical conception for the EM-field in EED}
As it was mentioned, the mathematical models in CED of the real
electromagnetic fields in vacuum are "infinite", or if they are finite, they
are strongly time-unstable. These models are not consistent with a number of
experimental facts. A deeper analysis resulted in the new conception for a
discrete character of the field. This physical understanding of the field is
the true foundation of EED and it shows clearly the principle differences
between CED and EED. For the sake of clarity we shall formulate our point of
view more explicitly.

{\it The electromagnetic field in vacuum is of discrete character and
consists of single, not-interacting (or very weakly interacting) finite
objects, called photons. All photons move uniformly as a whole by the same
velocity c, carry finite energy $E$, momentum {\bf p} and intrinsic angular
momentum. These features imply structure and internal periodic process of
period $T$, which may be different for the various photons. The quantity
$E.T$, called "elementary action", has the same value for all photons and is
numerically equal to the Planck constant $h$. The invariance of $c$ and $h$
means nondistinguishability of the photons, considered as invariant objects.
The integral value of the intrinsic angular momentum is equal to $\pm h$. For
the topology of the 3-dimensional region, occupied by the photon at any
moment, there are no experimental data, so it is desirable the
model-solutions to admit arbitrary initial data}.

We'd like to stress once more: EED considers photons as {\it real objects},
and {\it not as convenient theoretical concepts}, and it aims to build
adequate mathematical models of these entities. So, the first important
problem is to point out the algebraic character of the modeling mathematical
object for a single photon. The corresponding generalization for a number of
photons is easily done (subsec.2.4).

\vskip 0.5cm
\subsection{Choice of the modeling mathematical object}
According to the non-relativistic formulation of CED the electromagnetic
field has two aspects: "electric" and "magnetic". These two aspects of the
field are described by two 1-forms on ${\cal R}^3$ and a  parametric
dependence on time is admitted: the electric field $E$ and the magnetic field
$B$. The considerations made in (1.1.1) brought us to the conclusion that
these two fields can be considered as two {\it vector components} of a new
object, 1-form $\Omega$, {\it taking values in a real 2-dimensional vector
space}, naturally identified with ${\cal R}^2$. This mathematical object
unifies and, at the same time, distinguishes the two sides of the field:
there is a basis in ${\cal R}^2$, in which the electric and magnetic
components are delimited, but in an arbitrary basis the two components mix
(superimpose), so the difference between them is deleted. The physically
important quantity, energy density, is given by the sum $E^2+B^2$. This point
of view seems appropriate and relevant in view of the pointed out in (1.1.1)
invariance of Maxwell's equations with respect to some linear
transformations, mixing $E$ and $B$.  All unimodular such transformations keep
the energy-density unchanged.

In the relativistic formulation of CED the difference between the electric
and magnetic components of the field is already quite conditional, and from
invariant-theoretical point of view there is no any difference. However, the
2-component character of the field is kept in a new sense and manifests
itself at a different level. In fact, as we mentioned above, some linear
combinations of the electric and magnetic fields generate a new solution to
Maxwell's equations. In particular, such is the transformation, defined by
the matrix
\[
\left\|\matrix{0   &1\cr
	       -1  &0\cr} \right\|.
\]
This matrix, defining a complex structure in ${\cal R}^2$, transforms a field
of the kind $(E,0)$ into a new field of the kind $(0,E)$ and a field of the
kind $(0,B)$ into a field of the kind $(-B,0)$, i.e. the electric component
into magnetic and vice versa. This observation draws our attention to looking
for a complex structure $J,J^2=-id$ in the bundle of 2-forms on the Minkowski
space, such that if $F$ presents the first component of the field, then
$J(F)$ to present the second component of the same field. Such complex
structure truly exists and, according to (1.2.1), it coincides with the
restriction of the Hodge $*$-operator, defined by the pseudometric
$\eta$, to the space of 2-forms: $**_2=-id$. So, the non-relativistic vector
components $(E,B)$ are replaced by the relativistic vector components
$(F,*F)$. The following considerations support also such a choice.

The relativistic Maxwell's equations in vacuum ${\bf d}F=0,\ {\bf d}*F=0$
are, obviously, invariant with respect to the interchange $F\rightarrow *F$.
Moreover,if $F$ is a solution, then an arbitrary linear combination $aF+b*F$
is again a solution. More generally, if $(F,*F)$ defines a solution, then the
transformation $(F,*F)\rightarrow (aF+b*F,mF+n*F)$ defines a new solution for
an arbitrary matrix
\[
\left\|\matrix{a   &m\cr
	       b   &n\cr} \right\|.
\]
Now, using the old notation $\Omega$ for the new object
$\Omega=F\otimes e_1+*F\otimes e_2$, Maxwell's equations are written down
as ${\bf d}\Omega=0$, or equivalently $\delta \Omega=0$. Clearly, an
arbitrary linear transformation of the basis $(e_1,e_2)$ keeps $\Omega$ as a
solution.

Recall the energy-momentum tensor in CED, defined by (1.7)
\[
Q_\mu^\nu=\frac {1}{4\pi}\biggl[\frac 14 F_{\alpha\beta}F^{\alpha\beta}\delta_\mu^\nu-F_{\mu\sigma}F^{\nu\sigma}\biggr]=\\
\frac {1}{8\pi}\biggl[-F_{\mu\sigma}F^{\nu\sigma}-(*F)_{\mu\sigma}(*F)^{\nu\sigma}\biggr].
\]
It is quite clearly seen, that $F$ and $*F$ participate in the same way in
$Q_\nu^\mu$, and the full energy-momentum densities of the field are obtained
through summing up the energy-momentum densities, carried by $F$ and $*F$.
Since the two expressions $F_{\mu\sigma}F^{\nu\sigma}$ and
$(*F)_{\mu\sigma}(*F)^{\nu\sigma}$ are not always equal, the distribution of
energy-momentum between $F$ and $*F$ may change in time, i.e. energy-momentum
may be transferred from $F$ to $*F$, and vice versa. So we may interpret this
phenomenon as a special kind of interaction between $F$ and $*F$, responsible
for some internal redistribution of the field energy. Now, in vacuum it
seems naturally to expect, that the energy-momentum, carried from $F$ to
$*F$ in a given 4-volume, is the same as that, carried from $*F$ to $F$ in
the same volume. However, in presence of an active external field (medium),
exchanging energy-momentum with $\Omega$, it is hardly reasonable to trust
the same expectation just because of the specific structure the external
field (medium) may have. So, the external field (further  any such external
field will be called {\it medium} for short) may exchange energy-momentum
preferably by $F$ or $*F$, as well as it may support the internal
redistribution of the field energy-momentum between $F$ and $*F$, favouring
$F$ or $*F$. From the explicit form of the energy-momentum tensor it is seen
that the field may participate in this exchange by means of any of the two
terms $F_{\mu\sigma}F^{\nu\sigma}$ and $(*F)_{\mu\sigma}(*F)^{\nu\sigma}$.
Moreover, from the expression (1.8) for the divergence of the energy-momentum
tensor
\[
\nabla_\nu Q_\mu^\nu=\frac {1}{4\pi}\biggl[F_{\mu\nu}(\delta F)^\nu+(*F)_{\mu\nu}(\delta *F)^\nu\biggr]
\]
it is also clearly seen that the quantities of energy-momentum, which any of
the two components $F$ and $*F$ may exchange in a unit 4-volume are {\it
invariantly} separated and  given respectively by
$$
F_{\mu\nu}(\delta F)^\nu, \quad (*F)_{\mu\nu}(\delta *F)^\nu.
$$
But, in CED the exchange through $*F$ {\it is forbidden}, the expression
\linebreak $(*F)_{\mu\nu}(\delta *F)^\nu$ is always equal to zero. Of course,
we do not reject the existence of such media, but we do not share the
opinion that all media behave in this same way just because this can not be
verified.  On the other hand, in case of vacuum, we can not delimit $F$ from
$*F$, these are two solutions of the same equation and it is all the same
which one will be denoted by $F$ (or $*F$), i.e. CED does not give an
intrinsic criterion for a respective choice. Only in regions with non-zero
free charges and currents, when ${\bf d}F=0$ and $\delta F=j\neq 0$, the
choice can be made , but this presupposes (postulates) that the field is able
to interact, i.e. to exchange energy-momentum, {\it only} with charged
particles. This postulate we can not assume ad hoc.

Having in view these considerations we assume the following postulate in EED
in order to specify the algebraic character of the modeling mathematical
object:

\vskip 0.5cm
{\it In EED the electromagnetic field is described by a 2-form $\Omega$,
defined on the Minkowski space-time and valued in a real 2-dimensional vector
space ${\cal W}$ and such, that there is a basis $(e_1,e_2)$ of ${\cal W}$ in
which $\Omega$ takes the form}
\begin{equation}
\Omega = F\otimes e_1 + *F\otimes e_2.
\end{equation}

\vskip 0.5cm
Since ${\cal W}$ is isomorphic to ${\cal R}^2$, further we shall write only
${\cal R}^2$ and all relations obtained can be carried over to ${\cal W}$ by
means of the corresponding isomorphism. In particular, every ${\cal W}$ will
be considered as being provided with a complex structure $J$, so, the group of
automorphisms of $J$ is defined. Our purpose now is to prove that the set of
2-forms of the kind (1.37) is stable under the invariance group of $J$.

First we note, that the equation $aF+b*F=0$ requires $a=b=0$. In fact, if
$a\neq 0$ then $F=-\frac{b}{a}*F$. From $aF+b*F=0$ we get $a*F-bF=0$ and
substituting $F$, we obtain $(a^2+b^2)*F=0$, which is possible only if
$a=b=0$ since $*F\neq 0$. In other words, $F$ and $*F$ are linearly
independent.
Let now $(k_1,k_2)$ be another basis of ${\cal R}^2$ and consider the
2-form $\Psi =G\otimes k_1+*G\otimes k_2$. We express $(k_1,k_2)$ through
$(e_1,e_2)$ and take in view what we want
\[
G\otimes k_1+*G\otimes k_2=G\otimes (ae_1+be_2)+*G\otimes (me_1+ne_2)=
\]
\[
=(aG+m*G)\otimes e_1+(bG+n*G)\otimes e_2=(aG+m*G)\otimes e_1+
*(aG+m*G)\otimes e_2.
\]
Consequently, $bG+n*G=a*G-mG$, i.e. $(b+m)G+(n-a)*G=0$, which
requires $m=-b,\ n=a$, i.e. the transformation matrix is
\[
\left\|\matrix{a   &-b\cr
	       b   &a\cr} \right\|.
\]
Besides, if ${\Omega}_1$ and ${\Omega}_2$ are of the kind (1.37), it is
easily shown that the linear combination $\lambda\Omega_1+\mu\Omega_2$ is of
the same kind (1.37). These results show that the 2-forms of the kind (1.37)
form a stable with respect to the automorphisms of $({\cal R}^2,J)$ subspace
of the space $\Lambda ^2(M,{\cal R}^2)$.

We note that the following 2-forms: $F\otimes e_1+*F\otimes e_2$, \
 $F\otimes k_1+*F\otimes k_2$, are different, i.e. it is important which basis
will be used. In order to separate the class of bases we are going to use,
first we recall the product of 2 vector valued differential forms. If
$\Phi$ and $\Psi$ are respectively $p$ and $q$ forms on the same manifold $N$,
taking values in the vector spaces $W_1$ and $W_2$ with corresponding bases
$(e_1,...,e_m)$ and $(k_1,...,k_n)$, and
$\varphi :W_1\times W_2\rightarrow W_3$ is a bilinear map into the vector
space $W_3$, then a $(p+q)$-form $\varphi\left(\Phi,\Psi\right)$ on $N$ with
values in $W_3$ is defined by
\[
\varphi\left(\Phi,\Psi \right)=\sum_{i,j}\Phi^i\wedge \Psi^j \otimes \varphi(e_i,k_j).
\]
In particular, if $W_1=W_2$ and $W_3={\cal R}$, and the bilinear map is
scalar (inner) product $g$, we get
\[
\varphi\left(\Phi,\Psi \right)=\sum_{i,j}\Phi^i\wedge \Psi^j g_{ij}.
\]
Let now $X$ and $Y$ be 2 arbitrary vector fields on the Minkowski space $M$,
$\Omega$ be of the kind (1.37), $Q_{\mu\nu}$ be the energy tensor in CED and
$g$ be the canonical inner product in ${\cal R}^2$. Then the class of bases
we shall use will be separated by the following equation \begin{equation}
Q_{\mu\nu}X^\mu Y^\nu=\frac12 *g\Bigl(i(X)\Omega,*i(Y)\Omega\Bigr).
\end{equation}
We develop the right hand side of this equation and obtain
\[
\frac12 *g\Bigl(i(X)\Omega,*i(Y)\Omega\Bigr)=
\]
\[
\frac12 *g\Bigl(i(X)F\otimes e_1+i(X)*F\otimes e_2,*i(Y)F\otimes e_1+*i(Y)*F\otimes e_2\Bigr)=
\]
\[
=\frac12*\biggl[\Bigl(i(X)F\wedge*i(Y)F\Bigr)g(e_1,e_1)+
\Bigl(i(X)F\wedge*i(Y)*F\Bigr)g(e_1,e_2)+
\]
\[
+\Bigl(i(X)*F\wedge*i(Y)F\Bigr)g(e_2,e_1)+
\Bigl(i(X)*F\wedge*i(Y)*F\Bigr)g(e_2,e_2)\biggr]=
\]
\[
=-\frac12 X^\mu Y^\nu\biggl[F_{\mu\sigma}F_\nu^\sigma g(e_1,e_1)+
(*F)_{\mu\sigma}(*F)_\nu^\sigma g(e_2,e_2)+
\]
\[
+\Bigl(F_{\mu\sigma}(*F)_\nu^\sigma+(*F)_{\mu\sigma}F_\nu^\sigma\Bigr)g(e_1,e_2)\biggr]=
-\frac12 X^\mu Y^\nu\biggl[F_{\mu\sigma}F_\nu^\sigma +
(*F)_{\mu\sigma}(*F)_\nu^\sigma\biggr].
\]
In order this relation to hold it is necessary to have
\[
g(e_1,e_1)=1,\ g(e_2,e_2)=1,\ g(e_1,e_2)=0,
\]
i.e., we are going to use {\it orthonormal} bases. So, the stability group of
the subspace of forms of the kind (1.37) is reduced to $SO(2)$ or $U(1)$. So,
in this approach, the group $SO(2)$ appears in a pure algebraic way, while in
the gauge interpretation of CED this group is associated with the equation
${\bf d}F=0$, i.e. with the traditional and not shared bu us understanding,
that the $EM$-field can not exchange energy-momentum with some medium through
$*F$.

\vskip 0.5cm
\subsection{Differential equations for the field}
We proceed to the main purpose, namely, to write down differential equations
for our object $\Omega$, which was chosen to model the $EM$-field. We shall
work in the orthonormal basis $(e_1,e_2)$, where the field has the form
(1.37). The two vectors of this basis define two mutually orthogonal
subspaces $\{e_1\}$ and $\{e_2\}$, such that the space ${\cal R}^3$ is a
direct sum of these two subspaces: \ ${\cal R}^2=\{e_1\}\oplus\{e_2\}$. So,
we have the two projection operators $\pi_1:{\cal R}^2\rightarrow \{e_1\},\
\pi_2:{\cal R}^2\rightarrow \{e_2\}$. These two projection operators extend
to projections in the ${\cal R}^2$-valued differential forms on $M$:
\[
\pi_1 \Omega =\pi_1(\Omega^1\otimes k_1+\Omega^2 \otimes k_2)=
\Omega^1 \otimes \pi_1 k_1 +\Omega^2 \otimes \pi_1 k_2=
\]
\[
=\Omega^1 \otimes \pi_1(ae_1+be_2) + \Omega^2\otimes \pi_1(me_1+ne_2)=
(a\Omega^1 +m\Omega^2)\otimes e_1.
\]
Similarly,
\[
\pi_2\Omega=(b\Omega_1+n\Omega_2)\otimes e_2.
\]
In particular, if $\Omega$ is of the form (1.37), then
\[
\pi_1(F\otimes e_1+*F\otimes e_2)=F\otimes e_1,\
\pi_2(F\otimes e_1+*F\otimes e_2)=*F\otimes e_2.
\]

Let now our $EM$-field $\Omega$ propagates in a region, where some other
continuous physical object (external field, medium) also propagates and
exchanges energy-momentum with $\Omega$. We are going to define explicitly
the local law this exchange obeys.

First we note, that the external field is described by some mathematical
object(s). From this mathematical object, following definite rules,
reflecting the specific situation under consideration, a new mathematical
object  ${\cal A}_i$ is constructed and this new mathematical object
participates directly in the exchange defining expression. The $EM$-field
participates in this exchange defining expression directly through $\Omega$,
and since $\Omega$ takes values in ${\cal R}^2$, then ${\cal A}_i$ must also
take values in ${\cal R}^2$.

We make now two preliminary remarks. First, all operators, acting on the usual
differential forms, are naturally extended to act on vector valued
differential forms according to the rule $D\rightarrow D\times id$. In
particular,
\[
*\Omega=*(\sum_i \Omega^i\otimes e_i)=\sum_i (*\Omega^i)\otimes e_i,\
{\bf d}\Omega={\bf d}(\sum_i \Omega^i\otimes e_i)=\sum_i ({\bf d}\Omega^i)\otimes e_i,
\]
\[
\delta \Omega=\delta(\sum_i \Omega^i\otimes e_i)=\sum_i (\delta\Omega^i)\otimes e_i.
\]
Second, in view of the importance of the expression (1.8) for the divergence
of the CED energy-momentum tensor, we give its explicit deduction. Recall the
following algebraic relations on the Minkowski space:
\begin{equation}
\delta_p=(-1)^p *^{-1} {\bf d} * =*{\bf d}*,\ \delta *_p=*{\bf d}_p\ for\
p=2k+1,\ \delta *_p=-*{\bf d}_p\ for\  p=2k.
\end{equation}
If $\alpha$ is a 1-form and $F$ is a 2-form, the following relations hold:
\begin{equation}
*(\alpha\wedge *F)=-\alpha^\mu F_{\mu\nu}dx^\nu=*\left[(*F)\wedge *(*\alpha)\right]=
\frac12 (*F)^{\mu\nu}(*\alpha)_{\mu\nu\sigma}dx^\sigma.
\end{equation}
In particular,
\[
*(F\wedge *{\bf d}F)=\frac12 F^{\mu\nu}({\bf d}F)_{\mu\nu\sigma}dx^{\sigma}=
*[\delta *F\wedge *(*F)]=(*F)_{\mu\nu}(\delta *F)^{\nu}dx^{\mu}.
\]
Having in view these relations, we obtain consecutively:
\[
\nabla_\nu Q_\mu^\nu=\nabla_\nu\biggl[\frac 14 F_{\alpha\beta}F^{\alpha\beta}\delta_\mu^\nu-F_{\mu\sigma}F^{\nu\sigma}\biggr]=
\]
\[
=\frac12 F^{\alpha\beta}\nabla_{\nu}F_{\alpha\beta}\delta^{\nu}_{\mu}-
(\nabla_\nu F_{\mu\sigma})F^{\nu\sigma}-F_{\mu\sigma}\nabla_{\nu}F^{\nu\sigma}=
\]
\[
=\frac12 F^{\alpha\beta}\bigl[({\bf d}F)_{\alpha\beta\mu}-\nabla_\alpha F_{\beta\mu}-
\nabla_\beta F_{\mu\alpha}\bigr]-(\nabla_\nu F_{\mu\sigma})F^{\nu\sigma}-F_{\mu\sigma}\nabla_\nu F^{\nu\sigma}=
\]
\[
=\frac12 F^{\alpha\beta}({\bf d}F)_{\alpha\beta\mu}-F_{\mu\sigma}\nabla_\nu F^{\nu\sigma}=
-(*F)_{\mu\nu}\nabla_\sigma (*F)^{\sigma\nu}-F_{\mu\nu}\nabla_\sigma F^{\sigma\nu}=
\]
\[
=(*F)_{\mu\nu}(\delta *F)^{\nu}+F_{\mu\nu}(\delta F)^{\nu}.
\]

Let now our field $\Omega$ interact with some other field. This interaction,
i.e. energy-momentum exchange, is performed along 3 "channels". The first 2
channels are defined by the 2 (equal in rights) components $F$ and $*F$ of
$\Omega$. This exchange is {\it real} in the sense, that some part of the
$EM$-energy-momentum may be transformed into some other kind of
energy-momentum and assimilated by the external field or dissipated. Since
the two components $F$ and $*F$ are equal in rights it is naturally to expect
that the corresponding 2 terms, defining the exchange in a unit 4-volume,
will depend on $F$ and $*F$ similarly. The above expression for
$\nabla_\nu Q_\mu^\nu$ gives the two 1-forms
\[
F_{\mu\nu}(\delta F)^\nu dx^\mu,\quad  (*F)_{\mu\nu}(\delta* F)^\nu dx^\mu
\]
as natural candidates for this purpose. As for the third channel, it takes
into account a possible influence of the external field on the intra-field
exchange between $F$ and $*F$, which occurs without assimilation of
energy-momentum by the external field. The natural candidate, describing this
exchange is, obviously, the expression
\[
F_{\mu\nu}(\delta *F)^\nu dx^\mu + (*F)_{\mu\nu}(\delta F)^\nu dx^\mu.
\]
It is important to note, that these three channels are independent in the
sense, that any of them may occur without taking care if the other two work
or don't work. A natural model for such a situation is a 3-dimensional vector
space $K$, where the three dimensions correspond to the three exchange
channels. The non-linear exchange law requires some $K$-valued non-linear
map. Since our fields take values in ${\cal R}^2$ this 3-dimensional space
must be constructed from ${\cal R}^2$ in a natural way. Having in view the
bilinear character of $\nabla_\nu Q^\nu_\mu$ it seems naturally to look for
some bilinear construction with the properties desired. These remarks suggest
to choose the {\it symmetrized tensor product}
$Sym({\cal R}^2\otimes {\cal R}^2)\equiv {\cal R}^2\vee {\cal R}^2$.
So, from the point of view of the $EM$-field, the energy-momentum exchange
term should be written in the following way:
\begin{equation}
\vee(\delta \Omega,*\Omega).
\end{equation}
In fact, in the corresponding basis $(e_1,e_2)$ we obtain
\[
\vee(\delta \Omega,*\Omega)=
\vee (\delta F \otimes e_1 +\delta *F \otimes e_2, *F\otimes e_1+**F\otimes e_2)=
\]
\[
=(\delta F\wedge *F)\otimes e_1\vee e_1 +(\delta *F\wedge **F)\otimes e_2\vee e_2+
(-\delta F\wedge F + \delta*F\wedge *F)\otimes e_1\vee e_2.
\]
This expression determines how much of the $EM$-field energy-momentum has
been carried irreversibly over to the external field and how much has been
redistributed between $F$ and $*F$ by virtue of the external field influence
in a unit 4-volume.

Now, this same quantity of energy-momentum has to be expressed by new terms,
in which the external field "agents" should participate. Let's denote  by
$\Phi$ the first agent, interacting with $\pi_1\Omega$, and by $\Psi$ the
second agent, interacting with $\pi_2\Omega$. Since the corresponding two
exchanges are independent, we may write the exchange term in the following
way:
\begin{equation}
\vee(\Phi,*\pi_1\Omega)\ +\ \vee(\Psi,*\pi_2\Omega).
\end{equation}
According to the local energy-momentum conservation law these two
quantities have to be equal, so we obtain
\begin{equation}
\vee(\delta \Omega,*\Omega)=\vee(\Phi,*\pi_1\Omega)\ +\ \vee(\Psi,*\pi_2\Omega).
\end{equation}

This is the basic relation in EED. It contains the basic differential
equations for the $EM$-field components and requires additional equations,
specifying the properties of the external field, i.e. the algebraic and
differential properties of $\Phi$ and $\Psi$. The physical sense of this
equation is quite clear: local balance of energy-momentum. Further we shall
study this relation in various cases, and in the first place - the vacuum.

\chapter {\itshape \Large Extended Electrodynamics in Vacuum}
==========================================
\section {\bf Vacuum Equations in EED}
\vskip 0.05 cm
\subsection{Vacuum in EED}

In CED the term "vacuum" is used in the sense, that in the region, where we
consider an $EM$-field, there is no free or bound electric charges. This
notion of vacuum in CED comes from the conception, that the field may
exchange energy-momentum {\it only} with electric charge carriers and {\it
only} through the component $F$. EED extends the possibilities for
energy-momentum exchange assuming that the full quantity of exchanged
energy-momentum in a unit 4-volume is given by the general expression (1.42).
That's why in EED we talk about  a field $\Omega$ in vacuum every time when
this expression (1.42) is equal to zero. Formally this means that every
external field (medium), which does not exchange energy-momentum with
$\Omega$ can be treated as vacuum as far as the $EM$-field is concerned.
Talking about exchange, we mostly have in mind that some energy-momentum is
transferred from the field to the medium, however, we do not formally exclude
the reverse process.

Assume now that in some region we have an $EM$-field $\Omega$. In the
corresponding basis $(e_1,e_2)$ we can write
\[
\Omega=F\otimes e_1 +*F\otimes e_2,\
\Phi=\alpha^1\otimes e_1 +\alpha^2\otimes e_2,\
\Psi=\alpha^3\otimes e_1 +\alpha^4\otimes e_2.
\]
{\it Remark}.\ Further the 1-forms  $\alpha^i,\ i=1,...4$, as well as the
corresponding through the pseudometric $\eta$ vector fields, will be called
shortly {\it currents}. Of course, this terminology should not be associated
with some charged particles, or with some before given specific structure. In
the general case these currents are just the tools of the external
field to gain some energy-momentum from the field $\Omega $.
\vskip 0.5 cm
\noindent The expression (1.42) looks as follows:
\[
\vee(\Phi,*\pi_1\Omega)+\vee(\Psi,*\pi_2\Omega)=
\alpha^1\wedge*F\otimes e_1\vee e_1 +
\alpha^4\wedge**F\otimes e_2\vee e_2 +
\]
\[
+(\alpha^3\wedge**F +\alpha^2\wedge *F)\otimes e_1\vee e_2.
\]
Clearly. the necessary and sufficient condition for vacuum is
\begin{equation}
\alpha^1\wedge*F=0,\ \alpha^4\wedge**F=0,\
\alpha^3\wedge**F +\alpha^2\wedge *F=0,          
\end{equation}
or, in components
\begin{equation}
F_{\mu\nu}\alpha_1 ^\nu=0,\ (*F)_{\mu\nu}\alpha_4 ^\nu=0,
\ (*F)_{\mu\nu}\alpha_3 ^\nu + F_{\mu\nu}\alpha_2 ^\nu=0.   
\end{equation}
We see that there are various possibilities, i.e. relations among $F$ and
$\alpha_i$, to realize a vacuum situation. The strongest condition is, of
course,  $\Phi=\Psi=0$, i.e. all currents are equal to zero: $\alpha^i=0,
\ i=1,...,4$. If, at least one of the two currents $\alpha^1$ and $\alpha^2$
is different from zero, then the above relations (1.45) require
$det\|F_{\mu\nu}\|=0$, i.e. $F\wedge F=0$, or
orthogonality between $E$ and $B$. If $\alpha^4=\alpha^1=0$, but the other
two currents are different from zero then $F\wedge F\neq 0$ in general. In
accordance with our interpretation of the equations (1.42)this means that the
medium affects the exchange between $F$ and $*F$ without gaining and keeping
any energy-momentum. In such a case the orthogonality between $E$ and $B$ is
not needed. It seems important to note, that (2.1) requires every couple
$\alpha^i, \alpha^j$ to be in the kernel of $F$ and $*F$, i.e.
\begin{equation}
F(\alpha^i,\alpha^j)=(*F)(\alpha^i,\alpha^j)=0.           
\end{equation}
So, in case of vacuum, the currents are strongly dependent on the field
$\Omega$. The above relations define equations, connecting the 16 components
of the 4 currents with the components $F_{\mu\nu}$. Of course, in curvelinear
coordinates $\{y^\sigma\}$ these equations will depend strongly on the
metric coefficients $\eta_{\mu\nu}(y^\sigma)$ in these coordinates. Finally
we note, that $\alpha^1$ and $\alpha^4$ may be considered as eigen vectors
respectively for $F$ and $*F$ at zero eigen values, which, according to the
formulas in (1.1.2), is always possible if  $I_2=2E.B=0$.

\newpage
\subsection{Equivalent forms of the equations}
According to the last subsection an external field is called vacuum with
respect to the $EM$-field $\Omega$ if the right hand side of (1.43) is equal
to zero. Then the left hand side of (1.43) will also be equal to zero, so we
get the equations
\begin{equation}
\vee(\delta\Omega,*\Omega)=0.                           
\end{equation}
From this coordinate free compactly written expression we obtain the
following equations for the components of $\Omega$ in the basis $(e_1,e_2)$:

\begin{eqnarray}
&(\delta F\wedge *F)\otimes e_1\vee e_1 +(\delta *F)\wedge
**F)\otimes e_2\vee e_2 +\\ \nonumber
&+(\delta F\wedge **F+\delta*F\wedge *F)\otimes e_1\vee e_2=0.    
\end{eqnarray}
So, the field equations, expressed through the operator $\delta$, look as
follows
\begin{equation}
\delta F\wedge *F =0,\ \delta *F \wedge **F=0,\          
\delta *F\wedge *F-\delta F\wedge F=0.
\end{equation}
These equations, expressed through the operator ${\bf d}$, have the form
\begin{equation}
*F\wedge *{\bf d}*F=0,\ F\wedge *{\bf d}F=0,\
F\wedge *{\bf d}*F+*F\wedge *{\bf d}F=0.                  
\end{equation}
Using the components $F_{\mu\nu}$, we obtain for the equations (2.6)
\begin{equation}
F_{\mu\nu}(\delta F)^\nu=0,\ (*F)_{\mu\nu}(\delta*F)^\nu=0,\
F_{\mu\nu}(\delta*F)^\nu+(*F)_{\mu\nu}(\delta F)^\nu=0.       
\end{equation}
In the same way, for the equations (2.7) we get
\begin{equation}
(*F)^{\mu\nu}({\bf d}*F)_{\mu\nu\sigma}=0,
\ F^{\mu\nu}({\bf d}F)_{\mu\nu\sigma}=0,\
(*F)^{\mu\nu}({\bf d}F)_{\mu\nu\sigma}+
F^{\mu\nu}({\bf d}*F)_{\mu\nu\sigma}=0.                       
\end{equation}

Now we give the 3-dimensional form of the equations in the same order:
\begin{equation}
B\times\left(rotB-\frac{\partial E}{\partial \xi}\right)-EdivE=0,\
E.\left(rotB-\frac{\partial E}{\partial \xi}\right)=0,         
\end{equation}

\begin{equation}
E\times\left(rotE+\frac{\partial B}{\partial \xi}\right)-BdivB=0,\
B.\left(rotE+\frac{\partial B}{\partial \xi}\right)=0,         
\end{equation}

\[
\left(rotE+\frac{\partial B}{\partial \xi}\right)\times B+
\left(rotB-\frac{\partial E}{\partial \xi}\right)\times E+
BdivE+EdivB=0,\nonumber
\]
\begin{equation}
B.\left(rotB-\frac{\partial E}{\partial \xi}\right)-
E.\left(rotE+\frac{\partial B}{\partial \xi}\right)=0.          
\end{equation}
From the second equations of (2.10) and (2.11) the well known Poynting
relation follows
\[
div\left(E\times B\right)+\frac{\partial}{\partial \xi}\frac{E^2+B^2}{2}=0,
\]
and from the second equation of (2.12), if $E.B=g(x,y,z)$, we obtain the
known from Maxwell theory relation
\[
B.rotB=E.rotE.
\]
The explicit form of equations (2.10) end (2.11) should not make us conclude,
that the second (scalar) equations follow from the first (vector) equations.
Here is an example: let $divE=0,\ divB=0$ and the time derivatives of $E$ and
$B$ are zero. Then the system of equations reduces to
\[
E\times rotE=0,\ B\times rotB=0,\ B.rotE=0,\ E.rotB=0.
\]
As it is seen, the vector equations do not require any connection between $E$
and $B$ in this case, therefore, the scalar equations, which impose
such a connection, can not follow from the vector ones.
The third equations of (2.7) and (2.8) determine ( in equivalent way) the
energy-momentum quantities, transferred from $F$ to $*F$, and reversely,
in a unit 4-volume, with the expressions, respectively
\[
F_{\mu\nu}(\delta *F)^\nu=-(*F)_{\mu\nu}(\delta F)^\nu,\
F^{\mu\nu}({\bf d}*F)_{\mu\nu\sigma}=-(*F)^{\mu\nu}({\bf d}F)_{\mu\nu\sigma}.
\]
From these relations it is seen, that the 1-forms $\delta F$ and $\delta *F$
{\it play the role of "external" currents} respectively for $*F$ and $F$. In
the same spirit we could say, that the energy-momentum quantities
$F_{\mu\nu}(\delta F)^\nu$ and $(*F)_{\mu\nu}(\delta *F)^\nu$,
which $F$ and $*F$ exchange with themselves, are equal to zero. And this
corresponds fully to our former statements, concerning the physical sense of
the equations for the components of $\Omega$.

\vskip 0.5cm
\subsection{Conservation laws}
From the first two equations of (2.8) and from the earlier given expression
for the divergence $\nabla_\nu Q_\mu^\nu$ of the Maxwell's energy-momentum
tensor $Q_\mu^\nu$ in CED it is immediately seen that on the solutions of our
equations (2.8) this divergence is also zero. In view of this {\it we assume
the tensor} $Q_\mu ^\nu$, {\it defined by} (1.7) {\it to be the
energy-momentum tensor in} EED. We shall be interested in finding explicit
time-stable solutions of finite type, i.e. $F_{\mu\nu}$ to be finite
functions of the three spatial coordinates, therefore, if it turns out that
such solutions really exist, then integral conserved quantities can be easily
constructed and computed, making use of the 10 Killing vectors on the
Minkowski space-time.  We recall that in CED such finite and time-stable
solutions in the whole space are not allowed by the Maxwell's equations.

\section{\bf General Properties of the Equations and \\Their Solutions}
\vskip 0.05cm
\subsection{General properties of the equations}
We first note, that in correspondence with the requirement for {\it general
covariance}, equations (2.4), given above and presented in different but
equivalent forms, are written down in coordinate free manner. This
requirement is universal, i.e. it concerns all basic equations of a theory
and means simply, that the existence of real objects and the occurrence of
real processes {\it can not} depend on the local coordinates used in the
theory., i.e. on the convenient for us way to describe the local character of
the evolution and structure of the natural objects and processes. Of course,
in the various coordinate systems the equations and their solutions will look
differently. Namely the covariant character of the equations allows to choose
the most appropriate coordinates, reflecting most fully the features of
every particular case. A typical example for this is the usage of spherical
coordinates in describing spherically symmetric fields. Let's not forget
also, that the coordinate-free form of the equations permits an easy transfer
of the same physical situation onto manifolds with more complicated structure
and nonconstant metric tensor. Shortly speaking, {\it the coordinate free
form of the equations in theoretical physics reflects the most essential
properties of reality, called shortly objective character of the real
phenomena}.

Since the left hand sides of the equations are linear combinations of the
first derivatives of the unknown functions with coefficients, depending
linearly on these unknown functions, (2.8) present a special type system of
{\it quasilinear first order partial differential equations}. The number of
the unknown functions $F_{\mu\nu}=-F_{\nu\mu}$ is 6, and in general, the
number of the equations is $3.4=12$, but the number of the independent
equations depends strongly on if the two invariants
$I_{1}=\frac 12 F_{\mu\nu}F^{\mu\nu}$ and
$I_{2}=\frac 12 (*F)_{\mu\nu}F^{\mu\nu}$  are equal to zero or not equal to
zero. If $I_2\neq 0$ then $ det (F_{\mu\nu})\neq 0$ and the first two
equations of (2.8) are equivalent to $\delta F=0$ and $\delta *F=0$, which
automatically eliminates the third equation of (2.8), i.e. in this case our
equations reduce to Maxwell's equations.

It is clearly seen from the (2.9) form of the equations, that the metric
tensor essentially participates (through the $*$-operator applied to 2-forms
only) in the equations. If we use the $\delta$-operator, then the
metric participates also through the $*$-operator, applied to 3-forms, but
this does not lead to more complicated coordinate form of the equations. It
is worth to note that in nonlinear coordinates the metric tensor will
participate with its derivatives, therefore, the very solutions will depend
strongly on the metric tensor chosen. This may cause existence or
non-existence of solutions of a given class, e.g. soliton-like ones. In our
framework such additional complications do not appear because of the
opportunity to work in global coordinates with constant metric tensor.

We note 2 important invariance (symmetry) properties of our equations.

\vskip 0.5cm
{\bf Property 1}. {\it The transformation $F\rightarrow *F$ does
not change the system}.

The proof is obvious, in fact, the first two equations interchange, and the
third one is kept the same. In terms of $\Omega $ this means that if $\Omega $
is a solution, then $*\Omega $ is also a solution, which means, in turn, that
equations (2.4) are equivalent to equations
\begin{equation}
\vee(\Omega,*{\bf d}\Omega)=0.
\end{equation}

\vskip 0.05cm
{\bf Property 2}. {\it Under conformal change of the metric the
equations do not change}.

The proof of this property is also obvious and is reduced to the notice, that
as it is seen from (2.9), the $*$-operator participates only with its
reduction to 2-forms, and as we noted in subsec.(1.1.2), $*_2$ is conformally
invariant.

\vskip 0.5cm
Summing up the first two equations of (2.10) and (2.11) we obtain how the
classical Poynting vector changes in time in our more general approach:

\[
\frac {\partial}{\partial \xi}(E\times B)=EdivE+BdivB-E\times rotE-B\times rotB.
\]

In CED the first and the second terms on the right are missing.

Here is an example of static solutions of (2.4), which are not solutions to
Maxwell's equations.
\[
E=(asin\alpha z,\ acos\alpha z,\ 0),\  B=(b cos\alpha z,\ -b sin\alpha z,\ 0),
\]
where $a,b$ and $\alpha$ are constants. We obtain
\[
rotE=(a\alpha sin\alpha z,\ a\alpha cos\alpha z,\ 0),\
rotB=(b\alpha cos\alpha z,\ -b\alpha sin\alpha z,\ 0)
\]

Obviously, $E\times rotE=0,\ B\times rotB=0,\ E.rotB=0,\ B.rotE=0.$
For the Poynting vector we get $E\times B =(0,0,-ab)$, and for the energy
density \linebreak
$w=\frac12 (a^2+b^2)$. Considered in a finite volume, these solutions
could model some standing waves, but we do not engage ourselves with such
interpretations, since we do not accept seriously that static $EM$-fields may
really exist.

\vskip 0.5cm
\subsection{General properties of the solutions}
It is quite clear that the solutions of our equations are naturally divided
into two classes: {\it linear} and {\it nonlinear}. The first class consists
of all solutions to Maxwell's vacuum equations, where the name {\it linear}
comes from. These solutions are well known and won't be discussed here. The
second class, called {\it nonlinear}, includes all the rest solutions. This
second class is naturally divided into two subclasses. The first subclass
consists of all nonlinear solutions, satisfying the conditions
\begin{equation}
\delta F\neq 0,\ \delta *F\neq 0,                    
\end{equation}
and the second subclass consists of those nonlinear solutions, satisfying one
of the two couples of conditions:
\[
\delta F=0,\ \delta *F\neq 0 ;\ \delta F\neq 0,\ \delta *F=0 .
\]
Further we assume the conditions (2.14) fulfilled, i.e. the solutions of the
second subclass will be considered as particular cases of the first subclass.
Our purpose is to show explicitly, that among the nonlinear solutions there
are soliton-like ones, i.e. the components $F_{\mu\nu}$ of which at any
moment are {\it finite} functions of the three spatial variables with
{\it connected} support. We are going to study their properties and to
introduce corresponding characteristics. First we shall establish some of
their basic features, proving three propositions.

\vskip 0.5cm
{\bf Proposition 1.}\ {\it All nonlinear solutions have zero invariants}:
\newline
$$
I_1=\frac 12 F_{\mu\nu}F^{\mu\nu}=0,
\ I_2=\frac 12 (*F)_{\mu\nu}F^{\mu\nu}=2\sqrt{det(F_{\mu\nu})}=0.
$$
\indent{\bf Proof}.\ Recall the field equations in the form (2.8):
\[
F_{\mu\nu}(\delta F)^\nu=0,\ (*F)_{\mu\nu}(\delta*F)^\nu=0,\
F_{\mu\nu}(\delta*F)^\nu+(*F)_{\mu\nu}(\delta F)^\nu=0.
\]
It is clearly seen that the first two groups of these equations may be
considered as two linear homogeneous systems with respect to $\delta F^\mu $
and $\delta *F^\mu $ respectively. In view of the nonequalities (2.14) these
homogeneous systems have non-zero solutions, which is possible only if
$det(F_{\mu\nu})=det(*F)_{\mu\nu})=0$, i.e. if  $I_2=2E.B=0$. Further,
summing up these three systems of equations, we obtain
\[
(F+*F)_{\mu\nu}(\delta F+\delta *F)^\nu=0.
\]
If now $(\delta F+\delta *F)^\nu\neq 0$, then
\[
0=det(F+*F)_{\mu\nu}=\left [\frac 12 (F+*F)_{\mu\nu}(*F-F)^{\mu\nu}\right ]^2=
\frac 14\left [-2F_{\mu\nu}F^{\mu\nu}\right ]^2=(I_1)^2.
\]
If $\delta F^\nu=-(\delta *F)^\nu\neq 0$, we sum up the first two systems and
obtain \linebreak $(*F-F)_{\mu\nu}(\delta *F)^\nu=0$. Consequently,
\[
0=det(*F-F)_{\mu\nu}=\left [\frac 12 (*F-F)_{\mu\nu}(-F-*F)^{\mu\nu}\right ]^2=
\frac 14\left [2F_{\mu\nu}F^{\mu\nu}\right ]^2=(I_1)^2.
\]
This completes the proof.

Recall that in this case the energy-momentum tensor $Q_{\mu\nu}$ has just one
isotropic eigen direction and all other eigen directions are space-like.
Since all eigen directions of $F_{\mu\nu}$ and $*F_{\mu\nu}$ are eigen
directions of $Q_{\mu\nu}$ too, it is clear that $F_{\mu\nu}$ and
$(*F)_{\mu\nu}$ can not have time-like eigen directions. But the first two
systems  of (2.8) require $\delta F$ and $\delta *F$ to be eigen vectors of
$F$ and $*F$ respectively, so we obtain
\begin{equation}
(\delta F).(\delta F)\leq 0,\ (\delta *F).(\delta *F)\leq 0.    
\end{equation}

\vskip 0.5cm
{\bf Proposition 2.}\ {\it All nonlinear solutions satisfy the conditions}
\begin{equation}
(\delta F)_\mu (\delta *F)^\mu =0,\ \left|\delta F\right|=\left|\delta *F\right|  
\end{equation}
\indent{\bf Proof}.\  We form the inner product
$i(\delta *F)(\delta F\wedge *F)=0$  and get
\[
(\delta *F)^\mu(\delta F)_\mu(*F)-\delta F\wedge (\delta
*F)^\mu(*F)_{\mu\nu}dx^\nu=0.
\]
Because of the obvious nulification of the second term the first term will be
equal to zero (at non-zero $*F$) only if
$(\delta F)_\mu (\delta *F)^\mu=0$.

Further we form the inner product
$i(\delta *F)(\delta F\wedge F-\delta *F\wedge *F)=0$ and obtain

\[
(\delta *F)^\mu(\delta F)_\mu F-\delta F\wedge (\delta *F)^\mu F_{\mu\nu}dx^\nu-
\]
\[
-(\delta *F)^2(*F)+\delta *F\wedge(\delta *F)^\mu (*F)_{\mu\nu}dx^\nu=0.
\]
Clearly, the first and the last terms are equal to zero. So, the inner
product by $\delta F$ gives
\[
(\delta F)^2 (\delta *F)^\mu F_{\mu\nu}dx^\nu-
\left[(\delta F)^\mu (\delta *F)^\nu F_{\mu\nu}\right]\delta F
+(\delta *F)^2(\delta F)^\mu (*F)_{\mu\nu}dx^\nu=0.
\]
The second term of this equality is zero. Besides,
$(\delta *F)^\mu F_{\mu\nu}dx^\nu=\linebreak
-(\delta F)^\mu (*F)_{\mu\nu}dx^\nu.$ So,
\[
\left[(\delta F)^2-(\delta *F)^2\right](\delta F)^\mu(*F)_{\mu\nu}dx^\nu=0.
\]
Now, if $(\delta F)^\mu(*F)_{\mu\nu}dx^\nu\neq 0$, then the relation
$\left|\delta F\right|=\left|\delta *F\right|$ follows immediately.
If $(\delta F)^\mu(*F)_{\mu\nu}dx^\nu=0=-(\delta *F)^\mu F_{\mu\nu}dx^\nu$
according to the third equation of (2.8), we shall show that
 $(\delta F)^2=(\delta *F)^2=0$. In fact, forming the inner product
$i(\delta F)(\delta F\wedge *F)=0$ , we get
\[
(\delta F)^2*F-\delta F\wedge (\delta F)^\mu (*F)_{\mu\nu}dx^\nu=
(\delta F)^2*F=0.
\]
In a similar way, forming the inner product $i(\delta *F)\delta *F\wedge F=0$
we have
\[
(\delta *F)^2 F-\delta (*F)\wedge (\delta *F)^\mu F_{\mu\nu}dx^\nu=
(\delta *F)^2 F=0.
\]
This completes the proof. \newline
We just note that in this last case the isotropic
vectors $\delta F$ and $\delta *F$ are eigen vectors of $Q_{\mu\nu}$ too, and
since $Q_{\mu\nu}$ has just one isotropic eigen direction, we conclude that
$\delta F$ and $\delta*F$ are colinear.

In order to formulate the third proposition, we recall from subsec. (1.1.2)
that at zero invariants $I_1=I_2=0$ the following representation holds:
\[
F=A\wedge \zeta,\ *F=A^*\wedge \zeta,
\]
where $\zeta$ is the only (up to a scalar multiple) isotropic eigen vector of
$Q_\mu^\nu $. Also, the relations $A.\zeta=0,\ A^*.\zeta=0$ are in force.
Having this in view we shall prove the following

\newpage {\bf Proposition 3.}\ {\it All
nonlinear solutions satisfy the relations}
\begin{equation}
\zeta^\mu(\delta
F)_\mu=0,\ \zeta^\mu(\delta *F)_\mu=0.
\end{equation}
\indent{\bf Proof}.\ We form the inner product
$i(\zeta)(\delta F\wedge *F)=0$ :
\[
\left[\zeta^\mu(\delta F)_\mu\right]*F-
\delta F\wedge(\zeta)^\mu(*F)_{\mu\nu}dx^\nu=
\]
\[
=\left[\zeta^\mu (\delta F)_\mu\right]A^*\wedge\zeta-
(\delta F\wedge\zeta)\zeta^\mu (A^*)_\mu
+(\delta F\wedge A^*)\zeta^\mu \zeta_\mu=0.
\]
Since the second and the third terms are equal to zero and $*F\neq 0$, then
$\zeta^\mu(\delta F)_\mu=0$. Similarly, from the equation
$(\delta *F)\wedge F=0$ we get \linebreak
$\zeta^\mu(\delta *F)_\mu=0$. The proposition is proved.

\vskip 0.5cm
\subsection{Algebraic properties of the nonlinear \\ solutions}
Since all nonlinear solutions have zero invariants $I_1=I_2=0$ we can make a
number of algebraic considerations, which clarify considerably the structure
and make easier the study of the properties of these solutions. As we
mentioned earlier, all eigen values if $F$, $*F$ and $Q_{\mu\nu}$ in this
case are zero, and the eigen vectors can not be time-like. There is only one
isotropic direction, defined by the isotropic vectors $\pm\zeta$ and the
representations $F=A\wedge \zeta$,\linebreak $*F=A^*\wedge\zeta$ hold,
moreover, we have  $A.A^*=0,\ A^2=(A^*)^2\leq 0, A.\zeta=A^*.\zeta=0$. Recall
that the two 1-forms $A$ and $A^*$ are defined up to isotropic additive
factors, colinear to $\zeta$. The above representation of $F$ and $*F$
through $\zeta$ shows that these factors do not contribute to $F$ and $*F$,
therefore, we assume further that, {\it these additive factors are equal to
zero}.

We express now $Q_{\mu\nu}$ through $A$, $A^*$ and $\zeta$. First we normalize
the vector $\zeta$. This is possible, because it is an isotropic vector, so
its time-like component $\zeta_4$ is always different from zero. We divide
$\zeta_\mu$ by $\zeta_4$ and get the vector
${\bf V}=({\bf V}^1,{\bf V}^2,{\bf V}^3,1)$, defining, of course, the same
isotropic direction. Now we make use of the identity (1.25), where we put
$F_{\mu\nu}$ instead of $G_{\mu\nu}$. Having in view that $I_1=\frac 12
F_{\mu\nu}F^{\mu\nu}=0$, we obtain
$F_{\mu\sigma}F^{\nu\sigma}=(*F)_{\mu\sigma}(*F)^{\nu\sigma}$. So, the
energy-momentum tensor looks as follows
\[
Q_\mu^\nu=-\frac{1}{4\pi}F_{\mu\sigma}F^{\nu\sigma}=
-\frac{1}{4\pi}(*F)_{\mu\sigma}(*F)^{\nu\sigma}=
\]
\begin{equation}
=-\frac{1}{4\pi}(A)^2{\bf V}_\mu{\bf V}^\nu=
-\frac{1}{4\pi}(A^*)^2{\bf V}_\mu{\bf V}^\nu.
\end{equation}

This choice of  $\zeta={\bf V}$ determines the following energy density
\linebreak $4\pi Q_4^4=\left|A\right|^2=\left|A^*\right|^2$.

We consider now the influence of the conservation law $\nabla _\nu Q_\mu^\nu =0$
on ${\bf V}$.
\[
\nabla_\nu Q_\mu^\nu=-A^2{\bf V}^\nu \nabla_\nu{\bf V}_\mu-
{\bf V}_\mu\nabla_\nu\left(A^2{\bf V}^\nu\right)=0.
\]
This relation holds for every $\mu =1,2,3,4$. We consider it for $\mu=4$ and
get ${\bf V}^\nu\nabla_\nu (1)={\bf V}^\nu{\partial_\nu}(1)=0$.
Therefore,
${\bf V}_4\nabla_\nu\left(A^2 {\bf V}^\nu\right)=
\nabla_\nu\left(A^2 {\bf V}^\nu\right)=0$. Since $A^2\neq 0$, we obtain that
${\bf V}$ satisfies the equation
\[
{\bf V}^\nu\nabla_\nu{\bf V}^\mu=0,
\]
which means, that ${\bf V}$ is a {\it geodesic} vector field, i.e. the
integral trajectories of ${\bf V}$ are isotropic geodesics, or
isotropic straight lines. Hence, {\it every nonlinear solution $F$ defines
unique isotropic geodesic direction in the Minkowski space-time}. This
important consequence allows a special class of coordinate systems, called
further $F$-adapted, to be introduced. These coordinate systems are defined
by the requirement, that the trajectories of the unique ${\bf V}$, defined by
$F$, to be parallel to the $(z,\xi)$-coordinate plane. In such a coordinate
system we have ${\bf V}_\mu=(0,0,\varepsilon,1),\  \varepsilon=\pm1$. Further
on, we shall work in such arbitrary chosen but fixed $F$-adapted coordinate
system, defined by the corresponding $F$ under consideration.

We write down now the relations $F=A\wedge{\bf V},\ *F=A^*\wedge{\bf V}$
component-wise, take into account the values of ${\bf V}_\mu$ in the
$F$-adapted coordinate system and obtain the following explicit relations:

\[
F_{12}=F_{34}=0,\ F_{13}=\varepsilon F_{14},\ F_{23}=\varepsilon F_{24},
\]
\[
(*F)_{12}=(*F)_{34}=0,\ (*F)_{13}=\varepsilon (*F)_{14}=-F_{24},\ (*F)_{23}=\varepsilon (*F)_{24}=F_{14},
\]
\begin{equation}
A=\left(F_{14},F_{24},0,0\right),\ A^*=\left(-F_{23},F_{13},0,0\right)=
\left(-\varepsilon A_2,\varepsilon A_1,0,0\right).
\end{equation}
Clearly, the 1-forms $A$ and $-A^*$ can be interpreted as {\it electric} and
{\it magnetic} fields respectively. Only 4 of the components $Q_\mu^\nu$ are
different from zero, namely: $Q_4^4=-Q_3^3=\varepsilon Q_3^4=
-\varepsilon Q_4^3=\left|A^2\right|$. Introducing the notations
$F_{14}\equiv u,\ F_{24}\equiv p$, we can write
\[
F=\varepsilon udx\wedge dz + udx\wedge d\xi + \varepsilon pdy\wedge dz +
pdy\wedge d\xi
\]
\[
*F=-pdx\wedge dz - \varepsilon pdx\wedge d\xi + udy\wedge dz +
\varepsilon udy\wedge d\xi.
\]
In the important for us {\it spatially finite } case, i.e. when the functions
$u$ and $p$ are finite with respect to the spatial variables $(x,y,z)$, for
the integral energy $W$ and momentum ${\bf p}$ we obtain
\[
W=\int{Q_4^4}dxdydz=\int{(u^2+p^2)}dxdydz<\infty,
\]
\begin{equation}
{\bf p}=\left(0,0,\varepsilon \frac Wc\right),\ \rightarrow c^2|{\bf p}|^2-W^2=0.
\end{equation}

Now we show how the nonlinear solution $F$ defines at every point a
pseudoorthonormal basis in the corresponding tangent and cotangent spaces.
The nonzero 1-forms $A$ and $A^*$ are normed to ${\bf A}=A/|A|$ and
${\bf A^*}=A^*/|A^*|$. Two new unit 1-forms ${\bf R}$ and ${\bf S}$ are
introduced through the equations:
\[
{\bf R}^2=-1,\ {\bf A}^\nu {\bf R}_\nu=0,\  ({\bf A^*})^\nu {\bf R}_\nu=0,\
{\bf V}^\nu{\bf R}_\nu=\varepsilon,\ {\bf S}={\bf V}+\varepsilon{\bf R}.
\]
The only solution of the first 4 equations is ${\bf R}_\mu=(0,0,-1,0)$. Then
for ${\bf S}$ we obtain ${\bf S}_\mu=(0,0,0,1)$. Clearly, ${\bf R}^2=-1$ and
${\bf S}^2=1$. This pseudoorthonormal (co-tangent) basis is carried over to a
(tangent) pseudoorthonormal basis by means of the pseodometric $\eta$.

We proceed further to introduce the concepts of {\it amplitude} and {\it
phase} in a coordinate-free manner. Of course, we shall use the
considerations in subsec.(1.3.1). First, of course, we look at the
invariants, we have: $I_1=I_2=0$. But in our case we have got another
invariant, namely, the module of the 1-forms $A$ and $A^*$: $|A|=|A^*|$.
Let's begin with the {\it amplitude}, which shall be denoted by $\phi$. As
it's seen from the above obtained expressions, the magnitude of $|A|$
coincides with the square root of the energy density in any $F$-adapted
coordinate system. As we noted in (1.3.1) this is the sense of the quantity
amplitude. So, we define it by the module of $|A|=|A^*|$. We give now two
more coordinate-free ways to define the amplitude.

Recall first, that at every point, where the field is different from zero, we
have three bases: the pseudoopthonormal coordinate basis $(dx,dy,dz,d\xi)$,
the pseudoorthonormal basis $\chi^0=({\bf A},\varepsilon{\bf A^*},{\bf R},
{\bf S})$ and the pseudoorthogonal basis $\chi=(A,\varepsilon A^*,{\bf R},
{\bf S})$. The matrix $\chi_{\mu\nu}$ of  $\chi$ with respect to the
coordinate basis is
\[
\chi_{\mu\nu}=\left\|\matrix{
u  &-p  & 0  &0 \cr
p  & u  & 0  &0 \cr
0  & 0  &-1  &0 \cr
0  & 0  & 0  &1
\cr}\right\|.
\]
We define now the amplitude $\phi$ of the field by
\begin{equation}
\phi=\sqrt{|det(\chi_{\mu\nu})|}.
\end{equation}

We consider now the matrix ${\cal R}$ of 2-forms
\[
{\cal R}=\left\|\matrix{
udx\wedge d\xi  &-pdx\wedge d\xi  & 0            &0 \cr
pdy\wedge d\xi  & udy\wedge d\xi  & 0            &0 \cr
0               & 0               &-dy\wedge dz  &0 \cr
0               & 0               & 0            &dz\wedge d\xi
\cr}\right\|,
\]
or, equivalently:
\[
{\cal R}=udx\wedge d\xi \otimes(dx\otimes dx)-pdx\wedge d\xi \otimes(dx\otimes dy)+
pdx\wedge d\xi \otimes(dy\otimes dx)+
\]
\[
+udy\wedge d\xi \otimes(dy\otimes dy)-dy\wedge dz \otimes(dz\otimes dz)+
dz\wedge d\xi \otimes(d\xi\otimes d\xi).
\]
Now we can write
\[
\phi=\sqrt{\frac12 \left|R_{\mu\nu\alpha\beta}R^{\mu\nu\alpha\beta}\right|}.
\]

We proceed further to define the {\it phase} of the nonlinear solution $F$.
We shall need the matrix $\chi^0_{\mu\nu}$ of the basis $\chi^0$ with respect
to the coordinate basis. We obtain
\[
\chi^0_{\mu\nu}=\left\|\matrix{
\frac{u}{\sqrt{u^2+p^2}}  &\frac{-p}{\sqrt{u^2+p^2}}  & 0  &0 \cr
\frac{p}{\sqrt{u^2+p^2}}  &\frac{u}{\sqrt{u^2+p^2}}   & 0  &0 \cr
0                         & 0                         &-1  &0 \cr
0                         & 0                         & 0  &1
\cr}\right\|.
\]
The {\it trace} of this matrix is
$$
tr(\chi^0_{\mu\nu})=\frac{2u}{\sqrt{u^2+p^2}}.
$$
Obviously, the inequality $|\frac 12 tr(\chi^0_{\mu\nu})|\leq 1$ is
fulfilled. Now, by definition, the quantity
$\varphi=\frac 12 tr(\chi^0_{\mu\nu})$ will be called {\it phase function} of
the solution, and the quantity
\begin{equation}
\theta=arccos(\varphi)=arccos\left(\frac 12 tr(\chi^0_{\mu\nu})\right)
\end{equation}
will be called {\it phase} of the solution.

Making use of the amplitude $\phi$ and the phase function $\varphi $ we can
write
\begin{equation}
u=\phi.\varphi,\ p=\phi.\sqrt{1-\varphi^2}.
\end{equation}

We note that the couple of 1-forms $A=udx+pdy,\ A^*=-pdx+udy $ defines a
completely integrable Pfaff system, i.e. the following equations hold:
\[
{\bf d}A\wedge A\wedge A^* =0, \ {\bf d}A^*\wedge A\wedge A^* =0.
\]
In fact, $A\wedge A^*=(u^2+p^2)dx\wedge dy$, and in every term of
${\bf d}A$ and ${\bf d}A^*$ at least one of the basis vectors $dx$ and $dy$
will participate, so the above exteriour products will vanish.

\noindent{\it Remark}. These considerations stay in force also for those
linear solutions, which have zero invariants $I_1=I_2=0$. But Maxwell's
equations require $u$ and $p$ to be {\it running waves}, so the corresponding
phase functions will be also running waves. As we'll see further, the phase
functions for nonlinear solutions are arbitrary bounded functions.

We proceed further to define the new and important concept of {\it scale
factor} $L$ for a given nonlinear solution. It is defined by
\begin{equation}
L=\frac{|A|}{|\delta F|}=\frac{|A^*|}{|\delta*F|}.
\end{equation}
Clearly, $L$ can not be defined for the linear solutions, and in this sense it
is {\it new} and we shall see that it is really {\it important}.

From the expressions $F=A\wedge {\bf V}$ and $*F=A^*\wedge {\bf V}$ it
follows that the physical dimension of $A$ and $A^*$ is the same as that of
$F$. We conclude that the physical dimension of $L$ coincides with the
dimension of the coordinates, i.e. $[L]=length$. From the definition it is
seen that $L$ is an {\it invariant} quantity, and depends on the point, in
general. The invariance of $L$ allows to define a time-like 1-form (or
vector field) $f(L){\bf S}$, where $f$ is some real function of $L$. So,
every nonlinear solution determines a time-like vector field on $M$.

If the scale factor $L$, defined by the nonlinear solution $F$, is a
{\it finite} and {\it constant} quantity, we can introduce a {\it
characteristic} finite time-interval $T(F)$ by the relation
$$
cT(F)=L(F),
$$
as well, as corresponding {\it characteristic frequency} by
$$
\nu(F)=1/T(F).
$$
In these "wave" terms the scale factor $L$ acquires the sense of "wave
length", but this interpretation is arbitrary and we shall not make use of it.

It is clear, that the subclass of nonlinear solutions, which define constant
scale factors, factors over the admissible values of the invariant $f(L)$.
This makes possible to compare with the experiment. For example, at constant
scale factor $L$ if we choose $f(L)=L/c$, then the scalar product of
$(L/c){\bf S}$ with the integral energy-momentum vector, which in the
$F$-adapted coordinate system is $(0,0,\varepsilon W,W)$, gives the invariant
quantity $W.T$, having the physical dimension of action, and its numerical
value could be easily measured.

\vskip 0.5cm
\section {\bf Nonlinear Solutions. Description of  \\ photon-like objects}
\vskip 0.05cm
\subsection{Explicit solutions in canonical coordinates}

As it was shown in the preceding section with every nonlinear solution $F$ of
our nonlinear equations a class of $F$-adapted coordinate systems is
associated, such that $F$ and $*F$ acquire the form respectively
\[
F=\varepsilon udx\wedge dz + udx\wedge d\xi + \varepsilon pdy\wedge dz +
pdy\wedge d\xi
\]
\[
*F=-pdx\wedge dz -\varepsilon pdx\wedge d\xi + udy\wedge
dz + \varepsilon udy\wedge d\xi.
\]
After some elementary calculations we obtain
\[
\delta F=(u_\xi -\varepsilon u_z)dx +(p_\xi-\varepsilon p_z)dy +
\varepsilon(u_x + p_y)dz + (u_x + p_y)d\xi,
\]
\[
\delta
*F=-\varepsilon(p_\xi -\varepsilon p_z)dx +\varepsilon(u_\xi-\varepsilon
p_z)dy - (p_x - u_y)dz - (p_x - u_y)d\xi,
\]
\[
F_{\mu\nu}(\delta F)^\nu
dx^\nu= (*F)_{\mu\nu}(\delta *F)^\nu dx^\nu=
\]
\[
=\varepsilon\left[p(p_\xi-\varepsilon p_z)+u(u_\xi-\varepsilon u_z)\right]dz+
\left[p(p_\xi-\varepsilon p_z)+u(u_\xi-\varepsilon u_z)\right]d\xi,
\]
\[
(\delta F)^2=(\delta *F)^2=-(u_\xi-\varepsilon u_z)^2-(p_\xi-\varepsilon
p_z)^2
\]
A simple direct calculation shows, that the equation
\[
F_{\mu\nu}(\delta *F)^\nu +(*F)_{\mu\nu}(\delta F)^\nu =0
\]
is identically
fulfilled for any such $F$ with arbitrary $u$ and $p$. We obtain that our
equations reduce to only 1 equation, namely
\begin{equation}
p(p_\xi-\varepsilon p_z)+u(u_\xi-\varepsilon u_z)=
\frac 12\left[(u^2+p^2)_\xi-\varepsilon(u^2+p^2)_z\right]=0.
\end{equation}
The obvious solution of this equation is
\begin{equation}
u^2+p^2=\phi^2 (x,y,\xi+\varepsilon z).
\end{equation}
The solution obtained shows that the equations impose some limitations only on
the amplitude function $\phi$ and the phase function $\varphi$ is arbitrary
exept that it is bounded: $|\varphi |\leq 1$. The amplitude $\phi$ is a
running wave along the specially chosen coordinate $z$, which is common for
all $F$-adapted coordinate systems.Considered as a function of the spatial
coordinates, the amplitude $\phi$ is {\it arbitrary}, so it can be chosen
{\it spatially finite}. The time-evolution does not affect the initial form
of $\phi$, so it will stay the same in time. This shows, that {\it among the
nonlinear solutions of our equations there are (3+1) soliton-like
solutions}. The spatial structure is determined by the initial condition,
while the phase function $\varphi$ can be used to define {\it internal
dynamics} of the solution.

Recalling the substitutions (2.23)
\[
u=\phi.\varphi, \ p=\phi\sqrt{1-\varphi^2},
\]
and the equality $|A|=\phi$, we get
\begin{equation}
|\delta F|=|\delta *F|=\frac{|\phi||\varphi_\xi-\varepsilon\varphi_z|}{\sqrt{1-\varphi^2}},
\ L=\frac{\sqrt{1-\varphi^2}}{|\varphi_\xi-\varepsilon\varphi_z|}.
\end{equation}
For the induced pseudoorthonormal bases (1-forms and vector fields) we find
\[
{\bf A}=\varphi dx +\sqrt{1-\varphi^2}dy,\
{\bf \varepsilon A^*}=-\sqrt{1-\varphi^2}dx+\varphi dy,\ {\bf R}=-dz,\ {\bf S}=d\xi,
\]
\[
{\bf A}=-\varphi\frac{\partial}{\partial x}-\sqrt{1-\varphi^2}\frac{\partial}{\partial y},\
\varepsilon{\bf A^*}=\sqrt{1-\varphi^2}\frac{\partial}{\partial x}-\varphi\frac{\partial}{\partial y},\
{\bf R}=\frac{\partial}{\partial z},\ {\bf S}=\frac{\partial}{\partial \xi}.
\]

Hence, the nonlinear solutions in canonical coordinates are parametrized by
one function $\phi$ of 3 parameters and one {\it bounded} function of 4
parameters. Therefore, the separation of various subclasses of nonlinear
solutions is made by imposing additional conditions on these two functions.
Further in this subsection we are going to separate a subclass of solutions ,
the integral properties of which reflect well enough the well known from the
experiment integral properties and characteristics of the free photons. These
solutions will be called {\it photon-like} and will be separated through
imposing additional requirements on $\varphi$ and $L$ in a
coordinate-free manner.

We note first, that we have three invariant quantities at hand: $\phi$,
$\varphi$ and $L$. The amplitude function $\phi$ is to be determined by the
initial conditions, which have to be {\it finite}. So, we may impose
additional conditions on $L$ and $\varphi$. These conditions have to express
some intra-consistency among the various characteristics of the solution. The
idea, what kind of intra-consistency to use, comes from the observation that
the amplitude function $\phi$ is a first integral of the vector field
${\bf V}$, i.e.
\[
{\bf V}(\phi)=\left(-\varepsilon \frac{\partial}{\partial z}+\frac {\partial}
{\partial \xi}\right)(\phi)=-\varepsilon \frac{\partial}{\partial z}\phi(x,y,\xi+\varepsilon z)
+\frac {\partial}{\partial \xi}\phi(x,y,\xi+\varepsilon z)=0.
\]
We want to extend this available consistency between ${\bf V}$ and $\phi$,
i.e. we shall require the two functions $\varphi$ and $L$ to be first
integrals of some of the available vector fields. Explicitly, we require
the following:
\vskip 0.5cm
$1^0$. {\it The phase function}\  $\varphi$\  {\it is a first integral of the
three vector fields} ${\bf A,A^*}$ and ${\bf R}$: ${\bf A}(\varphi)=0,
{\bf A^*}(\varphi)=0, {\bf R}(\varphi)=0$.
\vskip 0.5cm
$2^0$. {\it The scale factor} $L$ {\it is a non-zero finite first integral of
the vector field} ${\bf S}$: ${\bf S}(L)=0$.
\vskip 0.5cm
The requirement ${\bf R}(\varphi)=0$ just means that in these coordinates
$\varphi$ does not depend on the coordinate $z$. The two other equations of
$1^0$ define the following system of differential equations for $\varphi$:
\[
-\varphi \frac{\partial \varphi}{\partial x}-\sqrt{1-\varphi^2}\frac{\partial \varphi}{\partial y}=0,\
\sqrt{1-\varphi^2}\frac{\partial \varphi}{\partial x}-\varphi\frac{\partial \varphi}{\partial y}=0.
\]
Noticing that the matrix
\[
\left\|\matrix{
-\varphi              &-\sqrt{1-\varphi^2}\cr
\sqrt{1-\varphi^2}    &-\varphi
\cr}\right\|
\]
has non-zero determinant, we conclude that the only solution of the above
system is the zero-solution:
\[
\frac{\partial \varphi}{\partial x}=\frac{\partial \varphi}{\partial y}=0.
\]
We obtain that in the coordinates used the phase function $\varphi$ depends
only on $\xi$. Therefore, in view of (2.27), for the scale factor $L$ we get
$$
L=\frac{\sqrt{1-\varphi^2}}{|\varphi_\xi|}.
$$
Now, the requirement $2^0$, which in these coordinates reads
\[
\frac{\partial L}{\partial \xi}=
\frac{\partial}{\partial \xi}\frac{\sqrt{1-\varphi^2}}{|\varphi_\xi|}=0,
\]
just means that the scale factor $L$ is a pure constant: $L=const$. In this
way we obtain the differential equation
\begin{equation}
\frac{\partial \varphi}{\partial \xi}=\mp \frac 1L \sqrt{1-\varphi^2}.
\end{equation}
The obvious solution to this equation reads
\begin{equation}
\varphi(\xi)=cos\left(\kappa\frac{\xi}{L}+const\right),
\end{equation}
where $\kappa=\pm 1$.
It is worth to note that the characteristic {\it frequency}
\begin{equation}
\nu=\frac cL
\end{equation}
has nothing to do with the frequency in CED. In fact, the quantity $L$ can
not be defined in Maxwell's theory.

Finally we note, that the so obtained phase function
$\varphi(\xi)$ leads to the
following. The 2-form $tr({\cal R}^0)$, where ${\cal R}^0$ is the matrix of
2-forms, formed similarly to the matrix ${\cal R}$, but using the basis
$({\bf A,\varepsilon A^*,R,S})$ instead of the basis
$(A,\varepsilon A^*,{\bf R,S})$, is closed. In fact,
\[
tr({\cal R}^0)=\varphi dx\wedge d\xi +\varphi dy\wedge d\xi -dy\wedge dz +dz\wedge d\xi
\]
and since $\varphi=\varphi(\xi)$, we get ${\bf d}tr({\cal R}^0)=0$.
Note also that the above explicit form of $tr({\cal R}^0)$ allows to define
the phase function by
\[
\varphi=\sqrt{\frac{|tr({\cal R}^0)|^2}{2}}.
\]
\noindent{\it Remark}.\ If one of the two functions $u$ and $p$, for example
$p$, is equal to zero: $p=0$, then formally we again have a solution, which
may be called {\it linearly polarized} by obvious reasons. Clearly, the phase
function of such solutions will be {\it constant}: $\varphi=const$, so, the
corresponding scale factor becomes infinitely large: $L\rightarrow \infty$,
therefore, condition $2^0$ is not satisfied. The reason for this is, that at
$p=0$ the function $u$ becomes a {\it running wave} and we get \linebreak
$|\delta F|=|\delta *F|=0$, so the scale factor can not be defined by (2.24).

\vskip 0.5cm
\subsection{Intrinsic angular momentum (spin, helicity)}
The problem for describing the {\it intrinsic angular momentum} (IAM), or in
short {\it helicity, spin} of the photon is of fundamental importance in
modern physics, therefore, we shall pay a special attention to it. In
particular, we are going to consider two approaches for its mathematical
description. But first, some preceding comments.

First of all, {\it there is no any doubt that every free photon carries such
an  intrinsic angular momentum}. Since the angular momentum is a
conserved quantity, the existence of the photon's intrinsic angular
momentum can be easily established and, in fact, its presence  has been
experimentally proved by an immediate observation of its mechanical action
and its value has been numerically measured.  Assuming this is so, we have to
understand its origin, nature and its entire meaning for the existence and
outer relations of those natural entities, called shortly photons somewhere
in the first quarter of this century.

So, we begin with the assumption:{\it every free photon carries an intrinsic
angular momentum with integral value equal to the Planck's constant
$h$}. According to our understanding, the photon's IAM comes from an intrinsic
{\it periodic process}. This point of view undoubtedly leads to the notion,
that photons {\it are not} point-like structureless objects, they have a
structure, i.e. they are {\it extended objects}. In fact, according to one
of the basic principles of physics {\it all free objects move as a whole
uniformly}.  So, if the photon is a point-like object any characteristic of a
periodic process, e.g. frequency,  should come from an outside force field,
i.e. it can not be free: a free point-like (structureless) object can not
have the characteristic frequency.

This simple, but true, conclusion sets the theoretical physics of the first
quarter of this century faced with a serious dilemma: to keep the notion of
structurelessness and to associate in a formal way the characteristic
frequency to the microobjects, or to leave off the notion of structuelesness,
to assume the notion of extendedness and availability of intrinsically
occurring periodic process and to build corresponding integral
characteristics, determined by this periodic process. A look back in time
shows that the majority of those days physicists had adopted the first
approach, which has brought up to life quantum mechanics as a computing
method , and the dualistic-probabilistic interpretation as a philosophical
conception.  If we set aside the wide spread and intrinsically controversial
idea that all microobjects are at the same time (point-like) particles and
(infinite) waves, and look impartially, in a fair-minded way, at the quantum
mechanical wave function for a {\it free} particle, we see that {\it the only
positive consequence} of its introduction is {\it the legalization of
frequency}, as an inherent characteristic of the microobject. In fact, the
probabilistic interpretation of the quantum mechanical wave function for a
free object, obtained as a solution of the free Schroedinger equation, is
impossible since its square is not an integrable quantity (the integral is
infinite). The frequency is really needed not because of the
dualistic-probabilistic nature of microobjects, it is needed because the
Planck's relation $E=h\nu$ turns out to be universally true in microphysics,
so there is no way to avoid the introduction of frequency. The question is,
if the introduction of frequency necessarily requires some (linear) "wave
equation" and the simple complex exponentials of the kind
$const.exp[i({\bf k.r}-\nu t)]$, i.e. running waves,
as "free solutions".  Our answer to this question is "no".  The
classical wave is something much richer and much more engaging concept, so it
hardly worths to use it just because of the attribute of frequency. In our
opinion, it suffices to have a periodic process at hand.

These considerations made us turn to the soliton-like objects, they realize
the two features of the microobjects ({\it localized spatial extendedness and
time-periodicity}), simultaneously, and, therefore, seem to be more adequate
theoretical models for those microobjects, obeying the Planck's relation
$E=h\nu$. Of course, if we are interested only in the behaviour of the
microobject as a whole, we can use the point-like notion, but any attempt to
give a meaning of its integral characteristics without looking for their
origin in the consistent intrinsic dynamics and structure, in our opinion, is
not a perspective theoretical idea. And the "stumbling point" of such an
approach is just the availability of an intrinsic mechanical angular
momentum, which can not be understood as an attribute of a free structureless
object.

\vskip 0.5cm
Having in view the above considerations, we are going to consider two ways to
introduce and define the intrinsic angular momentum as a local quantity and
to obtain, by integration, its integral value. So, these two approaches will
be of use only for the spatially finite nonlinear solutions of our equations.
The both approaches introduce in different ways 3-tensors (2-covariant and
1-contravariant). Although these two 3-tensors are built of quantities,
connected in a definite way with the field $F$, their nature is quite
different. The first approach is based on an appropriate tensor
generalization of the classical Poynting vector. The second approach makes
use of the concept of {\it torsion}, connected with the field $F$, considered
as 1-covariant and 1-contravariant tensor. The first approach is pure
algebraic, while the second one uses derivatives of $F_{\mu\nu}$. The
 spatially finite nature of the solutions $F$ allows to
build corresponding integral conserved quantities, naturally interpreted as
angular momentum. The scale factor $L$ appears as a multiple, so these
quantities go to infinity for all linear (i.e. for Maxwell's) solutions.

In the first approach we make use of the scale factor $L$, the isotropic
vector field ${\bf V}$ and the two 1-forms $A$ and $A^*$. By these four
quantities we build the following 3-tensor $H$:
\begin{equation}
H=\kappa \frac Lc {\bf V}\otimes(A\wedge A^*).
\end{equation}
The connection with the classical vector of Poynting comes through the
exteriour product of $A$ and $A^*$, the 3-dimensional sense of which is just
the Pointing's vector. In components we have
\[
H^\mu_{\nu\sigma}= \kappa\frac Lc {\bf V}^\mu(A_\nu A^*_\sigma-A_\sigma A^*_\nu).
\]
In our system of coordinates we get
\[
H=\kappa\frac Lc\left(-\varepsilon\frac{\partial}{\partial z}+
\frac{\partial}{\partial \xi}\right)\otimes(\varepsilon\phi^2dx\wedge dy),
\]
so, the only non-zero components are
\[
H^3_{12}=-H^3_{21}=-\kappa\frac Lc \phi^2,
\ H^4_{12}=-H^4_{21}=\kappa\varepsilon\frac Lc \phi^2.
\]
It is easily seen, that the divergence
$\nabla_\mu H^\mu_{\nu\sigma}\rightarrow\nabla_\mu H^\mu_{12}$
is equal to 0. In fact,
\[
\nabla_\mu H^\mu_{12}=\frac{\partial}{\partial z}H^3_{12}+
\frac{\partial}{\partial \xi}H^4_{12}=\kappa\frac Lc\left[-(\phi^2)_z+
(\varepsilon \phi^2)_\xi\right]=0
\]
because $\phi^2$ is a running wave along the coordinate $z$. Since the
tangent bundle is trivial we may construct the antisymmetric 2-tensor
\[
{\bf H}_{\nu\sigma}=\int_{R^3}{H_{4,\nu\sigma}}dxdydz,
\]
the constant components of which are conserved quantities.
\[
{\bf H}_{12}=-{\bf H}_{21}=\int_{R^3}{H_{4,12}}dxdydz
=\kappa\varepsilon\frac Lc W=\kappa\varepsilon WT=
\kappa\varepsilon \frac W\nu.
\]
The non-zero eigen values of  ${\bf H}_{\nu\sigma}$ are pure imaginary and
are equal to $\pm iWT$. This tensor has unique non-zero invariant $P(F)$,
\begin{equation}
P(F)=\sqrt{\frac12 {\bf H}_{\nu\sigma}{\bf H}^{\nu\sigma}}=WT.
\end{equation}
The quantity $P(F)$ will be called {\it Planck's invariant} for the finite
nonlinear solution $F$. All finite nonlinear solutions $F_1,F_2,...$,
satisfying the condition
\[
P(F_1)=P(F_2)=...=h,
\]
where $h$ is the Planck's constant, will be called further {\it photon-like}.
The tensor field $H$ will be called {\it intrinsic angular momentum tensor}
and the tensor ${\bf H}$ will be called {\it spin tensor} or {\it helicity
tensor}. The Planck's invariant $P(F)=WT$, having the physical dimension of
action, will be called {\it integral angular momentum}, or just {\it spin} or
{\it helicity}.

The reasons to use this terminology are quite clear: the time evolution of
the two mutually orthogonal vector fields $A$ and $A^*$ is a
rotational-advancing motion around and along the $z$-coordinate (admissible
are the right and the left rotations: $\kappa =\pm 1$) with the advancing
velocity of $c$ and the frequency of circulation $\nu=c/L$. We see the basic
role of the two features of the solutions: their soliton-like character,
giving finite value of all integral quantities, and their nonlinear
character, allowing to define the scale factor $L$ correctly. From this point
of view the intrinsic angular momentum $h$ of a free photon is far from being
{\it incomprihensible} quantity, connected with the even more
incomprihensible duality "wave-particle", and it looks as a quite normal
integral characteristic of a solution, presenting a model of our knowledge of
the free photon.

\vskip 0.5cm
We proceed to the second approach by recalling the definition of {\it torsion}
of two (1,1) tensors. If $G$ and $K$ are 2 such tensors
\[
G=G_\mu ^\nu dx^\mu\otimes \frac{ \partial}{\partial x^\nu },\quad
K=K_\mu ^\nu dx^\mu\otimes \frac{ \partial}{\partial x^\nu },
\]
their torsion is defined as a 3-tensor $S_{\mu\nu}^\sigma=-
S_{\nu\mu}^\sigma $ by the equation
\[
S(G,K)(X,Y)=[GX,KY]+[KX,GY]+GK[X,Y]+KG[X,Y]-
\]
\[
-G[X,KY]-G[KX,Y]-K[X,GY]-K[GX,Y],
\]
where $[,]$ is the Lie-bracket of vector fields,
\[
GX=G_\mu ^\nu X^\mu \frac {\partial}{\partial x^\nu},\quad
GK=G_\mu ^\nu K_\sigma ^\mu dx^\sigma \otimes \frac{\partial}{\partial x^\nu}
\]
and $X,Y$ are 2 arbitrary vector fields. If $G=K$, in general
 $S(G,G)\neq 0$ and
\[
S(G,G)(X,Y)=2\left\{[GX,GY]+GG[X,Y]-G[X,GY]-G[GX,Y]\right\}.
\]
This last expression defines at every point $x\in M$ the torsion $S(G,G)=S_G$
of $G$ with respect to the 2-dimensional plane, defined by the two vectors
$X(x)$ and $Y(x)$. Now we are going to compute the torsion $S_F$ of the
nonlinear solution $F$ with respect to the intrinsically defined by the two
unit vectors ${\bf A}$ and ${\bf \varepsilon A^*}$ 2-plane. In components we
have
\[
(S_F)_{\mu \nu }^\sigma =2\left[ F_\mu ^\alpha \frac{\partial F_\nu
^\sigma} {\partial x^\alpha }-F_\nu ^\alpha \frac{\partial F_\mu ^\sigma
}{\partial x^\alpha }-F_\alpha ^\sigma \frac{\partial F_\nu ^\alpha
}{\partial x^\mu } +F_\alpha ^\sigma \frac{\partial F_\mu ^\alpha }{\partial
x^\nu }\right].
\]
In our coordinate system
\[ {\bf A}=-\varphi
\frac{\partial} {\partial x}-\sqrt{1-\varphi ^2} \frac{\partial}{\partial
y},\quad {\bf \varepsilon A^*} = \sqrt{1-\varphi ^2}\frac{\partial}{\partial
x}- \varphi \frac{\partial} {\partial y},
\]
so,
\[ (S_F)_{\mu \nu }^\sigma
{\bf A}^\mu {\bf \varepsilon A^*}^\nu =(S_F)_{12}^\sigma ({\bf A}^1{\bf
\varepsilon A^*}^2- {\bf A}^2{\bf \varepsilon A^*}^1).
\]
For $(S_F)_{12}^\sigma $ we get
\[
(S_F)_{12}^1=(S_F)_{12}^2=0,\quad
(S_F)_{12}^3=-\varepsilon (S_F)_{12}^4=2\varepsilon \{p(u_\xi -\varepsilon
u_z)-u(p_\xi -\varepsilon p_z)\}.
\]
\vskip 0.5cm
{\it Remark}. In our case
$(S_F)_{12}^\sigma=(S_{*F})_{12}^\sigma $, so further we shall work with
$S_F$ only.
\vskip 0.5cm
It is easily seen that the following relation holds:
${\bf A}^1{\bf \varepsilon A^*}^2-{\bf A}^2{\bf \varepsilon A^*}^1=1.$
Now, for the case
\[
u=\phi (x,y,\xi +\varepsilon z)\cos\left(\kappa \frac \xi L +const\right),\quad
p=\phi (x,y,\xi +\varepsilon z)\sin \left(\kappa \frac \xi L +const\right)
\]
we obtain
\[
(S_F)_{12}^3=-\varepsilon (S_F)_{12}^4=-2\varepsilon \frac \kappa L \phi ^2,
\]

\[
(S_F)_{\mu \nu }^\sigma {\bf A}^\mu {\bf \varepsilon A^*}^\nu
=\left[0,0,-2\varepsilon
\frac \kappa L \phi ^2,2\frac \kappa L \phi ^2\right].
\]
Since $\phi^2$ is a running wave along the $z$-coordinate, the vector
field $S_F({\bf A,\varepsilon A^*})$ has zero divergence:
$\nabla_\nu \left[S_F({\bf A,\varepsilon A^*})\right]^\nu=0$.
Now we define the {\it helicity vector} for the solution $F$ by
\[
\Sigma_F=\frac{L^2}{2c}S_F({\bf A,\varepsilon A^*}).
\]
Since $L=const$, then $\Sigma_F$ has also zero divergence, so the integral
quantity
\[
\int{\left(\Sigma_F\right)_4}dxdydz
\]
does not depend on time and is equal to $\kappa WT$. The photon-like
solutions are separated in the same way by the condition $WT=h$. Here are
three more integral expressions for the quantity $WT$. We form the 4-form
\[
-\frac 1L {\bf S}\wedge*\Sigma_F=\frac \kappa c \phi^2\omega_\circ
\]
and integrate it over the 4-volume ${\cal R}^3\times L$, the result is $\kappa
WT$. Besides, we verify easily the relations
\[
\frac 1c \int_{R^3\times L}|A\wedge
A^*|\omega_\circ= \frac{L^2}{c}\int_{R^3\times L}|\delta F\wedge \delta
*F|\omega_\circ=WT.
\]

Since we separate the photon-like solutions by the relation $WT=h$, the last
expressions suggest the following interpretation of the Planck's constant
$h$.  Since $|A\wedge A^*|$ is proportional to the area of the square,
defined by the two mutually orthogonal vectors $A$ and $\varepsilon A^*$, the
above integral sums up all these areas over the whole 4-volume, occupied by
the solution $F$ during the intrinsically determined time period $T$, in
which the couple $(A,\varepsilon A^*)$ completes a full rotation. The same
can be said for the couple $(\delta F,\delta *F)$ with some different factor
in front of the integral. This shows quite clearly the "helical" origin of
the full energy $W=h\nu$ of the single photon.

\vskip 0.5cm
\subsection{Solutions in spherical cordinates}
The so far obtained soliton-like
solutions describe objects, "coming from infinity" and
"going to infinity". Of interest are also soliton like solutions "radiated"
from, or "absorbed", by some central "source" and propagating radially from
or to the center of this source. We are going to show, that our equations
admit such solutions too. We assume this central source to be a small ball
$R^0$ with radius $r_\circ$, and put the origin of the coordinate system at
the center of the source-ball. The standard spherical coordinates
$(r,\theta,\varphi,\xi)$ will be used and all considerations will be carried
out in the region out of the ball $R^0$. In these coordinates we have
\[
ds^2=-dr^2-r^2d\theta ^2-r^2 sin^2\theta d\varphi^2 +d\xi^2, \ \sqrt{|\eta|}=r^2sin\theta.
\]
The $*$-operator acts in these coordinates as follows:
\[
\begin{array}{ll}
*dr=r^2 sin\theta d\theta\wedge d\varphi\wedge d\xi  &*(dr\wedge d\theta\wedge d\varphi )=(r^2 sin\theta)^{-1} d\xi \cr
*d\theta=-sin\theta dr\wedge d\varphi\wedge d\xi     &*(dr\wedge d\theta\wedge d\xi)=sin\theta d\varphi  \cr
*d\varphi=(sin\theta)^{-1}dr\wedge d\theta d\xi      &*(dr\wedge d\varphi\wedge d\xi)=-(sin\theta)^{-1}d\theta     \cr
*d\xi=r^2 sin\theta dr\wedge d\theta d\varphi        &*(d\theta\wedge d\varphi\wedge d\xi)=(r^2 sin\theta)^{-1} dr
\end{array}
\]
\[
\begin{array}{ll}
*(dr\wedge d\theta)=-sin\theta d\varphi\wedge d\xi        &*(d\theta\wedge d\varphi)=-(r^2 sin\theta)^{-1}dr\wedge d\xi \cr
*(dr\wedge d\varphi)=(sin\theta)^{-1} d\theta\wedge d\xi  &*(d\theta\wedge d\xi)=-sin\theta dr\wedge d\varphi \cr
*(dr\wedge d\xi)=r^2 sin\theta d\theta\wedge d\varphi     &*(d\varphi\wedge d\xi=(sin\theta)^{-1}dr\wedge d\theta.
\end{array}
\]
We look for solutions of the following kind:
\begin{equation}
F=\varepsilon udr\wedge d\theta +ud\theta\wedge d\xi +\varepsilon pdr\wedge d\varphi
+pd\varphi \wedge d\xi,
\end{equation}
where $u$ and $p$ are spatially finite functions. We get
\[
*F=\frac{p}{sin\theta}dr\wedge d\theta +\varepsilon \frac{p}{sin\theta}d\theta\wedge d\xi-
usin\theta dr\wedge d\varphi-\varepsilon sin\theta d\varphi \wedge d\xi.
\]
The following relations hold:
\[
F\wedge F=2\varepsilon(up-up)dr\wedge d\theta\wedge d\varphi \wedge d\xi=0,
\]
\[
F\wedge *F= \left(-u^2 sin\theta +u^2 sin\theta-
\frac{p^2}{sin\theta}+\frac{p^2}{sin\theta}\right)dr\wedge d\theta\wedge d\varphi\wedge d\xi=0,
\]
i.e. the two invariants are equal to zero: $(*F)_{\mu\nu}F^{\mu\nu}=0,
\ F_{\mu\nu}F^{\mu\nu}=0$.

After some elementary computation we obtain
\[
\delta F\wedge F=\delta *F\wedge *F=
\varepsilon\left[u\left(\varepsilon p_r +p_\xi\right)-
p\left(\varepsilon u_r +u_\xi\right)\right]dr\wedge d\theta\wedge d\varphi+
\]
\[
+\left[u\left(\varepsilon u_r +u_\xi\right)-
u\left(\varepsilon p_r +p_\xi \right)\right]d\theta\wedge d\varphi\wedge d\xi,
\]
\[
F\wedge *{\bf d}F=\varepsilon\left[u\left(\varepsilon u_r +u_\xi\right)sin\theta+
\frac{p\left(\varepsilon p_r +p_\xi \right)}{sin\theta}\right]dr\wedge d\theta\wedge d\varphi-
\]
\[
-\varepsilon\left[u\left(\varepsilon u_r + u_\xi\right)sin\theta+
\frac{p\left(\varepsilon p_r + p_\xi \right)}{sin\theta}\right]d\theta\wedge d\phi\wedge d\xi,
\]
\[
(*F)\wedge *{\bf d}*F=\left[u\left(\varepsilon u_r +u_\xi\right)sin\theta+
\frac{p\left(\varepsilon p_r +p_\xi \right)}{sin\theta}\right]dr\wedge d\theta\wedge d\varphi-
\]
\[
-\left[u\left(\varepsilon u_r + u_\xi\right)sin\theta+
\frac{p\left(\varepsilon p_r + p_\xi \right)}{sin\theta}\right]d\theta\wedge
d\phi\wedge d\xi.
\]
So, the two functions $u$ and $p$ have to satisfy the equation
\begin{equation}
u\left(\varepsilon u_r + u_\xi\right)sin\theta+
\frac{p\left(\varepsilon p_r + p_\xi \right)}{sin\theta}=0,
\end{equation}
which is equivalent to the equation
\begin{equation}
\left(u^2 sin\theta +\frac{p^2}{sin\theta}\right)_\xi +
\varepsilon\left(u^2 sin\theta +\frac{p^2}{sin\theta}\right)_r=0.
\end{equation}
The general solution of this equation is
\begin{equation}
u^2 sin\theta +\frac{p^2}{sin\theta}=\phi^2(\xi - \varepsilon r,\theta,\phi).
\end{equation}
For the non-zero components of the energy-momentum tensor we obtain
\begin{equation}
-Q_{1}^{1}=-Q_{1}^{4}=Q_{4}^{1}=Q_4^4=
\frac{1}{4\pi r^2 sin\theta}\left(u^2 sin\theta +\frac{p^2}{sin\theta}\right).
\end{equation}
It is seen that the energy density is not exactly a running wave but when we
integrate to get the integral energy, the integrand is exactly a running
wave:
\[
W=\frac{1}{4\pi}\int_{R^3-R^0}{*\left(Q_\mu^4 d\xi\right)}=
\frac{1}{4\pi}\int_{R^3-R^0}{\left(u^2 sin\theta
+\frac{p^2}{sin\theta}\right)}dr\wedge d\theta\wedge d\phi.
\]
Since the functions $u$ and $p$ are spatially finite, the integral energy $W$
is finite, and from the explicit form of the energy-momentum tensor it
follows the well known relation between the integral energy and momentum:
$\ W^2-c^{2} {\bf p}^2=0$.

\section{\bf Interference, Nonlinearity and \\Superposition}
\vskip 0.05cm
\subsection{Introductory remarks}

Having at hand the photon-like solutions a natural next step is to try to
describe the situation when two photons occupy the same (or partially the
same) 3-region in some period of time. It is clear, that if these two photons
meet somewhere, i.e. their cilinder-like world-tubes  intersect
non-trivially, the interesting case is when they {\it move along the same
spatial straight line} and in {\it the same direction}. Since they move by
the same velocities they will continue to overlap each other until some
outer agent causes a change. What kind of an object is obtained in this way,
is it a photon or not, what kind of interaction takes place, what is its
integral energy , its momentum and its angular momentum? Many
challenging and still not answered questions may be set in this direction
before the theoretical physics. And this section is devoted to consideration
of some of these problems in the frame of Extended electrodynamics.

Almost all experiments set to find some immediate mutual
interaction of two (or more) electromagnetic fields in vacuum, causing some
observable effects (e.g.  frequency or amplitude changes), as far as we know,
have faild, exept when the two fields satisfy the so called {\it coherence
conditions}. In the frame of CED and working with plane waves this simply
means, that their {\it phase difference must be a constant quantity}. The
usual way of consideration is limited to {\it cosine}-like running waves with
the {\it same} frequency. The physical explanation is based on the linearity
of Maxwell's equations, which require any linear combination of solutions to
be again a solution, so the "building points" of the medium, subject to the
field pressure of the two independent fields, go out of their equilibrium
state obeying simoultaneously the two forces applied in the overlaping
3-region.  After getting out of this overlaping 3-region the fields stay what
they have been before the interaction. In order to describe the interaction,
i.e. the observed redistribution of the energy-momentum density inside the
overlaping 3-region, CED uses the corresponding mathematical expressions
in Maxwell's theory and gets comparetively good results. Most frequently the
Poynting vector $S\sim \left[(E_1+E_2)\times (B_1+B_2)\right]$ is used and
the cross-terms $(E_1\times B_2)+(E_2\times B_1)$ are held responsible for
the interaction, in fact, the very {\it interference} is defined by the
condition that these cross-terms, usually called "interference terms", are
different from zero.  But, the interference takes really place {\it only when
the coherence conditions are met}, while CED permits interference almost
always. In other words, from Maxell's theory we can not obtain these
coherence conditions as {\it necessary conditions} for some interference to
take place.

In EED we {\it have no superposition principle}, so we have to approach this
physical situation in a new way. First, let's specify the situation more
in detail and in terms of the notion for $EM$-field in EED. Roughly speaking,
this notion is based on the idea for discreteness, i.e. the real
electromagnetic fields consist of many noninteracting, or very weakly
interacting, photons, moving in {\it various} directions.  Because of the
great velocity of their straight line motion it is hardly possible to observe
and say what happens when two photons meet somewhere.  The experiment shows
that in the most cases they pass through each other and forget about the
meeting.  As we mentioned above, the interseting case is when they move along
the same direction and the regions, they occupy, overlap nontrivially.

The nonlinear solutions we obtained in the preceding section can not describe
such set of photons, moving in {\it various} directions. Even if we choose the
amplitude function $\phi$ to consist of many "3-bubles" (since $\phi$ is a
running wave, all these bubles, are parts of the same running wave), they all
have to move in the same direction, which is a special, but not the general,
case of the situation we consider here.  So, in order to incorporate for
description such situations, some perfection of EED is needed.  As before,
this perfection shall consist of two steps:  first, elaboration of the
algebraic character of the mathematical field, second, elaboration of the
equations.  The second step, besides its dynamical task, must define also the
necessary conditions for interference of photon-like solutions, which should
coincide with the above mentioned, experimentally established and repeatedly
confirmed coherence conditions. We shall see that this is easily achievable
in EED.

\vskip 0.5cm
\subsection{Elaborating the mathematical object}
Recall that our mathematical object that represents the field is a 2-form
with values in ${\cal R}^2$. We want to elaborate it in order to reflect more
fully the physical situation. The new moment is that inside the 3-region
under consideration we have {\it many} photons. Each of these photons,
considered as independent objects, is described by a pfoton-like solution as
given in the preceding section, i.e. each of them has its own spatial
structure, its own scale factor (or frequency) and its own direction of
motion as a whole. Of course, the velocity of motion is the same for all of
them. To this physical situation we have to juxtapose {\it one} mathematical
object, which have to generalize in a natural way our old object $\Omega$.
The idea for this generalization is very simple and consists in the
following. With every single photon, we associate its own ${\cal R}^2$-space,
so if the number of the presenting photons is $N$, we'll have $N$ such
spaces, or a new $N$-dimensional vector space. Denoting this vector space by
${\cal N}$, our object becomes a 2-form $\Omega$ with values in the vector
space
${\cal R}^2\otimes {\cal N}$:
$\ \Omega\in \Lambda^2\left(M,{\cal R}^2\otimes {\cal N}\right)$.
We recall now how this vector space ${\cal N}$ is explicitly built.

If ${\cal K}$ is an arbitrary set, finite or infinite, we consider those
mappings of this set into a given field, e.g. ${\cal R}$, which are different
from zero only for finite number of elements of ${\cal K}$. These mappings
will be the elements of the space ${\cal N}$. A basis of this space is built
in the following way. We consider the elements $f\in {\cal N}$, having the
property: if $a\in {\cal K}$ then $f(a)=1$ and $f$ has zero values for all
othe elements of ${\cal K}$. So, with every element $a\in {\cal K}$ we
associate the corresponding element $f_a \in {\cal N}$, therefore, an
arbitrary element $f\in {\cal N}$ is represented as follows:
\[
f=\sum_{i=1}^N \left(\lambda^i f_{a_i}\right),
\]
where $\lambda^i$, $i=1,2,...,N$, are the values, aquired by $F$, when $i$
runs from 1 to $N$ (of course, some of the $\lambda $'s may be equal to
zero). The linear structure in ${\cal N}$ is naturally introduced, making use
of the linear structure in ${\cal R}$ in the well known way. The linear
{\it independence} of $f_{a_i} $ is easily shown. In fact, assuming the
opposite, i.e. that there exist such $\lambda^i$, among which at least one is
not zero and the following relation holds
\[
\sum_{i=1}^{N}\lambda^{i}f_{a_i}=0,
\]
then for any $j=1,2,...,N$ we'll have
\[
\sum_{i=1}^{N}\lambda^{i}f_{a_i}(a_j)=\lambda^j=0,
\]
which contradicts the assumption. Hence, $f_{a_i}$ define really a basis of
${\cal N}$. Now we form the injective mapping $i_N :N\rightarrow {\cal N}$,
defined by
\[
i_N (a)=f_a,\ a\in N,
\]
so the set $N$ turns into a basis of ${\cal N}$. If such a construction is
made, then ${\cal N}$ is called a {\it free vector space over the set $N$}.
Further onthe corresponding basis of our set of photons will be denoted by
$E_a$. So, our mmathematical object will look as follows (summing up over the
repeating index $a$)
\begin{equation}
\Omega=\Omega^a \otimes E_a =
\left[F^a \otimes e_1^a + (*F)^a \otimes e_2^a\right]\otimes E_a,
\end{equation}
where $(e^a_1,e^a_2)$ is the associated with the field $F^a$ basis. If we
work in an arbitrary basis of ${\cal R}^2$, the full writing reads ($i$=1,2)
\begin{equation}
\Omega=\Omega^a \otimes E_a =\Omega_i^a\otimes k^i_a\otimes E_a.
\end{equation}
Following the several times used already method we define the product of two
2-forms of the kind (2.39). For erxample, if
$\Phi=\Phi^a_i\otimes k_a^i\otimes E_a,
\Psi=\Psi^b_j\otimes l_b^j\otimes E_b$, we'll have
\[
(\vee,\vee)(\Phi,\Psi)=(\vee,\vee)(\Phi^a_i\otimes k_a^i\otimes E_a,
 \Psi^b_j\otimes l_b^j\otimes E_b)=
\]
$$
 =\sum_{a=1}^{N}\Big[\Phi^a_1\wedge \Psi^a_1\otimes k^1_a\vee
 l^1_a+\Phi^a_2\wedge \Psi^a_2\otimes k^2_a\vee l^2_a+
$$
$$
 +(\Phi^a_1\wedge \Psi^a_2+\Phi^a_2\wedge \Psi^a_1)
 \otimes k^1_a\vee l^2_a\Big]\otimes E_a\vee E_a +
$$
\[
 +\sum_{a<b=1}^{N}\Big[(\Phi^a_1\wedge \Psi^b_1+\Phi^b_1\wedge \Psi^a_1)
 \otimes k_a^1\vee l_b^1+ (\Phi^a_2\wedge \Psi^b_2+\Phi^b_2\wedge \Psi^a_2)
 \otimes k_a^2\vee l_b^2+
\]
\[
+(\Phi^a_1\wedge \Psi^b_2+\Phi^a_2\wedge \Psi^b_1+
\Phi^b_1\wedge \Psi^a_2+\Phi^b_2\wedge \Psi^a_1)\otimes k_a^1\vee l_b^2\Big]
\otimes E_a\vee E_b.
\]

Let now $\Omega$ be of the kind $\Omega=(F^a\otimes e_a^1+*F^a \otimes e_a^2)
\otimes E_a$. Then, forming $*\Omega$ and $\delta \Omega$, for
$(\vee,\vee)(\delta\Omega,*\Omega)$ we obtain
\[
(\vee,\vee)(\delta\Omega,*\Omega)=
 \sum_{a=1}^{N}\Big[\delta F^a\wedge *F^a \otimes e^1_a\vee e^1_a+
\delta *F^a\wedge **F^a\otimes e^2_a\vee e^2_a+
\]
\[
 +(\delta F^a\wedge **F^a+\delta *F^a\wedge *F^a)\otimes e^1_a\vee e^2_a\Big]
\otimes E_a\vee E_a+
\]
\[
 +\sum_{a<b=1}^{N}\Big[(\delta F^a\wedge *F^b+\delta F^b\wedge *F^a)\otimes e^1_a\vee e^1_b+
(\delta *F^a\wedge **F^b+\delta *F^b\wedge **F^a)\otimes e^2_a\vee e^2_b
\]
\[
 +(\delta F^a\wedge **F^b+\delta *F^a\wedge *F^b+
 \delta F^b\wedge **F^a+\delta *F^b\wedge *F^a)\otimes e^1_a\vee e^2_b\Big]
\otimes E_a\vee E_b.
\]

\vskip 0.5cm
\subsection{Elaborating the field equations}
If we want to consider a set of independent solutions, then in the above
expression we take the trace $tr$ over the indeces of $E_a\vee E_b$. The
compact writing of this condition reads
\begin{equation}
tr(\vee,\vee)(\delta\Omega,*\Omega)=0,
\end{equation}
which is equivalent to the equations
\begin{equation}
\delta F^a\wedge *F^a=0,\ \delta *F^a\wedge **F^a=0,\
\delta F^a\wedge **F^a +\delta *F^a\wedge *F^a=0.
\end{equation}
Clearly, in this case the full energy-momentum tensor $Q_\mu^\nu$ will be a
sum of all energy tensors $(Q^a)_\mu^\nu$ of the single solutions.

The general equations are written down as follows:
\begin{equation}
(\vee,\vee)(\delta\Omega,*\Omega)=0.
\end{equation}
The equivalent (component-wise) form of (2.42) reads
\[
{\delta F^a\wedge *F^a=0,\ \delta *F^a\wedge **F^a=0,\ \delta F^a\wedge **F^a +\delta *F^a\wedge *F^a=0},
\]
\[
{\delta F^a\wedge *F^b +\delta F^b\wedge *F^a=0,\ \delta *F^a\wedge **F^b +\delta *F^b\wedge **F^a=0},
\]
\[
{\delta F^a\wedge **F^b +\delta *F^b\wedge **F^a+\delta *F^a\wedge *F^b +\delta *F^b\wedge *F^a=0}.
\]

Let now $F^a$, $a=1,2,...,N$ define a solution of the above system of
equations (2.42). We are going to show that the linear combination with
constant coefficients $\lambda_a$
\[
F=\sum_{a=1}^{N}\lambda_a F^a
\]
satisfies the equations:
$\delta F\wedge *F=0,\ \delta *F\wedge **F=0,\newline \delta
F\wedge **F+\delta*F\wedge *F=0$.
In fact
\[
\delta F\wedge
*F=\sum_{a=1}^{N}(\lambda_a)^2(\delta F^a\wedge *F^a)+ \sum_{a<b=1}^N
\lambda_a \lambda_b (\delta F^a\wedge *F^b + \delta F^b\wedge *F^a),
\]
\[
\delta *F\wedge **F=\sum_{a=1}^{N}(\lambda_a)^2(\delta *F^a\wedge **F^a)+
\sum_{a<b=1}^N \lambda_a \lambda_b (\delta *F^a\wedge **F^b +
\delta *F^b\wedge **F^a),
\]
\[
\delta F\wedge **F +\delta *F\wedge *F=
\sum_{a=1}^N (\lambda_a)^2 (\delta F^a\wedge **F^a)+
\]
\[
+\sum_{a<b=1}^N \lambda_a \lambda_b(\delta F^a\wedge **F^b+
\delta F^b\wedge **F^a)+
\]
\[
+\sum_{a=1}^N (\lambda_a)^2 (\delta *F^a\wedge *F^a)+
\sum_{a<b=1}^N \lambda_a \lambda_b(\delta *F^a\wedge *F^b+
\delta *F^b\wedge *F^a)=
\]
\[
=\sum_{a=1}^N(\lambda_a)^2(\delta F^a\wedge **F^a +\delta *F^a\wedge *F^a)+
\]
\[
+\sum_{a<b=1}^N \lambda_a\lambda_b(\delta F^a\wedge **F^b +
\delta F^b\wedge **F^a+\delta *F^a\wedge *F^b +\delta *F^b\wedge *F^a).
\]
Obviously, the component-wise writing down of the equations (2.42) shows that
every addend is equal to zero. This result can be interpreted as some
particular "superposition principle", i.e. if we have finite number of
solutions $F^a$ of the system
\begin{equation}
\delta F\wedge *F=0,\ \delta *F\wedge **F=0,\
\delta F\wedge **F+\delta*F\wedge *F=0,
\end{equation}
which solutions satisfy additionally the equations
\begin{equation}
{\delta F^a\wedge *F^b +\delta F^b\wedge *F^a=0,\ \delta *F^a\wedge **F^b +\delta *F^b\wedge **F^a=0},
\end{equation}
\begin{equation}
{\delta F^a\wedge **F^b +\delta *F^b\wedge **F^a+\delta *F^a\wedge *F^b +\delta *F^b\wedge *F^a=0},
\end{equation}
then the 2-form $F=\sum_{a=1}^{N}\lambda_a F^a$ is again a solution of
(2.43). Then, clearly, if $F$ and $G$ are 2 solutions of (2.43) and satisfy
(2.44) and (2.45), the new solution $(F+G)$ of (2.43) is naturally endowed
with the following energy-momentum tensor
\[
Q_{\mu\nu}=\frac{1}{4\pi}\Big[-(F+G)_{\mu\sigma} (F+G)_\nu^\sigma \Big].
\]
In the general case we'll have
\[
Q_{\mu\nu}=\frac{1}{4\pi}\left[-\left(\sum_{a=1}^{N}\lambda_a F^a \right)_{\mu\sigma}
\left(\sum_{a=1}^{N}\lambda_a F^a \right)_\nu^\sigma \right]
\]
In this way we can compute the corresponding "interference terms". In
\linebreak
particular, the "interference" energy density is obtained proportional
\linebreak to $-2F_{4\sigma}G^{4\sigma}$.

\vskip 0.5cm
\subsection{Coherence and interference}
We consider now two photon-like solutions $F_1$ and $F_2$ of equations
(2.43), propagating along the same direction. We choose this direction for
the $z$-axis of our coordinate system. We are going to find what additional
conditions on these solutions come from the additional equations (2.44) and
(2.45). We assume also, that the 3-regions, where the two amplitudes $\phi_1$
and $\phi_2$ are different from zero have non-empty intersection, because
otherwise, the interference term is equal to zero.
\[
F_1=\varepsilon_1 u_1 dx\wedge dz + u_1 dx\wedge d\xi + \varepsilon_1 p_1
dy\wedge dz +p_1dy\wedge d\xi
\]
\[
F_2=\varepsilon_2 u_2 dx\wedge dz + u_2 dx\wedge d\xi + \varepsilon_2 p_2
dy\wedge dz +p_2 dy\wedge d\xi,
\]
where
\[
u_1=\phi_1cos\left(\frac{\kappa_1\nu_1}{c}\xi+b_1\right),\
p_1=\phi_1sin\left(\frac{\kappa_1\nu_1}{c}\xi+b_1\right),\
\]
\[
u_2=\phi_2cos\left(\frac{\kappa_2\nu_2}{c}\xi+b_2\right),\
p_2=\phi_2sin\left(\frac{\kappa_2\nu_2}{c}\xi+b_2\right).
\]
After an elementary computation we obtain
\[
\delta F_1\wedge *F_2+\delta F_2\wedge *F_1=\left(\frac{\kappa_1\nu_1}{c}-\frac{\kappa_2\nu_2}{c}\right)(u_1p_2-u_2p_1)dx\wedge dy\wedge dz+
\]
\[
+\left(\varepsilon_1\frac{\kappa_1\nu_1}{c}-\varepsilon_2\frac{\kappa_2\nu_2}{c}\right)(u_1p_2-u_2p_1)dx\wedge dy\wedge d\xi+
\]
\[
+\left[p_1\left(u_{1x}+p_{1y}\right)+p_2\left(u_{2x}+p_{2y}\right)\right](\varepsilon_1\varepsilon_2 -1)dx\wedge dz\wedge d\xi+
\]
\[
+\left[u_1\left(u_{1x}+p_{1y}\right)+u_2\left(u_{2x}+p_{2y}\right)\right](1-\varepsilon_1\varepsilon_2)dy\wedge dz\wedge d\xi.
\]
Since
\[
u_1p_2-u_2p_1=\phi_1\phi_2sin\left[\left(\frac{\kappa_2\nu_2}{c}-\frac{\kappa_1\nu_1}{c}\right)\xi +b_2-b_1\right]\neq 0,
\]
the coefficient before $dx\wedge dy\wedge dz$ will be equal to zero only if
$\kappa_1\nu_1=\kappa_2\nu_2$. But $\nu_1$ and $\nu_2$ are positive
quantities, so it is necessary $\kappa_1=\kappa_2,\ \nu_1=\nu_2$.
Now, the coefficient in front of $dx\wedge dy\wedge d\xi$ will become zero
if $\varepsilon_1=\varepsilon_2$. From this last relation it follows that the
other two coefficients, obviously, are also zero. A corresponding computation
shows that the so obtained conditions
\begin{equation}
\nu_1=\nu_2,\ \varepsilon_1=\varepsilon_2,\ \kappa_1=\kappa_2
\end{equation}
are sufficient for $F_1$ and $F_2$ to satisfy the rest two equations of
(2.44) and (2.45). Hence, {\it if the 2-form}
\[
\Omega=(F_1\otimes e_1 +*F_1\otimes e_2)\otimes E_1+
(F_2\otimes k_1 +*F_2\otimes k_2)\otimes E_2
\]
{\it satisfies equations} (2.42), {\it then the 2-form} $F=F_1+F_2$ {\it
is a solution of our initial equations}
\[
\delta F\wedge *F=0,\ \delta *F\wedge **F=0,\
\delta F\wedge **F+\delta *F\wedge *F=0,
\]
{\it but not of photon-like type, the coherence conditions} (2.42) {\it are
satisfied} {\it and the interference of the two fields} $F_1$ and $F_2$ {\it
is possible}.  As for the "interference" energy density we obtain the well
known from CED expression \[
 W_{12}=\phi_1^2+\phi_2^2+2\phi_1\phi_2cos(b_2-b_1) \] from which the
classical interference picture is readily obtained.

\chapter {\itshape \Large  Extended Electrodynamics in Media}
==========================================
\section {\bf Basic equations}
\vskip 0.05 cm
\subsection{Preliminary remarks}
Recall from subsec.(1.5.3) that when we talk about a {\it medium} in EED, we
mean any continuous, i.e. spatially distributed, physical object, exchanging
energy-momentum with the available in the same region $EM$-field $\Omega $.
Formally, the medium is described by some mathematical object and, when this
object is chosen, we talk about {\it external} or {\it outer} field.
When interaction between $\Omega $ and the outer field takes place, of
basic importance for the theory is how $\Omega $ and the external field
participate in the expression, defining the exchanged energy-momentum in an
unit 4-volume. According to the hypotheses we made in subsec.(1.5.3) the
$EM$-field $\Omega $ participates directly in this expression, while the
exteriour field participates in this expression through specially constructed
two ${\cal R}^2$-valued 1-forms, and these 1-forms may depend on the
derivatives of the external field too. The vector-components of these ${\cal
R}^2$-valued 1-forms were called {\it currents}, since the classical {\it
current}, considered as a 1-form, may be considered as a particular case.
Taking into account the mathematical model of the $EM$-field, as well as the
experience with Maxwell's theory, we postulated the expression (1.42)

\[
\vee(\Phi,*\pi_1\Omega)\ +\ \vee(\Psi,*\pi_2\Omega)
\]
as sufficiently general and adequate to describe large enough class of
exchange processes, i.e. interaction of $\Omega $ with outer fields.

Such an approach, when the algebraic character and the differential equations
for the exteriour field are not known, needs a new, general enough and
sufficiently adaptable viewpoint, expressing a definite comprehension for
the character of the physical processes considered,
as well as general enough and
adequate enough mathematical facts. Such an adequacy must give reasons for
definite hypotheses, and, finally, to result in writing down definite
equations for the currents, no matter what the particular nature of the
currents is. We note, that we do not require and do not forbid the four
currents $\alpha ^i$ to have zero divergence. In our opinion, the new facts
to be used as fully as possible, are that their number is {\it more than one}
, that every current realizes a separate energy exchange channel, and that
there should be some correlations among them. It is naturally to expect such
correlations among the four currents to exist
since the exchange processes occur
locally and on the other side of this exchange stays just one physical object
- the $EM$-field $\Omega $. These correlations must be organized in such a
way, that to incorporate the special case of only one current different from
zero, as it is in CED. Of course, we have to remember that in CED the current
is a {\it vector field}, while our currents are 1-forms. These are different
objects although the available isomorphism through the pseudometric $\eta$,
and this difference will be explicitly taken into account in our approach.

\vskip 0.5 cm
\subsection{Maxwell's theory as a particular case}

Let's recall some important features of CED. First, {\it the $EM$-field
exchanges energy-momentum only through $F$ and does not exchange
energy-momentum through $*F$}. Formally, this is accounted by the equation
$\delta *F=0$, which contains (in differential form) the Faraday's
induction law.  In other words, CED assumes, that this experimentally
established fact for {\it some} media, holds for {\it all} media. As we know,
this assumption legalizes the $U(1)$-gauge interpretation of CED, where the
equation $\delta *F=0$ acquires the geometrical interpretation of Bianchi's
identity for the abelean group $U(1)$.

Second, an energy-momentum exchange may be realized {\it only with free or
bound electric charges}. The formal description of this exchange was
explained and commented in subsec. (1.1.3). The second equation of CED,
$$
\delta F=4\pi (j+j_{b}),
$$
identifies the pure field quantity $\delta F$
with the outer quantity {\it current}, which characterizes the distribution
and mechanical behaviour of the charge carriers. To what extent such an
identification is admissible is a personal view, and we are not going to
comment it. According to us much more natural is to write down a relationship
having the sense of local energy-momentum balance. In other words, the same
(exchanged) quantity of energy-momentum to be  written down in two ways: in
the first way, by means of the components $F_{\mu\nu}$ and their derivatives
only, and in the second way, through expressions where the outer field
components necessarily take part. Then, according to the local conservation
law, the two expressions are equalized. Namely such an approach we have
realized in EED.

The third feature we recall is the lack of some general enough and common
approach for determination of the full current $(j_{free}+j_{bound})$.
As we mentioned earlier, the
series developments of the polarization vector $P$ and magnetization vector
${\cal M}$ (or the 2-form $S$) with respect to $E$ and $B$ and their
derivatives may be a felicitous working skill, but it is not a perspective
theoretical idea. For example, this approach is not applicable for
strongly nonhomogeneous media, while a local energy-momentum balance equation
can be written down always and to be particularized and made more
precise in the course of work.

We see now how, from formal point of view, CED is incorporated in EED. The
mathematical expression for the exchanged energy-momentum in CED is
\[
4\pi F_{\mu\nu}(\delta S+J)^\nu dx^\mu.
\]
In our approach, using the defined by the field $\Omega $ basis $(e_1,e_2)$,
this expression is represented consecutively as follows:
\[
4\pi F_{\mu\nu}(\delta S+J)^\nu dx^\mu \otimes e_1 \vee e_1=
*\bigl[4\pi(\delta S+J)\wedge *F\bigr]\otimes e_1 \vee e_1=
\]
\[
=*\vee\bigl[4\pi(\delta S+J)\otimes e_1, *F\otimes e_1\bigr]=
*\vee\bigl[\pi_1\Phi, \pi_1*\Omega\bigr],
\]
where $\pi_1\Phi=4\pi(\delta S+J)\otimes e_1$. In this way, using the
notations of EED, we get
\[
\vee(\delta \Omega, *\Omega)=\vee(\pi_1\Phi, \pi_1*\Omega)+
\vee(\pi_1\Psi,\pi_2 *\Omega),\quad \pi_1\Phi =\pi_1\Psi.
\]
We see that there are no terms, describing an exchange through $*F$,
and the redistribution energy-momentum equation reduces to one of the real
exchange equations. So, CED reflects the following additional requirements to
the equations of EED: \ $\Phi=\alpha^1\otimes e_1,\ \Psi=\alpha^3 \otimes
e_1,\ \alpha ^1=\alpha ^3=4\pi(\delta S+J),\ \alpha ^2=\alpha ^4=0$. The very
Maxwell's equations $\delta *F=0,\ \delta F=4\pi(\delta S+J)$ may be written
also as: $\pi_2\delta \Omega =0,
\ \pi_1\delta \Omega=4\pi(\delta S+J)\otimes e_1$.

\vskip 0.5cm
\subsection{Component form of the equations}
The coordinate free written relationship (1.43)
\[
\vee(\delta\Omega,*\Omega)=\vee(\Phi,*\pi_1\Omega)\ +\ \vee(\Psi,*\pi_2\Omega)
\]
is equivalent to the following relations:
\begin{eqnarray}
&\delta F\wedge *F=\alpha^1\wedge*F,\ \delta *F\wedge **F=\alpha^4\wedge**F,\\ \nonumber
&\delta F\wedge **F+\delta *F\wedge *F=\alpha^3\wedge**F +\alpha^2\wedge *F,          
\end{eqnarray}
or, in components
\begin{eqnarray}
&F_{\mu\nu}(\delta F)^\nu=F_{\mu\nu}(\alpha ^1)^\nu,\ (*F)_{\mu\nu}(\delta*F)^\nu=(*F)_{\mu\nu}(\alpha ^4) ^\nu,\\ \nonumber
&F_{\mu\nu}(\delta*F)^\nu+(*F)_{\mu\nu}(\delta F)^\nu=(*F)_{\mu\nu}(\alpha^3) ^\nu + F_{\mu\nu}(\alpha^2) ^\nu.
\end{eqnarray}
Moving everything on the left, we get
\[
(\delta F-\alpha^1)\wedge *F=0,\ (\delta *F-\alpha^4)\wedge **F=0,\
\]
\[
(\delta F-\alpha^3)\wedge **F+(\delta *F-\alpha^2)\wedge *F=0,
\]
or in components
\[
F_{\mu\nu}(\delta F-\alpha^1)^\nu=0,\ (*F)_{\mu\nu}(\delta*F-\alpha^4)^\nu=0,\
\]
\[
F_{\mu\nu}(\delta*F-\alpha^2)^\nu+(*F)_{\mu\nu}(\delta F-\alpha^3)^\nu=0.
\]
Summing up the two equations
\[
F_{\mu\nu}(\delta F)^\nu=F_{\mu\nu}(\alpha ^1)^\nu,
\ (*F)_{\mu\nu}(\delta*F)^\nu=(*F)_{\mu\nu}(\alpha ^4) ^\nu
\]
we obtain
\[
F_{\mu\nu}(\delta F)^\nu+(*F)_{\mu\nu}(\delta*F)^\nu=\nabla_\nu Q_\mu^\nu=
F_{\mu\nu}(\alpha ^1)^\nu+(*F)_{\mu\nu}(\alpha ^4) ^\nu.
\]
This relation shows that the sum
\[
F_{\mu\nu}(\alpha ^1)^\nu+(*F)_{\mu\nu}(\alpha ^4) ^\nu
\]
is a divergence of a 2-tensor, which we denote by  $-P_\mu^\nu$. In this way
we obtain the local conservation law
\begin{equation}
\nabla_\nu (Q_\mu^\nu +P_\mu^\nu)=0.
\end{equation}
Thus, we get the possibility to introduce the full energy-momentum tensor
\[
T_\mu^\nu = Q_\mu^\nu +P_\mu^\nu,
\]
where $P_\mu^\nu$ is interpreted as {\it interaction energy-momentum tensor}.
Clearly, $P_\mu^\nu $ can not be determined uniquely in this way.

So, according to (3.2), for the 22 functions  $F_{\mu\nu}, (\alpha^i)_\mu$
we have 12 equations, and these 12 equations are differential with respect to
$F_{\mu\nu}$ and algebraic with respect to $(\alpha^i)_\mu$. Our purpose now
is to try to write down differential equations for the components of the 4
currents. The leading idea in pursuing this goal will be to establish a
correspondence between the physical concept of {\it non-dissipation} and the
mathematical concept of {\it integrability of Pfaff systems}. The suggestion
to look for such a correspondence comes from the following considerations.

Recall from the theory of the ordinary differential equations (or vector
fields), that every solution of a system of ordinary differential equations
(ODE) defines a local (with respect to the parameter on the trajectory) group
of transformations, frequently called {\it local flow}. This means, in
particular, that the motion along the trajectory is admissible in the two
directions: we have a {\it reversible} phenomenon, which has the physical
interpretation of {\it lack of losses} (energy-momentum losses are meant).
Assuming this system of ODE describes {\it fully} the process of motion of a
small piece of matter (particle), we assume at the same time, that {\it all}
energy-momentum exchanges between the particle and the outer field are taken
into account, i.e. we have assumed that {\it there is no dissipation}. In
other words, {\it the physical assumption for the lack of dissipation is
mathematically expressed by the existence of solution - local flow, having
definite group properties}. The existence of such a local flow is guaranteed
by the corresponding theorem for existence and uniqueness of a solution at
given initial conditions. This correspondence between the mathematical fact
{\it integrability} and the physical fact {\it lack of dissipation} in the
simple case "motion of a particle" , we want to generalize in an appropriate
way, having in view possible applications in more complicated physical
systems, in particular, the physical situation we are going to describe:
interaction of the field $\Omega $ with some outer field, represented in
the exchange process by the four 1-forms $\alpha ^i$. This will allow to
write down equations for $\alpha^i$ in a direct way. Of course, in the real
world there is always dissipation, and following this idea we are going to
take into account its neglecting as conditions (i.e. equations) on the
currents. As it is well known,
the mathematicians have made serious steps towards studding and formulation of
criteria for integrability of partial differential equations, so it looks
unreasonable to close eyes before the available and represented in an
appropriate form mathematical results.

On the other hand it is interesting, and may be suggesting, to note the
following . In physics we have two {\it universal} things: {\it dissipation}
and {\it gravitation}. We are going now to establish a correspondence between
the physical notion of {\it dissipation} and the mathematical concept of {\it
non-integrability}. As we know, the mathematical  non-integrability is
measured by the concept of {\it curvature}. General theory of Relativity
describes gravitation by means of Riemannian curvature. The circle will be
closed if we connect the universal property of any real physical  process
to dissipate energy-momentum with the only known so far universal
interaction in nature, the gravitation.

\vskip 0.5cm
\section{\bf The Frobenius Integrability and \\Dissipation}
\subsection{Integrable distributions and curvature}

The problem for integration of a system of partial differential equations of
the kind
\begin{equation}
\frac{\partial y^a}{\partial x^i}=f^a_i(x^k,y^b),       
\ i,k=1,...,p;\ a,b=1,...,q,
\end{equation}
where $f^a_i(x^k,y^b)$ are given functions, obeying some definite smoothness
conditions, has brought about  to the formulation of a number of concepts,
which in turn have become generators of ideas and directions, as well as have
shown an wide applicability in many branches of modern mathematics. A
particular case of the above system (nonlinear in general) of equations is
when there is only one independent variable, i.e. when all $x^i$ are reduced
to $x^1$, which is usually denoted by $t$ and the system acquires the form
\begin{equation}
\frac{dy^a}{dt}=f^a (y^b,t),\ a,b=1,...,q.               
\end{equation}
We are going to give now the system of concepts used in considering the
integrability problems for the equations (3.4) and (3.5), making use of the
geometric language of manifolds theory.

Let $X$ be a vector field on the $q$-dimensional manifold $M$ and the map
$c:I\rightarrow M$, where $I$ is an open interval in ${\cal R}$, defines a
smooth curve in $M$. Then if $X^a$ are the components of $X$ with respect to
the local coordinates $(y^1, ..., y^q)$ and the equality $c'(t)=V(c(t))$
holds for every $t\in I$, or in local coordinates,
\[
\frac{dy^a}{dt}=X^a (y^b),
\]
$c(t)$ is called {\it integral curve} of the vector field $X$. As it is seen,
the difference with (3.5) is in the additional dependence of the right side
of (3.5) on the independent variable $t$. Mathematics approaches these
situations in an unified way as follows. The product ${\cal R}\times M$ is
considered and the important theorem for uniqueness and existence of a
solution is proved: For every point $p\in M$ and point $\tau \in {\cal R}$
there exist a vicinity $U$ of $p$, a positive number $\varepsilon $ and a
smooth map $\Phi:(\tau-\varepsilon,\tau +\varepsilon)\times U\rightarrow M$,
$\Phi:(t,y)\rightarrow \varphi_t (y)$, such that for every point $y\in U$ the
following conditions are met: $\varphi_\tau (y)=y,\
t\rightarrow \varphi_t (y)$ is an integral curve of $X$, passing through the
point $y\in M$; besides, if two such integral curves of $X$ have at least one
common point, they coincide. Moreover, if $(t',y),\ (t+t',y)$ and
$(t,\varphi(y))$  are points of a vicinity $U'$ of $\{0\}\times {\cal R}$
in ${\cal R}\times M$, we have
$\varphi_{t+t'}(y)=\varphi_t (\varphi_{t'}(y))$. This last relation gives the
local group action: for every $t\in I$ we have the local diffeomorphism
$\varphi_t :U\rightarrow \varphi_t (U)$. So, through every point of $M$ there
passes only one trajectory of $X$ and in this way the manifold $M$ is
foliated to non-crossing trajectories - 1-dimensional manifolds, and these
1-dimensional manifolds define all trajectories of the defined by the vector
field $X$ system of ODE. This fibering of $M$ to nonintersecting
submanifolds, the union of which gives the whole manifold $M$, together
with uniting $t$ and $y(t)$ in one manifold, is the leading idea
in treating the system of partial differential equations (3.4), where the
number of the independent variables is more than 1, but finite. For example,
if we consider two vector fields on $M$, then through every point of $M$ two
trajectories will pass and the question: when a 2-dimensional surface,
passing through a given point can be built, and such that the representatives
of the two vector fields at every point of this 2-surface to be tangent to
the surface, naturally arises. Mathematics sets this question for
$p$-dimensional surfaces, builds the necessary concepts and proves the
corresponding theorems. These theorems formulate the criteria for
integrability of (3.4), and are known in the literature as {\it Frobenius
theorems}. For simplicity, further we consider regions of the space
${\cal R}^p \times {\cal R}^q$, but this is not essentially important since
the Frobenius theorems are local statements, so the results will hold for any
$(p+q)$-dimensional manifold.

Let $U$ be a region in ${\cal R}^p\times{\cal R}^q$, and
$(x^1,...,x^p,y^1=x^{p+1},...,y^q =x^{p+q})$ are the canonical coordinates.
We set the question: for which points $(x_0,y_0)$ of $U$ the system of
equations (3.4) has a solution $y^a=\varphi ^a(x^i)$, defined for points $x$,
sufficiently close to $x_0$ and satisfying the initial condition
$\varphi (x_0)=y_0$? The answer to this question is: for this it is necessary
and sufficient the functions $f_i^a$ on the right hand side of (3.4) to
satisfy the following conditions:
\begin{equation}
\frac{\partial f^a_i}{\partial x^j}(x,y)+
\frac{\partial f^a_i}{\partial y^b}(x,y).f^b_j(x,y)=
\frac{\partial f^a_j}{\partial x^i}(x,y)+                           
\frac{\partial f^a_j}{\partial y^b}(x,y).f^b_i(x,y).
\end{equation}
This relation is obtained as a consequence of two basic steps: first,
equalizing the mixed partial derivatives of $y^a$ with respect to $x^i$ and
$x^j$, second, replacing the obtained first derivatives of $y^a$ with
respect to $x^i$ on the right hand side of (3.4) again from the system (3.4).
This second step means, that everywhere in (3.6) $y$ are considered as
functions of $x$, i.e. there is no explicit dependence on $y$. If the
functions $f^a_i$ satisfy the equations (3.6), the system (3.4) is called
{\it completely integrable}. In order to give a coordinate free formulation
of (3.6) and to introduce the {\it curvature, as a measure for
non-integrability} of (3.4), we shall first sketch the necessary
terminology.

Let $M$ be an arbitrary $n=p+q$ dimensional manifold. At every point $x\in M$
the tangent space $T_x(M)$ is defined. The union of all these spaces with
respect to the points of $M$ defines the {\it tangent bundle}. On the other
hand, the union of the co-tangent spaces $T^*_x(M)$ defines the {\it
co-tangent bundle}. At every point now of $M$ we separate a $p$ dimensional
subspace $\Delta_x(M)$ of $T_x(M)$ in a smooth way, i.e. the map
$x\rightarrow \Delta_x$ ix smooth. If this is done we say that a
$p$-dimensional {\it distribution} $\Delta$ on $M$ is defined.
From the elementary linear algebra we know that every $p$-dimensional
subspace $\Delta_x$ of $T_x(M)$ defines unique $(n-p)=q$ dimensional subspace
$\Delta^*_x$ of the dual to $T_x(M)$ space $T^*_x(M)$, such that all elements
of $\Delta^*_x$ annihilate (i.e. send to zero) all elements of $\Delta_x$. In
this way we get a $q$-dimensional {\it co-distribution} $\Delta^*$ on $M$.
We consider those vector fields, the representatives of which at every point
are elements of the distribution $\Delta$, and those
1-forms, the representatives of which at every point are elements of the
co-distribution $\Delta^*$. Clearly, every system of $p$ independent vector
fields, belonging to $\Delta$, defines $\Delta$ equally well, and in this
case we call such a system a {\it differential $p$-system} ${\cal P}$ on $M$.
The corresponding system ${\cal P}^*$ of $q$ independent 1-forms is called
$q$-dimensional {\it Pfaff system}. Clearly, if $\alpha \in {\cal P}^*$
and $X\in {\cal P}$, then $\alpha (X)=0$.

Similarly to the integral curves of vector fields, the concept of {\it
integral manifold} of a $p$-dimensional differential system is introduced.
Namely, a $p$-dimensional submanifold $V^p$ of $M$ is called {\it integral
manifold} for the $p$-dimensional differential system ${\cal P}$, or for the
$p$-dimensional distribution $\Delta$,to which ${\cal P}$ belongs, if the
tangent spaces of $V^p$ at every point coincide with the subspaces of the
distribution $\Delta$ at this point. In this case $V^p$ is called also
integral manifold for the $q$-dimensional Pfaff system ${\cal P}^*$. If
through every point of $M$ there passes an integral manifold for ${\cal P}$,
then ${\cal P}$ and ${\cal P}^*$ are called {\it completely integrable}.

Now we shall formulate the Frobenius theorems for integrability.
\vskip 0.5cm
{\it A differential system ${\cal P}$ is completely integrable if and only if
the Lie bracket of any two vector fields, belonging to ${\cal P}$, also
belongs to ${\cal P}$}.
\vskip 0.5cm
So, if $(X_1, ..., X_p)$ generate the completely integrable
differential system ${\cal P}$, then
\begin{equation}
\left[X_i,X_j\right]=C_{ij}^k X_k,                     
\end{equation}
where the coefficients $C_{ij}^k$ depend on the point.

This criterion is not quite convenient to use because its usage presupposes
the knowledge of the functions $C_{ij}^k$. It turns out that the corresponding
criterion for Pfaff systems does not require any additional information. In
fact, let the 1-forms $(\alpha^1, ..., \alpha^q)$ define the $q$-dimensional
Pfaff system ${\cal P}^*$. The following criterion holds (the dual Frobenius
theorem):

\vskip 0.5cm
{\it The Pfaff system ${\cal P}^*$ is completely integrable if and only if}
\begin{equation}
{\bf d}\alpha^a=K^a_{bc}\alpha^b\wedge \alpha^c, \ b<c.    
\end{equation}
\vskip 0.5cm
It is easily shown, that the above equations are equivalent to the following
equations:
\begin{equation}
({\bf d}\alpha^a)\wedge \alpha^1\wedge\ .\ .\ .\wedge\alpha^a\wedge\ .\ .\ .\wedge\alpha^q=0,\ a=1,...,q,  
\end{equation}
which, obviously, do not depend on any coefficients.

When a given Pfaff system ${\cal P}^*$, or the corresponding differential
system ${\cal P}$, are not integrable, then the relations (3.7)-(3.9) {\it
are not fulfilled}.  From formal point of view this means that there is at
least one couple of vector fields $(X,Y)$, belonging to ${\cal P}$, such
that the Lie bracket $\left[X,Y\right]$ {\it does not belong} to ${\cal P}$.
Therefore, if at the corresponding point $x\in M$ we choose a basis of
$T_x(M)$ such, that the first $p$ basis vectors to form a basis of
$\Delta_x$, then $\left[X,Y\right]$ will have nonzero components with respect
to those basis vectors, which belong to some complimentary to $\Delta_x$
subspace $\Gamma_x:\ T_x(M)=\Delta_x \oplus \Gamma_x$.
If the spaces $\Gamma_x, x\in M$ are beforehand given
 we may consider the {\it projection} of the Lie bracket $\left[X,Y\right]$
onto these complimentary subspaces. Then this projection is defined uniquely
by the choice of the distribution $\Delta$, so it is a natural measure for
the non-integrability of $\Delta$. If $M$ has the structure of {\it bundle
space}, which means that a base-space $B$, $dimB=p$, is given, and a smooth
map $\pi:\ M\rightarrow B$ of maximal rank, i.e. $rank(d\pi)=p<n$, is given,
then the
subspaces $V_x=Ker(d\pi)_x\subset T_x(M)$ are naturally separated. It is
easily shown that these subspaces, called usually {\it vertical}, form an
integrable distribution. If we orient our interest towards distributions
$\Delta(M)$, which are complimentary to vertical distributions and are
usually called {\it horizontal}, then the non-integrability of $\Delta (M)$
will be determined entirely by the vertical projection $v:T(M)\rightarrow
V(M)$, defined by the definition of $\Delta (M)$ and considered on the
various Lie brackets of horizontal vector fields. It is clear now, that the
{\it curvature} ${\cal K}$ of the horizontal distribution $\Delta$ is defined
by
\begin{equation}
{\cal K}(X,Y)=v([X,Y]), \ X,Y\in \Delta.                     
\end{equation}
It is also clear, that the curvature ${\cal K}$ is a 2-form on $M$ with
values in the tangent bundle $T(M)$, and ${\cal K}$ is reduced to identity on
Lie brackets of vertical vectors. In fact, if $f$ is a smooth function and
$X,Y$ are two horizontal vector fields, then
\[
{\cal K}(X,fY)=v([X,fY])=v(f[X,Y]+X(f)Y)=
\]
\[
=fv([X,Y])+X(f)v(Y)=f{\cal K}(X,Y)
\]
because $v(Y)=0,\ Y-horizontal$.

The corresponding $q$-dimensional co-distribution $\Delta^*$ (or Pfaff system
${\cal P}^*)$ is defined locally by the 1-forms $(\theta^1,...,\theta^q)$,
such that $\theta^a(X)=0$ for all horizontal vector fields $X$. In this case
the 1-forms $\theta^a$ are called {\it vertical}, and clearly, they depend on
the choice of the horizontal distribution $\Delta$. In these terms the
non-integrability of $\Delta$ means
\[
{\bf d}\theta^a \neq K^a_{bc}\theta^b\wedge \theta^c, \ b<c.
\]
This non-equality means that at least one of the 2-forms ${\bf d}\theta^a$ is
not vertical, i.e. it has a nonzero horizontal projection
$H^*{\bf d}\theta^a$, which means that it does not annihilate all horizontal
vectors. In these terms it is naturally to define the curvature by
\[
H^*{\bf d}\theta ^a={\bf d}\theta ^a-K^a_{bc}\theta^b\wedge \theta^c, \ b<c.
\]
Further we shall see how this picture is defined by the equations (3.4).

Let's begin with the remark that the consideration of bundle spaces only,
is not a limitation and does not bounds the results, because we want through
every point of $M$ to pass locally only one integral manifold of a given
horizontal distribution and this integral manifold will be diffeomorphic to an
open set in the base manifold $B$. And this is just what is guaranteed by the
bundle structure of $M$: every point $b\in B$ has an open vicinity $U$, and
$\pi^{-1}(U)$ is diffeomorphic to the direct product $U\times N$, where $N$
is a $q$-dimensional manifold, called {\it standard fiber}. This bundle
structure allows canonical (or bundle adapted) local coordinates $(x^i,y^a)$
to be introduced, reflecting the local-product nature of $M$: $x^i=z^i\circ
\pi$, where $z^i$ are local coordinates on $U\subset B$, and $y^a$
are local coordinates on $N$. In these coordinates a local basis of the
vertical vector fields is given by
\[
\frac{\partial}{\partial y^a},\ a=p+1,...,p+q=n.
\]
Now, making use the equations of the system (3.4), i.e. the functions
$f^a_i$, we have to define the horizontal spaces at every point of
$\pi^{-1}(U)\subset M$, i.e. local linearly independent vector fields $X_i,
i=1,...,p$. The definition is
\begin{equation}
X_i=\frac{\partial}{\partial x^i}+f^a_i\frac{\partial}{\partial y^a}. 
\end{equation}
The corresponding Pfaff system shall consist of 1-forms $\theta^a,
a=p+1,...,p+q$ and is defined by
\begin{equation}
\theta^a=dy^a-f^a_i dx^i.                            
\end{equation}
In fact,
\[
\theta^a (X_i)=dy^a\left(\frac{\partial}{\partial x^i}\right)
+dy^a\left(f^b_i \frac{\partial}{\partial y^b}\right)-
f^a_j dx^j \left(\frac{\partial}{\partial x^i}\right)-
f^a_j dx^j\left(f^b_i\frac{\partial}{\partial y^b}\right)=
\]
\[
=0+f^b_i\delta_b^a -f^a_j\delta_i^j-0=0.
\]
{\it Remark}. The coordinate 1-forms $dy^a$ are not vertical with respect to
the so defined horizontal distribution.

In this way in the bundle-adapted coordinate systems the system (3.4) defines
unique horizontal distribution. On the other hand, if a horizontal
distribution is given and the corresponding vertical Pfaff system admits at
least i basis, then this basis may be chosen of the kind (3.12) always.
Moreover, in this kind it is unique. In fact, let $(\theta'^1,...,\theta'^q)$
be any local basis of $\Delta^*$. Then in the adapted coordinates we'll have
\[
\theta'^a=A^a_b dy^b+B^a_i dx^i,
\]
where $A^a_b, B^a_i$ are functions on $M$. We shall show that the matrix
$A^a_b$ has non-zero determinant, i.e. it is non-degenerate. Assuming the
opposite, we could find scalars $\lambda_a$, not all of which are equal to
zero, and such that the equality $\lambda_aA^a_b=0$  holds. We multiply now
the above equality by $\lambda^a$ and sum up with respect to $a$. We get
\[
\lambda_a \theta'^a=\lambda_a B^a_idx^i.
\]
Note now that on the right hand side of this last relation we have a
horizontal 1-form, while on the left hand side we have a vertical 1-form.
This is impossible by construction, so our assumption is not true, i.e. the
inverse matrix $(A^a_b)^{-1}$ exists, so multiplying on the left
$\theta'^a$ by $(A^a_b)^{-1}$ and putting $(A^a_b)^{-1} \theta'^b=\theta^a$
we obtain
\[
\theta^a=dy^a+(A^a_b)^{-1}B^b_i dx^i.
\]
We denote now $(A^a_b)^{-1}B^b_i=-f^a_i$ and get what we need. The uniqueness
part of the assertion is proved as follows. Assume there is another basis
$(\theta^a)''$ of the same kind. So, there must be a non-degenerate matrix
$C^a_b$, such that $(\theta^a)''=C^a_b\theta^b$. We get
\[
dy^a-(f^a_i)''dx^i=C^a_b dy^b-C^a_b f^b_idx^i,
\]
and from this relation it follows that the matrix $C^a_b$ is the unit one.

Let's see now the explicit relation between the integrability condition (3.6)
of the system (3.4) and the curvature  $H^*{\bf d}\theta^a$ of the defined by
this system horizontal distribution. We obtain
\[
{\bf d}\theta^a=-\frac{\partial f^a_i}{\partial x^j} dx^j\wedge dx^i
-\frac{\partial f^a_i}{\partial y^b}dy^b\wedge dx^i.
\]
In order to define the horizontal projection of ${\bf d}\theta^a$ it is
necessary to separate the horizontal part of $dy^a$. Since $\theta^a$ are
vertical, from their explicit form is seen that the horizontal part of $dy^a$
is just $f^a_idx^i$. That's why for the curvature $H^*{\bf d}\theta^a$ we get
\begin{equation}
H^*{\bf d}\theta^a=\left(\frac{\partial f^a_i}{\partial x^j}-
\frac{\partial f^a_j}{\partial x^i}+                                   
\frac{\partial f^a_i}{\partial y^b}f^b_j-
\frac{\partial f^a_j}{\partial y^b}f^b_i\right)dx^i\wedge dx^j,\ i<j.
\end{equation}
It is clearly seen that the integrability condition (3.6) coincides with the
requirement for zero curvature. The replacing of $dy^a$, making use of the
system (3.4), in order to obtain (3.6), acquires now the status of "horizontal
projection".

We verify now that the curvature, defined by $v(\left[X_i,X_j\right])$ gives
the same result.
\[
v([X_i,X_j])=v\left(\left[\frac{\partial}{\partial x^i}+
f^a_i\frac{\partial}{\partial y^a},\frac{\partial}{\partial x^j}+
f^b_j\frac{\partial}{\partial y^b}\right]\right)=
\]
\[
=v\left(\frac{\partial f^b_j}{\partial x^i}\frac{\partial}{\partial y^b}-
\frac{\partial f^a_i}{\partial x^j}\frac{\partial}{\partial y^a}+
f^a_i\frac{\partial f^b_j}{\partial y^a}\frac{\partial}{\partial y^b}-
f^b_i\frac{\partial f^a_i}{\partial y^b}\frac{\partial}{\partial y^a}\right)=
\]
\[
=\left(\frac{\partial f^a_j}{\partial x^i}-\frac{\partial f^a_i}{\partial x^j}+
\frac{\partial f^a_j}{\partial y^b}f^b_i -
\frac{\partial f^a_i}{\partial y^b}f^b_j\right)\frac{\partial}{\partial y^a}.
\]
The interchange of the indices $i$ and $j$ does not impact the equivalence
of this result to the above obtained for $H^*{\bf d}\theta^a$.

\vskip 0.5cm
\subsection{Physical interpretation}
As it was noticed in the preceding subsection we want to connect the
physical concept of {\it dissipation} with the mathematical concept of {\it
Frobenius non-integrability of Pfaff systems}. The availability of a well
defined mathematical quantity as the {\it curvature}, which has been
extensively used from the beginning of this century in
mathematics (differential geometry and differential topology) and theoretical
physics (General Relativity (GR) and Gauge theories) makes the things more
attractive in view of its wider use in physics. In General Relativity
curvatures of Riemannean connections are used, and because of the stress on
the metric tensor as a potential of the curvature, there has not been paid
enough attention to the original meaning of the curvature, namely as a
{\it measure for non-integrability}. Moreover, the definitive and physically
not motivated assumption of the Riemannean curvature as a mathematical
adequate of the gravitational field in GR does not contribute
to a full comprehension of why the computations in the theory meet the
experiment in the Solar system (and even out of it) so well. This
circumstance, being so charming in the early days of the theory, may generate
some hesitations, because 80 years seem to be long enough time for the
clarification of this fundamental for GR problem. Together with the well known
"energy-momentum problems", this may lower the authority of GR.

In gauge theories the curvature, considered as generated by  connections on
principle bundles for some groups and their representations, leading to
linear connections in vector bundles, is also a {\it leading concept}. For
example, the energy-momentum tensor in these theories is a quadratic
expression of the curvature, and the significance of energy and momentum in
modern classical and quantum gauge theories and in the whole physics at all
is out of any doubt. Except electrodynamics, where we have much enough
experience, in other gauge theories there is also no enough motivation for
using namely the curvature as a mathematical adequate of a physical field.
The consideration of connections as basic mathematical objects in physical
theories we do not consider as sufficiently legalizing move for the
introduction of curvatures, although the connections define derivative laws
for the sections of vector bundles. In our opinion, a more basic analysis of
the question why the curvature works well in physical theories from the point
of view of the Frobenius integrability theory would contribute to a more
complete and detailed  understanding of this important moment in the field
theories.

In mathematics the curvature defines those conditions, at which, given
differential relations determine integral objects, or what are the obstacles
for building these integral objects. In physics, from Newton's time on, a
basic quantity is the {\it force}, i.e. the quantity of transferred
energy-momentum from the outer field to the object under observation
(usually test particle(s)) in an unit 4-volume. From the point of view of the
outer field this means {\it loss} of energy-momentum, and that's why it does
not seem reasonable to expect a Frobenius integrability for the equations,
describing the outer field, if these equations do not take into account these
losses. It is quite illogical and unreasonable to expect that the
expressed through differential relationships properties of a physical object
should define it entirely (or integrally) if there is some "flowing out"
of substance towards other physical objects, if  this "flowing out" takes
some energy-momentum from it and carries it to the other objects. In other
words, the interaction, if it is not fully described, may violate the
existing for any extended object connection between its differential (local)
and integral properties. And a fully described interaction means to say, and
to take into account mathematically, where the energy-momentum losses go to.
These losses are just the {\it force}, applied onto the other object(s).

The curvature of a given differential system ${\cal P}$ measures
mathematically (at a given point and in a differential way) something very
close to a "flowing out" of ${\cal P}$, namely, it determines {\it what parts
of the Lie brackets of vector fields in ${\cal P}$ belong no more to
${\cal P}$}. We may say, that these parts of the Lie brackets, namely the
vertical projections of the Lie brackets, define local flows, directed out of
${\cal P}$. Therefore, from physical point of view it is natural to choose
curvatures as measuring quantities of the external exertion on test
particles in outer fields. The physical measure of this exertion is a change
in the {\it universally conserved quantities energy and momentum} (of the
particle).

As far as test (point-like) particles in outer fields are considered the
usage of curvature as a direct participant in describing the energy-momentum
transfer seems natural, since we are not interested in the dynamics of the
external fields. "No change in the energy-momentum of the particle" is equal
to "no presence of external fields", i.e. the force is zero. This approach
works no more in case of local interaction of two (or more) continuous
objects, i.e.  two fields, where together with the local energy-momentum
transfer we are interested also in the dynamics of the two interacting
fields. So, the local dynamical changes of any of the two interacting fields
should depend on 2 things: the proper dynamical character of any of the
fields and the kind of interaction. These two components of the physical
system should be consistent, and when there is no interaction, the
mathematical expression for zero energy-momentum transfer should also be
consistent with the proper dynamics of any of the fields, considered now as
free. Because of the local character of the interaction, derivatives of the
fields' components participate in the corresponding mathematical expression.
In this way, putting the interaction expression equal to zero, we obtain
general enough differential equations for the free fields. And if the fields
are mathematically described by curvatures we obtain differential equations
for the curvatures.

Let's consider now a more complicated situation, when the interaction
energy-momentum flows along {\it several} channels. Every such channel may
or may not produce dissipation. The produced by some of the channels
dissipation of energy-momentum could be utilized or not by some of the rest
channels. Our proposal to interpret mathematically the non-zero dissipation
by Frobenius non-integrability would imply in such cases, that some of the
subdistributions of the initial distribution may be nonintegrable, but these
nonintegrable subdistributions may participate in integrable subdistributions
of a higher dimension. The same can be said about the corresponding Pfaff
systems. This establishes some kind of hierarchy among the subsystems of the
Pfaff systems and leads to a more complete study of the initially assumed
Pfaff system, defining the energy-momentum interaction channels.

We shall see
in the next section that the considered examples-solutions of our equations,
representing a (3+1)-interpretation of all known (1+1)-soliton solutions
illustrate the above outlined idea: availability of differential and integral
conservation laws, non-integrability of 1-dimensional Pfaff subsystem,
integrability of all 2-dimensional Pfaff systems. This is in the general
spirit of our approach, in which the leading ideas are the extended character
of the real objects and the interrelation between their local and integral
properties.

We are not going to consider here non-utilized dissipations at the highest
possible level. This would lead us far behind our purposes.

After these mathematical and interpretational deviations we go back to EED.
First we note that our base manifold, where all fields and operations are
defined, is the simple 4-dimensional Minkowski space. According to our
equations (1.43) the medium reacts to the field $\Omega $ by means of the two
${\cal R}^2$-valued 1-forms $\Phi$ and $\Psi$. So, we obtain four
${\cal R}$-valued 1-forms $\alpha^1, \alpha^2, \alpha^3, \alpha^4$. Because
of the 4-dimensions of Minkowski space and the kind of the equations (3.9) it
is easily seen that only 1-dimensional and 2-dimensional Pfaff systems may be
of interest from the Frobenius integrability point of view. All Pfaff systems
of higher dimension are trivially integrable. Note also that closed 1-forms
and linearly dependent 1-forms define always integrable Pfaff systems in an
obvious way.

The integrability equations for 1-dimensional Pfaff systems are
\begin{equation}
{\bf d}\alpha^i\wedge \alpha^i=0,\ i=1,2,3,4.                  
\end{equation}
Every of the 4 equations (3.14) is equivalent to 4 scalar nonlinear equations
for the components of the corresponding 1-form. We note, that the solutions
of (3.14), as well as the solutions of the general integrability equations
for a $p$-dimensional Pfaff system are determined up to a scalar multiplier,
i.e. if $\alpha^i$ are solutions, then $f_i.\alpha^i$ (no summation over $i$),
where $f_i$ are smooth functions, are also solutions.

In case of 2-dimensional Pfaff systems $(\alpha^i,\alpha^j)$, defined by four
1-forms, their maximal number is 4.3=12. The Frobenius equations read
\begin{equation}
{\bf d}\alpha^i\wedge \alpha^i\wedge \alpha^j=0,\ i\neq j.     
\end{equation}
We have here 12 nonlinear equations for the 16 components of $\alpha^i$.
Clearly, these equations (3.15) make some interest only if the corresponding
$\alpha^i$, the exteriour differential ${\bf d}\alpha^i$ of which
participates in (3.15), does not satisfy (3.14).

After these general remarks we pass to finding explicit solutions of (1.43)
with non-zero currents.

\vskip 0.5cm
\section{\bf Explicit Solutions with Nonzero Currents}
\subsection{Choice of the anzatz and finding the solutions}
From purely formal point of view finding of a solution, whatever it is,
legitimizes the equations (1.43) and (3.15) as a consistent system. Our
purpose, however, is not purely formal, we consider it as physically
meaningful, namely, we are interested in solutions, which are {\it
physically interpretable} as models of real objects in the above commented
sense. That's why we have to meet the following. First, the solutions must
be {\it physically clear}, which means that the anzatz assumed should be
comparatively simple and its choice should be made on the base of a
preliminary analysis of the physical situation in view of the mathematical
model used.  Second, {\it it is absolutely obligatory the solutions to have
well defined integral energy and momentum}. Third, {\it to be in the spirit
of the soliton-like comprehension of the real natural objects} when the
solution is interpreted as a model of such an object. Fourth, to realize the
above given {\it physical interpretation of the Frobenius integrability
equations} and the mentioned {\it dimensional hierarchy of Pfaff systems as a
possible model of suitably chosen and intrinsically structured interrelated
processes}. Finally, it would be nice, the solutions found to be
comparatively simple and interesting as corresponding generalizations, or
extensions, of "popular" and well known solutions of "well liked" equations.
Probably not all solutions will satisfy these requirements, but there must be
such solutions, since the physical significance of our equations depends
strongly on this circumstance, to admit solutions with the above mentioned
features. Let's now get started.

The first, we take into account, is the necessary time-dependence of the
solutions, therefore, the "electric" and the "magnetic" components should
present. The simplest $\Omega $, or $(F,*F)$, meeting this requirement,
looks as follows (we use the above assumed notations):
\begin{equation}
F=-udy\wedge dz -vdy\wedge d\xi,\ *F=vdx\wedge dz + udx\wedge d\xi.
\end{equation}

When choosing the 1-forms $\alpha^i$ we shall obey the requirement, that the
"medium" {\it does not involve} in the $F\leftrightarrow *F$ exchange, so we
put
\begin{equation}
\alpha^2=\alpha^3=0.
\end{equation}
The rest two 1-forms , $\alpha^1$ and $\alpha^4$ must be {\it linearly
independent}. This would be guaranteed if we choose one of them to be {\it
time-like}, and the other to be {\it space-like}. The simplest time-like
1-form is of course $\alpha=A(x,y,z,\xi)d\xi$, moreover, any such 1-form
defines {\it integrable} 1-dimensional Pfaff system:
\[
{\bf d}\alpha\wedge \alpha=\left(A_x dx\wedge d\xi +A_y dy\wedge d\xi +
A_zdz\wedge d\xi\right)\wedge Ad\xi=0.
\]
The choice of the last one, denoted by $\beta$, reads
$\beta=\beta_1dx+\beta_2dy+\beta_3dz$. Clearly, $\beta^2<0 $, and
$\alpha.\beta=0$, so they are independent. Now, $\beta$ participates in the
equations through the expression $\beta\wedge *F$, and from the explicit form
of $*F$ is seen, that the coefficient $\beta_1$ in front of $dx$ does not
take part in the equations, therefore, we put $\beta_1=0$. So, we get
\begin{equation}
\alpha^1\equiv \beta=bdy-Bdz,\ \alpha^4\equiv \alpha=Ad\xi,\
\alpha^2=\alpha^3=0.
\end{equation}
We note, that the so chosen $\beta$ does {\it not} define in general an
integrable 1-dimensional Pfaff system, so the requirement for
{\it hierarchy} of
the exchange processes, i.e.   ${\bf d}\alpha\wedge \alpha=0,\
{\bf d}\beta\wedge \beta\neq 0,\ {\bf d}\beta\wedge \beta\wedge \alpha=0$,
is obeyed at a definite level.

At these conditions our equations
\[
\delta *F\wedge F=\alpha\wedge F,\ \delta F\wedge* F=\beta\wedge * F,\
\delta *F\wedge *F-\delta F\wedge F=0,
\]
\[
{\bf d}\alpha\wedge \alpha\wedge \beta=0,\
{\bf d}\beta\wedge \alpha\wedge \beta=0
\]
take the form: $\delta *F\wedge *F-\delta F\wedge F=0$ is reduced to
\[
-vu_y+uv_y=0,\  -uv_x +vu_x=0,
\]
the Frobenius equations  ${\bf d}\alpha\wedge \alpha\wedge \beta=0,\
{\bf d}\beta\wedge \alpha\wedge \beta=0$ reduce to
\[
\left(-b_xB+B_xb\right).A=0,
\]
$\delta *F\wedge F=\alpha\wedge F$ is reduced to
\[
u\left(u_\xi-v_z\right)=0,\ v\left(u_\xi-v_z\right)=0,\ uu_x-vv_x=Au,
\]
$\delta F\wedge* F=\beta\wedge *F$ reduces to
\[
v\left(v_\xi-u_z\right)=-bv,\ u\left(v_\xi-u_z\right)=-bu,\ uu_y-vv_y=Bu.
\]
In this way we obtain 7 equations for 5 unknown functions $u,v,A,B,b$. The
equations $-vu_y+uv_y=0,\  -uv_x +vu_x=0$ have the following solution:
\[
u(x,y,z,\xi)=f(x,y)U(z,\xi),\ v(x,y,z,\xi)=f(x,y)V(z,\xi).
\]
That's why
\[
AU=f_x\left(U^2-V^2\right),\ BU=f_y\left(U^2-V^2\right),\
f.\left(U_z-V_\xi\right)=b,\ U_\xi-V_z=0.
\]
Now the equation $B_xb-Bb_x=0$ takes the form
\[
ff_{xy}=f_xf_y,
\]
and the general solution of this equation is $f(x,y)=g(x)h(y)$.
Besides, the equation $gh\left(V_\xi-U_z\right)=-b$ requires
$b(x,y,z,\xi)=g(x)h(y)b^{o}(z,\xi)$, so we get
\[
V_\xi-U_z=-b^o.
\]
The relations obtained show how to build a solution of this class. Namely,
first, we choose the function $V(z,\xi)$, then we determine the function
$U(z,\xi)$ by
\[
U(z,\xi)=\int{V_z d\xi} +l(z),
\]
where $l(z)$ is an arbitrary function, which may be assumed equal to $0$.
After that we define $b^{o}=U_z-U_\xi$. The functions $g(x)$ and $h(y)$ are
arbitrary, and for $A$ and $B$ we find
\[
A(x,y,z,\xi)=g'(x)h(y)\frac{U^2-V^2}{U},\ \ B(x,y,z,\xi)=g(x)h'(y)\frac{U^2-V^2}{U}.
\]
In this way we obtain a family of solutions, which is parametrized by one
function $V$ of the two variables $(z,\xi)$ and two functions $g(x),\ h(y)$,
each depending on one variable.

For $*(\alpha\wedge F+\beta\wedge *F)$ we obtain
\[
*(\alpha\wedge F+\beta\wedge *F)=Audx-Budy-budz-bvd\xi=
\]
\[
=\frac12 (U^2-V^2)\left[(gh)^2\right]_x dx-
\frac12 (U^2-V^2)\left[(gh)^2\right]_y dy-
\]
\[
-(gh)^2\left(\int{Ub^{o}dz}\right)_{z}dz-(gh)^2\left(\int{Vb^{o}d\xi}\right)_{\xi}d\xi=
=-\left\{\frac{\partial}{\partial x^\nu}P_\mu^\nu\right\}dx^\nu,
\]
where the interaction energy-momentum tensor is defined by the matrix
\[
P_\mu^\nu=\left\|\matrix{
-\frac12(gh)^2 Z  &0                &0                     &0                      \cr
0                 &\frac12(gh)^2 Z  &0                     &0                      \cr
0                 &0                &(gh)^2\int{Ub^{o}dz}  &0                      \cr
0                 &0                &0                     &(gh)^2\int{Vb^{o}d\xi}
\cr}\right\|,
\]
and the notation $Z\equiv U^2-V^2$ is used. For the full energy tensor
\newline
$T_\mu^\nu=Q_\mu^\nu +P_\mu^\nu$ we obtain
\[
T_3^3=(gh)^2\left[\int{Ub^{o}dz} - \frac12(U^2+V^2)\right],
\]
\[
T_3^4=-T_4^3=(gh)^2 UV,
\]
\[
T_4^4=(gh)^2\left[\int{Vb^{o}d\xi}+\frac12(U^2+V^2)\right],
\]
and all other components are zero.

\vskip 0.5cm
\subsection{Examples}

In this subsection we consider some of the well known and well studied
(1+1)-dimensional soliton equations as generating procedures for choosing the
function $V(z,\xi)$, and only the 1-soliton solutions will be explicitly
elaborated. Of course, there is no anything standing in our way to consider
other (e.g. multisoliton) solutions. We do not give the corresponding
formulas just for the sake of simplicity.

We turn to the soliton equations mainly because of two reasons. First, most
of the solutions have a clear physical sense in a definite part of physics
and, according to our opinion, they are sufficiently attractive for models of
real physical objects with internal structure. Second, the soliton solutions
describe free and interacting objects with {\it no dissipation} of energy and
momentum, which corresponds to our interpretation of the Frobenius
integrability equations, as we explained in the preceding (sub)sections.

\vskip 0.5cm
\noindent{\it 1. Nonlinear equation Klein-Gordon}.
In this example we define our functions $U$ and $V$ through the derivatives
of the function $f(z,\xi)$ in the following way: $U=f_z,\ V=f_\xi$. Then the
equation $U_\xi-V_z=f_{z,\xi}-f_{\xi z}=0$ is satisfied automatically, and
the equation $U_z-V_\xi=b^{o}$ takes the form $f_{zz}-f_{\xi\xi}=b^{o}$. Since
$b^o$ is unknown, we may assume $b^o=b^o(f)$, which reduces the whole
problem to solving the general nonlinear Klein-Gordon equation when $b^o$
depends nonlinearly on $f$. Since in this case $V=f_\xi$ we have
\[
\int{Vb^o(f)d\xi}=\int{f_\xi b^o(f)d\xi}=
\int{\left[\frac{\partial}{\partial \xi}\int{b^o(f)df}\right]d\xi}=
\int{b^o(f)df}.
\]
For the full energy density we get
\[
T_4^4=\frac12(gh)^2\left\{f_z^2+f_\xi^2+2\int{b^o(f)df}\right\}.
\]
Choosing $b^o(f)=m^2 sin(f) $ we get the well known and widely used in physics
Sine-Gordon equation, and together with this, we can use {\it all} solutions
of this nonlinear equation. When we consider the (3+1) extension of
the soliton solutions of this equation, the functions $g(x)$ and $h(y)$ have
to bo localized too. The determination of the all 5 functions in our approach
is straightforward, so we obtain a (3+1)-dimensional version of the soliton
solution chosen. As it is seen from the above given formula, the integral
energy of the solution differs from the energy of the corresponding
(1+1)-dimensional solution just by the $(x,y)$-localizing factor $(gh)^2$.

For the 1-soliton solution (kink) we have:
\[
f(z,\xi)=4arctg\left\{exp\left[\pm \frac{m}{\gamma}(z-\frac wc
\xi)\right]\right\},\ \gamma=\sqrt{1-\frac{w^2}{c^2}}
\]
\[
U(z,\xi)=f_z=\frac{\pm 2m}{\gamma
ch\left[\pm\frac{m}{\gamma}\left(z-\frac{w}{c} \xi \right)\right]},\
V(z,\xi)=f_\xi=\frac{\pm 2mw}{c\gamma
ch\left[\pm\frac{m}{\gamma}\left(z-\frac{w}{c} \xi \right)\right]},
\]
\[
A=g'(x)h(y)\frac{\pm 2m\gamma}{ch\left[\pm\frac{m}{\gamma}\left(z-\frac{w}{c} \xi \right)\right]},\
B=g(x)h'(y)\frac{\pm 2m\gamma}{ch\left[\pm\frac{m}{\gamma}\left(z-\frac{w}{c} \xi \right)\right]},\
\]
\[
b^o=U_z-V_\xi=\frac{-2m^2 sh\left[\pm\frac{m}{\gamma}\left(z-\frac{w}{c} \xi \right)\right]}
{ch\left[\pm\frac{m}{\gamma}\left(z-\frac{w}{c} \xi \right)\right]},\
T_4^4=\frac{(gh)^2 4m^2}{\gamma^2 ch^2\left[\pm\frac{m}{\gamma}\left(z-\frac{w}{c} \xi \right)\right]}
\]
and for the 2-form $F$ we get
\[
F=-\frac{\pm2mg(x)h(y)}{\gamma ch\left[\pm\frac{m}{\gamma}(z-\frac wc \xi)\right]}dy\wedge dz+
\frac wc \frac{\pm2mg(x)h(y)}{\gamma ch\left[\pm\frac{m}{\gamma}(z-\frac wc \xi)\right]}dy\wedge d\xi .
\]
In its own frame of reference this soliton looks like
\[
F=-\frac{\pm2mg(x)h(y)}{ ch(\pm mz)}dy\wedge dz.
\]
From this last expression and from symmetry considerations, i.e. at
homogeneous and isotropic medium, we come to the most natural choice of the
functions $g(x)$ and $h(y)$:
$$
g(x)=\frac{1}{ch(mx)},\quad h(y)=\frac{1}{ch(my)}.
$$

\vskip 0.5cm
2.{\it Korteweg-de Vries equation.} This nonlinear equation has the following
general form:
\[
f_\xi+a_1ff_z+a_2f_{zzz}=0,
\]
where $a_1$ and $a_2$ are 2 constants. The well known 1-soliton solution is
\[
f(z,\xi)=\frac{a_o}{ch^2\left[\frac zL -\frac{w}{cL}\xi\right]},\
L=2\sqrt{\frac{3a_2}{a_o a_1}},\ w=\frac{ca_oa_1}{3}.
\]
We choose $V(z,\xi)=f(z,\xi)$ and get
\[
U=-\frac{a_oc}{w}\frac{1}{ch^2\left[\frac zL -\frac{w}{cL}\xi\right]},\
b^o=U_z-V_\xi=\left(\frac{c}{Lw}-\frac wc\right)\frac{2a_o}{ch^3\left[\frac zL -\frac{w}{cL}\xi\right]},
\]
\[
T_4^4=(gh)^2\frac{a_o^2c^2(1+L)}{2w^2Lch^4\left[\frac zL -\frac{w}{cL}\xi\right]}.
\]
\vskip 0.5cm
3. {\it Nonlinear Schroedinger equation}. In this case we have an equation
for a complex-valued function, i.e. for two real valued functions. The
equation reads
\[
if_\xi+f_{zz}+|f|^2 f=0,
\]
and its 1-soliton solution, having oscillatory character, is
\[
f(z,\xi)=2\beta^2\frac{exp\left[-i\left(2\alpha z+4(\alpha^2-\beta^2)\xi -\theta\right)\right]}
{ch\left(2\beta z+8\alpha\beta \xi -\delta\right)}.
\]
The natural substitution $f(z,\xi)=\sqrt{\rho}.exp(i\varphi)$ brings this
equation to the following two equations
\[
\rho_\xi + (2\rho \varphi_z)_z=0,\
4\rho + \frac{2\rho\rho_{zz}-\rho_z^2}{\rho^2}=4(\varphi_\xi+\varphi_z^2).
\]
For the 1-soliton solution we get
\[
\rho=\frac{4\beta^4}{ch^2\left(2\beta z+8\alpha\beta \xi -\delta\right)},\
\varphi=-\left[2\alpha z+4(\alpha^2-\beta^2)\xi-\theta\right].
\]
We put $U=\rho, \ V=-2\rho\varphi_z$ and obtain
\[
U=\rho=\frac{4\beta^4}{ch^2\left(2\beta z+8\alpha\beta \xi -\delta\right)},\
V=-2\rho\varphi_z=\frac{16\alpha\beta^4}{ch^2\left(2\beta z+8\alpha\beta \xi -\delta\right)},
\]
\[
A=g'(x)h(y)\frac{4\beta^4(1-16\alpha^2)}{ch^2\left(2\beta z+8\alpha\beta \xi -\delta\right)},\
B=g(x)h'(y)\frac{4\beta^4(1-16\alpha^2)}{ch^2\left(2\beta z+8\alpha\beta \xi -\delta\right)},
\]
\[
b^o=\frac{16\beta^3(16\alpha^2\beta^2-1)sh(2\beta z+8\alpha\beta \xi -\delta)}
{ch^3(2\beta z+8\alpha\beta \xi -\delta)},
\]
\[
T_4^4=(gh)^2\frac{16\beta^8}{ch^4(2\beta z+8\alpha\beta \xi -\delta)}.
\]
We note that the solution of our equations obtained has no the oscillatory
character of the original Schroedinger 1-soliton solution, it is a
(3+1)-localized running wave and moves as a whole with the velocity
$4c\alpha$.

\vskip 0.5cm
4.{\it Boomerons}. The system of differential equations, having soliton
solutions, known as {\it boomerons}, is defined by the following functions:
${\bf K}:{\cal R}^2\rightarrow {\cal R}^3,\ H:{\cal R}^2\rightarrow {\cal
R}$, and besides, two constant 3-dimensional vectors ${\bf r}$ and ${\bf s}$,
where ${\cal s}$ is a unit vector: $|{\bf s}|=1$. The equations have the form
\[
H_\xi-{\bf s}.{\bf K}_z =0,\  {\bf K}_{z\xi}=H_{zz}{\bf s}+{\bf r}\times {\bf K}_z
-2{\bf K}_z\times({\bf K}\times{\bf s}).
\]
Now we have to define our functions $U(z,\xi),\ V(z,\xi)$ and $b^o(z,\xi)$.
The defining relations are:
\[
U=H_z=|{\bf s}|^2 H_z,\ V={\bf s}.{\bf K}_z=({\bf s.K})_z,\
b^o=-{\bf s}.\left[{\bf r}\times {\bf K}_z-2{\bf K}_z\times({\bf K}\times {\bf s})\right].
\]
Under these definitions our equations $U_\xi-V_z=0,\ V_\xi-U_z=-b^o$ look as
follows:
\[
\left[H_z-{\bf s}.{\bf K}_z\right]_z=0,\
{\bf s}.\left[{\bf K}_{z\xi}-H_{zz}{\bf s}-{\bf r}\times{\bf K}_z+
2{\bf K}_z\times({\bf K}\times {\bf s})\right]=0.
\]
It is clear that every solution of the "boomeron" system determines a
solution of our system of equations according to the above given rules and
with the multiplicative factors $g(x)$ and $h(y)$. The two our functions
$A(z,\xi)$ and $B(z,\xi)$ are easily then computed.

\vskip 0.5cm
Following this procedure we can generate a solution to our system of
equations by means of every solution to any (1+1)-soliton equation, as well
to compute the corresponding conserved quantities. It seems senseless to give
here these easily obtainable results. The richness of this comparatively
simple family of solutions, as well as the availability of corresponding
correctly defined integral conserved quantities, are obvious and should not
be neglected. In particular, it would be interesting to analyze the abilities
of the {\it breather}-solutions of some soliton equations as possible models
of bounded systems of the type of {\it hydrogen atom}. It is worth to note
that in this approach the classical quantity {\it potential} is out of need.
The proton and the electron participate equally in rights as relatively
isolated subsystems of a more general dynamical system. The discrete character
of the energy spectra trivially follows from the fact, that the transitions
among the various stationary states of the more general system are caused by
creation or annihilation of photons, taking into the system or out of it
the corresponding conserved quantities. As for the quantitative
description of this spectra, probably, it will be achieved by a
corresponding choice of the integrability constants.

\newpage
{\Large Literature}
\vskip 0.5cm

Historical analysis of the development of CED:\\

{\bf Sir Edmund Whittaker}, {\it A HISTORY OF THE THEORIES OF \newline
AETHER AND ELECTRICITY}, vol.1: The Classical Theories, vol.2: The Modern
Theories 1900-1926, Dover Publication Inc., New York, 1989.\\

A good book on CED:\\

{\bf J.D.Jackson}, {\it CLASSICAL ELEKTRODYNAMICS}, John Wiley and Sons, Inc.,
New York-London, 1962., also its new addition, and all cited therein
monographs on CED.\\

A systematic review of the standard nonlinearization of CED through an
analogy with the electrodynamics of the continuous media:\\

{\bf J.Plebansky}, {\it LECTURES ON NONLINEAR ELECTRODYNAMICS},
Nordita, 1970.\\

Attempts for soliton-like approaches in CED:\\

{\bf A.Lees}, Phyl.Mag., 28 (385), 1939

{\bf N.Rosen}, Phys.Rev., 55 (94), 1939

{\bf P.Dirac}, Proc.Roy.Soc., A268 (57), 1962

{\bf H.Schiff}, Proc.Roy.Soc., A269 (277), 1962

{\bf H.Schiff}, Phys.Rev., 84 (1), 1951

{\bf J.Synge}, RELATIVITY:THE SPECIAL THEORY, North Holl.,1958

{\bf D.Finkelstein, C.Misner}, Ann.Phys., 6 (230), 1959

{\bf D.Finkelstein}, Journ.Math.Phys., 7 (1218), 1966

{\bf J.P.Vigier}, Found. of Physics, vol.21 (1991), 125.

{\bf T.Waite}, Annales de la Fondation Louis de Broglie, 20 No.4, 1995.\\

A specially recommended paper (together with the citations therein by the
choice of the reader):\\

{\bf G.Hunter, R.L.P.Wadlinger}, "Photons and Neutrinos as Electromagnetic
Solitons", Physics Essays,vol.2,No.2,p.158,(1989)\\

Frobenious Integrability:\\

{\bf H.Cartan}, {\it CALCUL DIFFERENTIEL. FORMES DIFFERENTIELLES},
Herman, Paris, 1967

{\bf C.Godbillon}, {\it GEOMETRY DIFFERENTIELLE ET MECANIQUE\\ ANALYTIQUE},
Herman, Paris 1969\\

Solitons in classical and quantum field theory:\\

{\bf R.Rajaraman}, {\it SOLITONS AND INSTANTONS},
North-Holland, \\ Amsterdam-New York-Oxford, 1982

{\bf S.Coleman}, in {\it NEW PHENOMENA IN SUBNUCLEAR PHYSICS}-part A,
p.297, ed. A.Zichichi, Geneva, Switzerland.

{\bf C.Rebbi, C.Soliani}, /editors/, {\it SOLITONS AND PARTICLES}, World
Scientific, 1984\\

Publications of the author on EED:\\

{\bf S.Donev}, {\it A particular non-linear generalization of Maxwell
equations admitting spatially localized wave solutions},Compt.Rend.Bulg.
Acad.Sci., \newline vol.34, No.4, 1986

{\bf S.Donev}, {\it A covariant generalization of sine-Gordon equation on \newline
Minkowski space-time}, Bulg.Journ.Phys., vol.13, (295), 1986

{\bf S.Donev}, {\it Autoclosed differential forms and (3+1)-solitary waves},\newline
Bulg.Journ.Phys., vol.15, (419), 1988

{\bf S.Donev}, {\it On the description of single massless quantum objects},
Helvetica Physica Acta, vol.65, (910), 1992

{\bf S.Donev, M.Tashkova}, {\it Energy-momentum directed nonlinearization
of Maxwell's pure field equations}, Proc.Roy.Soc. of London, A 443, (301),
1993

{\bf S.Donev, M.Tashkova}, {\it Energy-momentum directed nonlinearization
of Maxwell's equations in the case of a continuous medium}, Proc.Roy.Soc.
of London, A 450, (281), 1995

{\bf S.Donev, M.Tashkova}, {\it Extended Electrodynamics I, II, III};
submitted for publication.

\newpage
The author does not know about any other published or not published studies
and results following the same or similar line of nonlinear
extension of Maxwell's equations. The author would be kindly grateful for
any indication of results, having something in common with those given in
this review. Any comments and remarks, please, send to

\vskip 0.5cm
	Dr.Stoil G.Donev\\
	\indent Institute for Nuclear Research and Nuclear Energy,\\
	\indent Bulgarian Academy of Sciences,\\
	\indent 1784 Sofia, blvd. Tzarigradsko chaussee 72, BULGARIA\\
	\indent tel. office: (+359 2) 7431375; home: (+359 2) 732352\\
	\indent e-mail: sdonev@inrne.acad.bg

\tableofcontents
\addcontentsline{toc}{chapter}{\protect\numberline{4}
{\sc Literature}}

\end{document}